\documentclass[useAMS,usenatbib]{mn2e}

\usepackage[letterpaper,margin=0.635in]{geometry}
\usepackage[ansinew]{inputenc}
\usepackage{amsfonts}
\usepackage{amsmath}
\usepackage{enumerate}
\usepackage{enumitem}
\usepackage{graphicx}
\usepackage{epsfig}
\usepackage{color}
\usepackage[normalem]{ulem}
\usepackage{comment}
\usepackage{amssymb}
\usepackage{pdflscape}

\newcommand{\pz}{photo-$z$\ }
\newcommand{\pzs}{photo-$z$'s\ }

\newcommand{\magauto}{{\tt MAGAUTO}\ }
\newcommand{\magdetmodel}{{\tt MAGDETMODEL}\ }

\title{Photometric redshift analysis in the Dark Energy Survey Science Verification data}
\author[C. S\'{a}nchez et al.]{C. S\'{a}nchez$^{1}$\thanks{Direct your inquiries to csanchez@ifae.es. Author affiliations are listed at the end of this paper.}, M. Carrasco Kind$^{2}$, H. Lin$^{3}$, R. Miquel$^{1,4}$, F. B. Abdalla$^{5}$, 
\newauthor
A. Amara$^{6}$, M. Banerji$^{5}$, C. Bonnett$^{1}$, R. Brunner$^{2}$, D. Capozzi$^{7}$, A. Carnero$^{8,9}$,   
\newauthor
F. J. Castander$^{10}$, L. A. N. da Costa$^{8,9}$, C. Cunha$^{11}$, A. Fausti$^{9}$, D. Gerdes$^{12}$,  
\newauthor
N. Greisel$^{13,14}$, J. Gschwend$^{8,9}$, W. Hartley$^{6,15}$, S. Jouvel$^{5}$, O. Lahav$^{5}$, M. Lima$^{16,9}$,
\newauthor
M. A. G. Maia$^{8,9}$, P. Mart\'{i}$^{1}$, R. L. C. Ogando$^{8,9}$, F. Ostrovski$^{8,9}$, P. Pellegrini$^{8}$,
\newauthor
M. M. Rau$^{13,14}$, I. Sadeh$^{5}$, S. Seitz$^{13,14}$, I. Sevilla-Noarbe$^{17}$, A. Sypniewski$^{12}$, 
\newauthor
J. de Vicente$^{17}$, T. Abbot$^{18}$, S. S. Allam$^{19,3}$, D. Atlee$^{20}$, G. Bernstein$^{21}$, 
\newauthor
J. P. Bernstein$^{22}$, E. Buckley-Geer$^{3}$, D. Burke$^{23,11}$, M. J. Childress$^{24,25}$, 
\newauthor
T. Davis$^{26,25}$, D. L. DePoy$^{27}$, A. Dey$^{20,28}$, S. Desai$^{29,30}$, H. T. Diehl$^{3}$, P. Doel$^{5}$,  
\newauthor
J. Estrada$^{3}$, A. Evrard$^{12,31,32}$, E. Fern\'andez$^{1}$, D. Finley$^{3}$, B. Flaugher$^{3}$,  
\newauthor
J. Frieman$^{3}$, E. Gaztanaga$^{10}$, K. Glazebrook$^{33}$, K. Honscheid$^{34}$, A. Kim$^{35}$, 
\newauthor
K. Kuehn$^{36}$, N. Kuropatkin$^{3}$, C. Lidman$^{36}$, M. Makler$^{37}$, J. L. Marshall$^{27}$, 
\newauthor
R. C. Nichol$^{7}$, A. Roodman$^{23,11}$, E. S\'{a}nchez$^{17}$, B. X. Santiago$^{38,9}$, M. Sako$^{21}$, 
\newauthor
R. Scalzo$^{24}$, R. C. Smith$^{18}$, M. E. C. Swanson$^{39}$, G. Tarle$^{12}$, D. Thomas$^{7,40}$, 
\newauthor
D. L. Tucker$^{3}$, S. A. Uddin$^{33,25}$, F. Vald\'es$^{20}$, A. Walker$^{18}$,  F. Yuan$^{24,25}$, J. Zuntz$^{41}$
}

\begin{document}


\pagerange{\pageref{firstpage}--\pageref{lastpage}} \pubyear{0000}

\maketitle

\label{firstpage}

\begin{abstract}
We present results from a study of the photometric redshift
performance of the Dark Energy Survey (DES), using the early data from
a Science Verification (SV) period of observations in late 2012 and early 2013 that
provided science-quality images for almost 200
sq.~deg.~at the nominal depth of the survey. We assess
the photometric redshift (photo-$z$) performance using about 15000 galaxies with
spectroscopic redshifts available from other
surveys. These galaxies are used,
in different configurations, as a calibration sample, and photo-$z$'s
are obtained and studied using most of the existing photo-$z$ codes. A
weighting method in a multi-dimensional color-magnitude space is
applied to the spectroscopic sample in order to evaluate the photo-$z$
performance with sets that mimic the full DES photometric sample,
which is on average significantly deeper than the calibration sample
due to the limited depth of spectroscopic surveys. Empirical \pz
methods using, for instance, Artificial Neural Networks or Random
Forests, yield the best performance in the tests, achieving core photo-$z$
resolutions $\sigma_{68} \sim 0.08$.
Moreover, the results from most of the codes, including template fitting methods, comfortably meet the DES requirements on photo-$z$ performance, therefore, providing an excellent precedent for future DES data sets. 
\end{abstract}


\section{Introduction}
\label{sec:intro}

Large galaxy surveys provide detailed information on the large-scale structure of the Universe, which, in turn, helps understand its geometry, composition, evolution and fate. On one hand, spectroscopic surveys like 2dF \citep{Colless2001}, VVDS \citep{LeFevre2005}, WiggleZ \citep{Drinkwater2010} or BOSS \citep{Dawson2013} provide a three-dimensional picture of the galaxy distribution, but they are costly in time and resources, and may suffer from limited depth, incompleteness and selection effects. On the other hand, photometric surveys such as SDSS \citep{York:2000gk}, PanSTARRS \citep{Kaiser2000}, KiDS \citep{Jong2013}, HSC\footnote{http://www.naoj.org/Projects/HSC/index.html} or LSST \citep{Tyson:2002nh} are more efficient and usually deeper, more complete and nearly unbiased, but do not provide a complete 3D view of the Universe, due to their limited resolution in the galaxy positions along the line of sight, which are computed by measuring the photometric redshift (photo-$z$) of each galaxy from the fluxes measured through a set of broadband filters. Even with their limited resolution along the line of sight, photometric surveys, because of their larger volume, are extremely useful for cosmology and, furthermore, uniquely provide some of the most stringent probes of dark energy, such as weak lensing.

There are two main approaches for measuring photometric redshifts: template fitting methods~(e.g. {\tt Hyperz}, \cite{Bolzonella2000}; {\tt BPZ}, \cite{Benitez2000,Coe2006}; {\tt LePhare}, \cite{Arnouts2002,Ilbert2006}; {\tt EAZY}, \cite{Brammer2008}), in which the measured broadband galaxy spectral energy distribution (SED) obtained from the fluxes is compared to a set of redshifted galaxy templates until a best match is found, thereby determining both the galaxy spectral type and its redshift; and training methods~(e.g. {\tt ANNz}, \cite{Collister2004}; {\tt ArborZ}, \citet{Gerdes2010}; {\tt TPZ}, \citet{CarrascoKind2013a}), in which a set of galaxies with known spectroscopic redshifts is used to train a machine-learning algorithm (an artifitial neural network, for example), which is then applied over the galaxy set of interest. Each technique has its own advantages and disadvantages, as we will discuss in this paper, and a combination of them can fully exploit this fact \citep{CarrascoKind2014}. 

In order for photo-$z$'s to be useful for cosmological studies, it is necessary to calibrate them, by understanding the statistical properties of the distribution of the differences between the true galaxy redshifts and their photo-$z$ estimates: its mean value (for the bias), its width (for the resolution), and its tails (for the fraction of outliers, with grossly misestimated photo-$z$'s). To accomplish this, a sample of galaxies with spectroscopic redshifts is required, ideally with a galaxy population that reproduces the population in the photometric survey.

The Dark Energy Survey (DES, \citet{Flaugher2005}) is one such photometric redshift survey, and will cover about one eighth of the sky (5000 sq. deg.) to an unprecedented depth ($i_{AB} < 24$), imaging about 300 million galaxies in 5 broadband filters ($grizY$) up to redshift $z=1.4$. The DES camera (DECam, \citet{Flaugher2012,Diehl2012}) was installed and commissioned in the second semester of 2012, and a Science Verification (SV) period of observations followed, lasting from November 2012 to February 2013. The survey officially started in late August 2013.

The SV observations provided science-quality data for almost 200 sq.~deg.~at close to the nominal depth of the survey. The SV footprint was chosen to contain areas already covered by several deep spectroscopic galaxy surveys, including VVDS (\cite{LeFevre2005}), ACES (\cite{Cooper2012}), and zCOSMOS (\cite{Lilly2007}), which together provide a suitable calibration sample for the DES photometric redshifts. This paper presents a study of the photo-$z$ precision achieved by DES during the SV period, by taking advantage of the available spectroscopic data in its footprint, and by using a large number of photo-$z$ algorithms of different nature.

It has been pointed out \citep{Cunha2012} that cosmic variance in the spectroscopic samples used for photo-$z$ calibration may bias the results of an analysis such as the one we present here, which uses spectra in four relatively small (1 sq.~deg.~each) patches of sky. A robust photo-$z$ calibration requires galaxy spectra distributed all over the photometric survey's footprint, calling for as many as {\cal O}(50--100) patches~\citep{Cunha2012} . While 
the plan for the ultimate photo-$z$ calibration of the whole DES data will need such a spectroscopic calibration sample, and steps are being taken towards the acquisition of the relevant data, the currently available spectroscopic data set is good enough for a first analysis of the photo-$z$ precision that can be achieved with the early DES data. Analogously, the ultimate DES \pz calibration will have to worry about the effects of the possible incompleteness of the spectroscopic calibration samples, effects that we can safely ignore here, given the scope of this first study.

Many studies have been performed in the past comparing in detail several photo-$z$ codes over the same, real or simulated, data set~\citep{Hogg1998, Abdalla2008, Hildebrandt2008, Hildebrandt2010, Dahlen2013}. Particularly comprehensive is the work by~\citet{Hildebrandt2010}, which compares the performance of 19  photo-$z$ codes both over simulated and real (including HST) observations taken in 18 optical and near-infrared bands. Similarly, \citet{Dahlen2013} compares 11 codes over real data in 14 bands, including also some HST data. On the other hand, \citet{Abdalla2008}, analyzed the performance of six photo-z algorithms on the MegaZ Luminous Red Galaxy sample extracted from the SDSS Data Release 7 five-band photometry, with a magnitude limit around $i_{AB} = 20$. This paper differs from these previous studies in that, on the one hand, it uses solely DECam five-band photometry ($grizY$), and on the other, it studies all kinds of galaxies up to the DES nominal limiting magnitude $i_{AB} = 24$.
Furthermore, in the present study, rather than trying to carry out a thorough comparison of all the photo-$z$ codes available in the literature, we concentrate on assessing the performance of the early DES data with respect to the photometric redshift determination, and, in order to do so, we try the codes in which members of the DES collaboration have a certain degree of expertise, without attempting to be complete or even necessarily fair in the comparison. 
Beyond providing a snapshot of the quality of the DES-SV data regarding \pz estimation and accuracy, a secondary goal of this work is to tune these photo-$z$ codes to the particular characteristics of the DES data: filter set, depth, etc, in preparation for the upcoming larger data sets.

Since even the deep spectroscopic samples mentioned above fail to reproduce exactly the depth and colors of the DES-SV photometric galaxy sample, a multi-dimensional weighting technique (\cite{Lima2008,Cunha2009}) was used in order to bring the spectroscopic and photometric samples into better agreement. Matching the galaxies in the spectroscopic samples with those in the DES-SV photometric sample and comparing their spectroscopic redshifts with the DES photo-$z$'s, we will show that, even at this early stage, the DES-SV data fulfill the set of photo-z requirements on bias, resolution and outlier fraction that were defined prior to the start of the survey.

The outline of the paper is as follows. Section 2 describes the DES-SV photometric galaxy sample, whereas the spectroscopic galaxy samples are presented in Section 3, together with the weighting technique that has been used to match their depth and colors to those of the DES-SV sample. Section 4 describes briefly the conditions in which the 13 different photo-$z$ codes studied were run, and
contains the bulk of the results of the paper, including the comparison between the results obtained with the different photo-$z$ codes, the dependence of the results on both the depth of the DES-SV data and the specific spectroscopic calibration samples used, and an in-depth presentation of the results obtained with four representative photo-$z$ codes, in particular with respect to the set of requirements of the DES survey, which we set up at the beginning of Section 4. A discussion of the main results in the paper can be found in Section 5. Finally, we present our conclusions in Section 6, while we confine to an appendix the detailed description of the metrics used to characterize the photo-$z$ distributions.

\section{DES-SV photometric sample}
\label{sec:photometric_sample}

\begin{figure*}
\centering
\includegraphics[width=150mm]{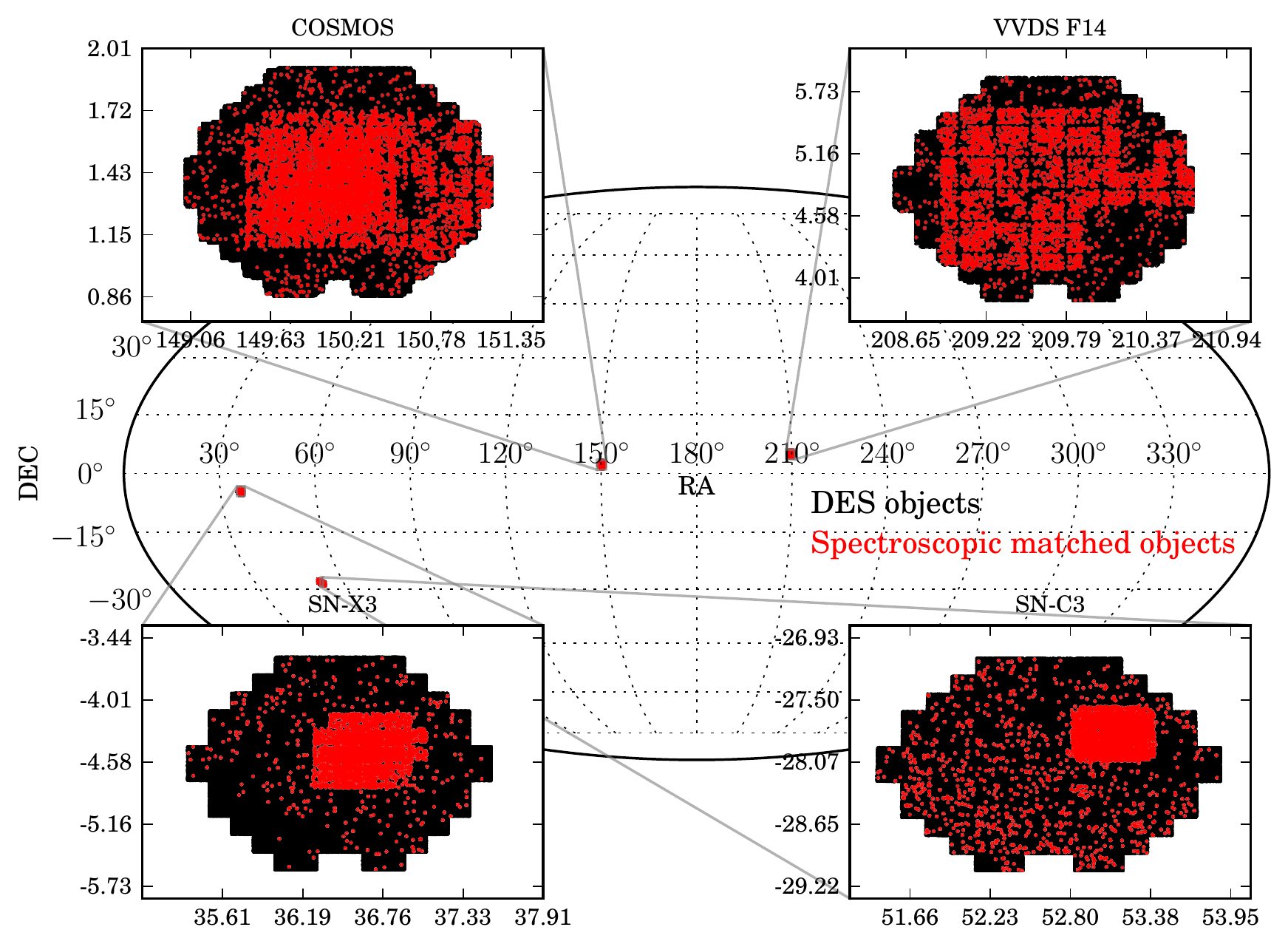}
\caption{Positions in the sky of the four calibration fields. In the zoomed-in inset panels it is possible to observe the spectroscopic matched galaxies, in red, in front of all the DES galaxies detected in the fields, in black.}
\label{fig:fields}
\end{figure*}

DECam imaging on fields overlapping those from deep spectroscopic redshift surveys 
were obtained for the following four DES fields: SN-X3, SN-C3, 
VVDS F14, and COSMOS, whose positions in the sky are shown in Fig. \ref{fig:fields}. SN-X3 and SN-C3 are the two deep fields in the DES
supernova survey, and dithered observations of these fields were obtained
routinely during the DES SV period. The SN-X3 field includes the VVDS-02hr field of the VVDS Deep survey \citep{LeFevre2005,LeFevre2013}, while SN-C3 overlaps with the CDFS (Chandra Deep Field South) area of the ACES survey \citep{Cooper2012}. The VVDS F14 field was 
centered on the VVDS-Wide redshift survey 14hr field \citep{Garilli2008},
and dithered imaging to DES main survey depth of this field was likewise 
obtained during DES SV.  Deep dithered imaging data for the COSMOS field, 
centered on the Cosmological Evolution Survey (COSMOS) area \citep{Lilly2007,Lilly2009} were obtained 
during February 2013 by a DECam community program.\footnote{Proposal 2013A-0351 
Made available for DES photo-$z$ calibration use by arrangement with 
PI Arjun Dey.} Each one of the four fields covers about the area of a single 
DECam pointing, or about 3~deg$^2$. See Section \ref{sec:spectroscopic_sample} for a detailed description of the spectroscopic data matched in each of the fields.

\begin{figure}
\centering
\includegraphics[width=90mm]{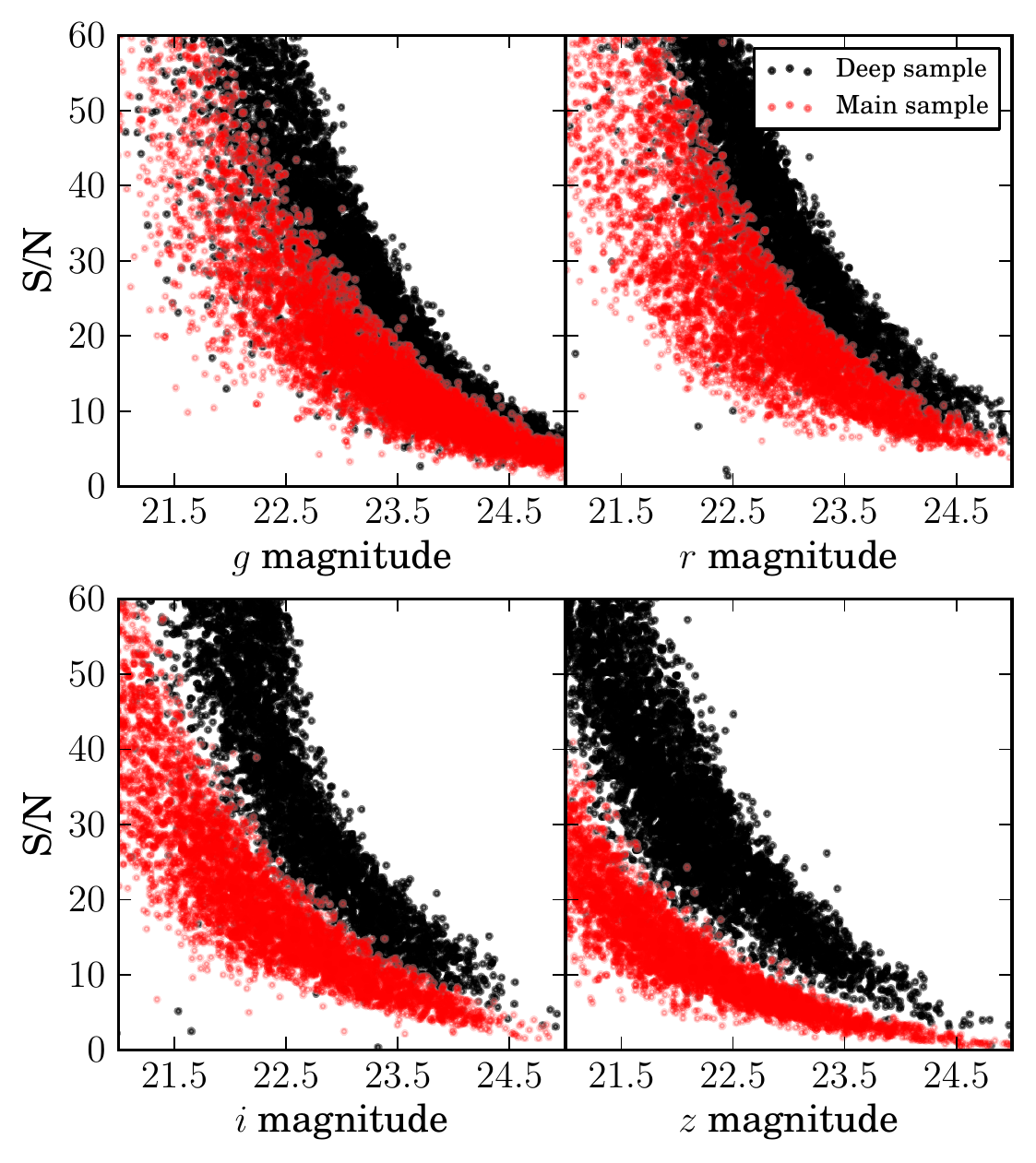}
\caption{S/N vs. magnitude for $g,r,i$ and $z$ DES bands, and for Main (red dots) and Deep (black dots) samples.}
\label{fig:ston}
\end{figure}

All fields include imaging in the 5 DES filters $grizY$, and additionally in the $u$ band, which is part of DECam but not used by the DES survey.
The data have been processed to two imaging depths: Main, corresponding
to approximately DES main survey exposure times, and Deep, corresponding 
to about 3 times the exposure of a single visit to a DES supernova 
deep field (for SN-X3 and SN-C3) or deeper (for COSMOS).
Differences in S/N between the Main and Deep samples can be appreciated in Fig \ref{fig:ston}; details of the data, the exposure times used and the magnitude depths are given in Table \ref{table:exposures}. Similar to DES science requirements convention, the 10$\sigma$
magnitude limit is defined to be the \magauto value (see definition below in this section) at which
the flux in a 2-arcsec diameter aperture is measured at 10$\sigma$.
Note that for the SN-X3 and SN-C3 fields, we selected those SV observations 
that approximately met DES main survey sky background and seeing criteria
in constructing the processed data used for this paper.

\begin{table}
 \centering
 \begin{minipage}{80mm}
  \caption{Imaging exposure times and depths for photo-$z$ calibration fields.}
  \label{table:exposures}
  \begin{tabular}{@{}lccc@{}}
  \hline
   Field & Filter & Tot. Exp. Time (sec) & 10$\sigma$ Depth \\
 \hline
SN-X3 Main & $u$ & 900  & 22.6 \\
 SN-X3 Main & $g$ & 800  & 24.1 \\
 SN-X3 Main & $r$ & 1200 & 24.3 \\
 SN-X3 Main & $i$ & 1080 & 23.6 \\
 SN-X3 Main & $z$ & 990  & 22.7 \\
 SN-X3 Main & $Y$ & 500  & 21.0 \\
\hline
 SN-C3 Main & $u$ & 900  & 22.9 \\
 SN-C3 Main & $g$ & 800  & 24.3 \\
 SN-C3 Main & $r$ & 1200 & 24.3 \\
 SN-C3 Main & $i$ & 1080 & 23.6 \\
 SN-C3 Main & $z$ & 990  & 22.9 \\
 SN-C3 Main & $Y$ & 500  & 20.9 \\
\hline
 VVDS F14 Main & $u$ & 900  & 22.5 \\
 VVDS F14 Main & $g$ & 900  & 24.0 \\
 VVDS F14 Main & $r$ & 900  & 23.6 \\
 VVDS F14 Main & $i$ & 900  & 23.1 \\
 VVDS F14 Main & $z$ & 900  & 22.4 \\
 VVDS F14 Main & $Y$ & 750  & 21.6 \\
 \hline
 \hline
 SN-X3 Deep & $u$ & 900   & 22.6 \\
 SN-X3 Deep & $g$ & 2000  & 24.5 \\
 SN-X3 Deep & $r$ & 3600  & 24.9 \\
 SN-X3 Deep & $i$ & 5400  & 24.5 \\
 SN-X3 Deep & $z$ & 10890 & 24.0 \\
 SN-X3 Deep & $Y$ & 1000  & 21.0 \\
\hline
 SN-C3 Deep & $u$ & 900   & 22.8 \\
 SN-C3 Deep & $g$ & 1800  & 24.6 \\
 SN-C3 Deep & $r$ & 3600  & 24.9 \\
 SN-C3 Deep & $i$ & 5400  & 24.5 \\
 SN-C3 Deep & $z$ & 10890 & 24.3 \\
 SN-C3 Deep & $Y$ & 500   & 20.8 \\
\hline
 COSMOS Deep & $u$ & 33600 & 25.2 \\
 COSMOS Deep & $g$ & 4500  & 24.8 \\
 COSMOS Deep & $r$ & 4800  & 24.9 \\
 COSMOS Deep & $i$ & 12000 & 24.8 \\
 COSMOS Deep & $z$ & 7000  & 23.5 \\
 COSMOS Deep & $Y$ & 2400  & 21.6 \\
\hline
\end{tabular}
\end{minipage}
\end{table}

The data were processed using the same routines used by DES
Data Management (DESDM) in their processing pipeline \citep{Mohr2012,Desai2012}, in particular 
for image detrending, astrometric calibration (SCAMP, \citet{Bertin2006}), image remapping and 
coaddition (SWarp, \cite{Bertin2002}), point spread function modeling (PSFEx, \citet{Bertin2011}),
and object detection and photometry (SExtractor, \citet{Bertin1996}).
The data were processed by running these codes in 
standalone mode at Fermilab, rather than by running them within the 
DESDM processing framework at NCSA.  Running standalone was needed as
the DESDM framework was not yet fully setup at the time (Spring 2013)
to process and calibrate the data for these isolated fields all the way 
through to image coaddition.

Though we basically used the DESDM codes, there were some detailed
differences in processing and photometric calibration that we highlight here.
For image detrending we did not include corrections for CCD nonlinearity,
pupil ghost, and illumination that are now used by DESDM, as these corrections
were not available at the time.  Image coaddition was done using a median
coadd rather than by using a weighted mean as in DESDM.  
Photometric calibration in the $ugriz$ filters for the SN-X3, VVDS F14, 
and COSMOS fields was done by matching against overlapping bright stars from 
the SDSS Data Release 9 database \citep{Ahn2012}.  This was done to calibrate each individual 
CCD on each separate DECam exposure, before image coaddition.
In the $Y$ band for all fields, and in all filters for the SN-C3 field
(which did not overlap SDSS), 
we picked a fiducial exposure for each field, adopted the typical DECam 
CCD-by-CCD photometric zeropoints as determined from DES SV data,
and then tied the photometry for subsequent exposures/CCDs to the fiducial
exposure by matching overlapping bright objects.
In addition, we also applied a further relative photometric
calibration step, by selecting bright $r = $~18--22 galaxies in each field
and offsetting the zeropoints in the other 5 filters so that the 
median galaxy colors relative to $r$ (i.e., $g-r$, $r-i$, etc.) 
would match those for fiducial DECam data of the VVDS-Deep 02hr field 
(part of SN-X3).  This additional step was intended to match up the median galaxy colors
among the different fields by using photometry of galaxies directly,
as the earlier calibration steps use photometry of stars,
and there can be small (percent level) systematic differences between
the stellar and galaxy photometric zeropoints, in particular due to
seeing. We also applied corrections for Milky Way extinction
based on the \cite{Schlegel1998} dust maps evaluated at
the center of each field.

As for the use of imaging data by photo-$z$ algorithms, either {\tt MAGAUTO} or {\tt MAGDETMODEL} magnitudes (or both) were employed by the different photo-$z$ codes. \magauto magnitudes come from the flux (counts) measured in an elliptical aperture defined as in \citet{Bertin1996}. It provides an estimation of the total magnitude of the object. \magdetmodel magnitudes are measured from the shape (a Sersic profile model \citep{Sersic1963}) fit to the object in the SExtractor detection image (either the $r$ band or the $i$ band for our data), and the flux is then measured separately in each band using that same model shape. Also available are {\tt MAGMODEL} magnitudes, which fit the shape of the object independently in each of the bands. However, \magdetmodel magnitudes, which result from one unique best-fit shape for the object, are in general better suited for color measurement and hence more appropriate to use for photo-$z$ estimation.

We want to emphasize here that because of the differences mentioned 
above between the reductions of SV data used in this paper and the
improved DESDM reductions of SV data (to be released and described
elsewhere), the results of this paper are meant to reflect
the photo-$z$ quality achievable from early DES data, rather than
from final DES data or even from SV data.
We expect that final DESDM reductions
of the calibration field data will be better in terms of photometric
quality and consequently of photo-$z$ quality, so the results in this
paper will serve as a lower bound on the photo-$z$ quality that may
be achieved by final DES data.  Nonetheless, as we will show later in 
this paper, the photo-$z$ quality achieved in these early DES data is 
good and already sufficient to meet the basic DES science requirements 
on photo-$z$ scatter and outlier fractions.

\section{DES-SV spectroscopic sample}
\label{sec:spectroscopic_sample}
In general, to exploit a galaxy photometric survey to its maximum scientific potential, it is necessary to be able to calibrate or control the performance of the photo-$z$ estimation by using data from a spectroscopic survey. To accomplish this, it is necessary to have, for a subset of galaxies, both the spectroscopic redshifts and the estimated photo-$z$'s. With this information in hand, the characterization of the behavior of the photometric redshifts is possible, and it becomes a crucial step for cosmological probes such as galaxy clustering or weak lensing. In particular, among other quantities, it is very important to characterize the true redshift distribution of a set of galaxies after a selection in photo-$z$ space. 

A photometric survey like DES will therefore need to observe one or several regions of the sky that have been previously covered by a spectroscopic survey, and then match the galaxies in the catalog of the spectroscopic survey to galaxies observed photometrically by DES. In this paper, the matching between DES and spectroscopic galaxies is performed by using the positions of the galaxies in the sky plane, with a matching radius of 1 arcsec.

Four regions of the sky included in the DES-SV footprint  have been used for photo-$z$ calibration in this study (Fig. \ref{fig:fields}):  
\begin{itemize}
\item \textbf{SN-X3 field}: This area, centered at RA $\sim$ 36$\degr$, DEC $\sim$ -5$\degr$, overlaps with the VIMOS (\cite{LeFevre2003}) VLT Deep Survey (VVDS) 02hr field. DES photometry has been matched in this field with spectroscopic redshift data from VVDS Deep (\cite{LeFevre2005,LeFevre2013}).
\item \textbf{VVDS F14 field}: This area, centered at RA $\sim$ 209$\degr$, DEC $\sim$ 5$\degr$, overlaps with the VIMOS VLT Deep Survey (VVDS) 14hr field. DES photometry has been matched in this field with spectroscopic redshift data from VVDS Wide (\cite{Garilli2008}).
\item \textbf{SN-C3 field}: This area, centered at RA $\sim$ 52$\degr$, DEC $\sim$ -28$\degr$, overlaps with the Chandra Deep Field South. DES photometry has been matched in this field with spectroscopic redshift data from both VVDS Deep and ACES (\cite{Cooper2012}).  
\item \textbf{COSMOS field}: This area, centered at RA $\sim$ 150$\degr$, DEC $\sim$ -1.4$\degr$, overlaps with the Cosmic Evolution Survey field. DES photometry has been matched in this field with spectroscopic redshift data from both VVDS Wide and zCOSMOS (\cite{Lilly2007,Lilly2009}).  
\end{itemize}

\begin{figure}
\centering
\includegraphics[width=90mm]{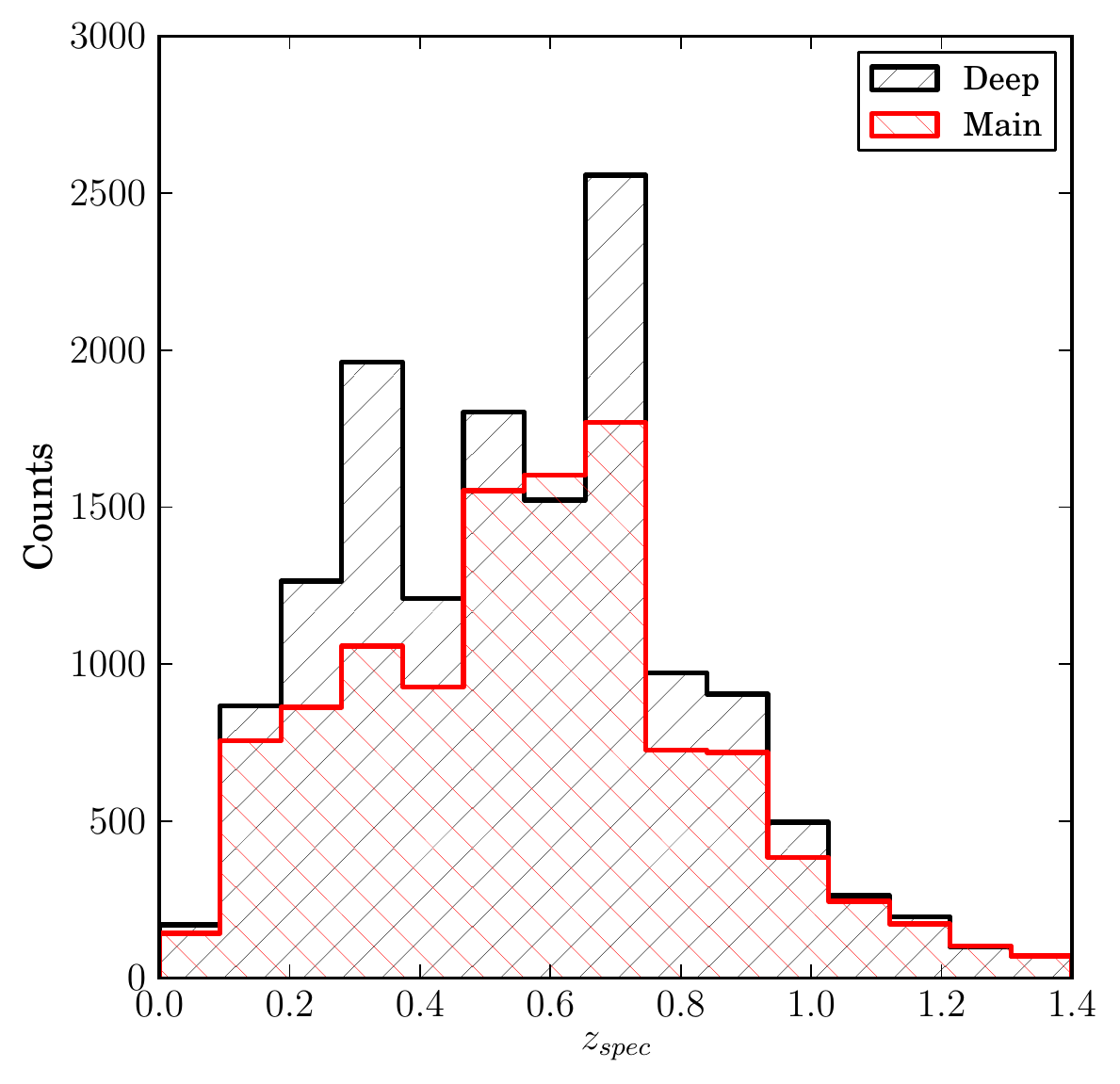}
\caption{Redshift distributions for the spectroscopic matched galaxies in the Main and Deep calibration samples.} 
\label{fig:redshift_distributions}
\end{figure}

We have also used data from brighter, shallower spectroscopic surveys available in these four regions: OzDES (\citet{Kuehn2014}, Yuan et al., in preparation); SDSS-I/II \citep{Strauss2002,Eisenstein2001}; SDSS-III BOSS Data Release 9 \citep{Ahn2012} and 2dF Galaxy Redshift Survey \citep{Colless2001}. Galaxies matched to these surveys help to increase the statistics on the brighter tail of the distribution for galaxies used for this study. 

Next we build a set of training and testing samples, using DES photometry from the Main and Deep samples and spectroscopy from different surveys. Note that only high-confidence spectroscopic redshifts, i.e., having redshift flags between 3 and 5, corresponding to secure and very secure ($>95\%$ accuracy) redshift determinations, have been selected to construct these samples. Spectroscopic failures can bias cosmological results, as studied in \citet{Cunha2012b}, where they showed how for a final DES analysis such failures need to be reduced to the percent level. While a complete study concerning spectroscopic failures will have to take place for the analysis of the final DES data set, here we rely on the high-confidence redshift flags for the \pz calibration of the early DES data. Below we describe how the data is distributed on each of the four calibration samples (training and testing for both Main and Deep catalogs):
\begin{itemize}
\item Main training sample: 5859 galaxies, photometry from Main catalogs, spectroscopic redshifts from the following data sets: 

- One randomly selected half of VVDS Deep, SDSS/BOSS in the SN-X3 field, ACES, 2dFGRS, OzDES in the SN-C3 field, and VVDS Wide, SDSS/BOSS in the VVDS F14 field.
\item Main testing sample: 6381 galaxies, photometry from Main catalogs, spectroscopic redshifts from the following data sets: 

- The other half left out from the samples in the Main training set. 

- All VVDS Deep in the SN-C3 field.
\item Deep training sample: 7249 galaxies, photometry from Deep catalogs, spectroscopic redshifts from the following data sets: 

- One randomly selected half of VVDS Deep, SDSS/BOSS in the SN-X3 field, ACES, 2dFGRS, OzDES in the SN-C3 field and zCOSMOS, SDSS/BOSS, 2dFGRS in the COSMOS field.
\item Deep testing sample: 8358 galaxies, photometry from Deep catalogs, spectroscopic redshifts from the following data sets: 

- The other half left out from the samples in the Deep training sample.

- All VVDS Deep in the SN-C3 field.

- All VVDS Wide in the VVDS F14 field.
\end{itemize}

The spectroscopic redshift distributions of the Main and Deep calibration samples defined above, spanning all the redshift range of interest for DES ($0<z<1.4$), are shown in Fig. \ref{fig:redshift_distributions}.

\subsection{The weighting procedure}
\label{sec:weights}
In order to assess the photo-$z$ performance of the DES-SV data we would ideally need a calibration sample being representative of the DES-SV full sample, i.e. having exactly the same photometric properties (magnitude and colour distributions). However, spectroscopic galaxy samples are shallower, and suffer from selection effects. A weighting procedure, which assigns a weight to each of the galaxies in the calibration sample so that the distributions of their photometric observables reproduce the distributions of the same observables in the full sample, can be used provided there is enough overlap between the photometric spaces of the calibration and full samples \citep{Lima2008,Cunha2009}. 

Different algorithms can be used to compute the weights, but basically all compare local densities in the photometric spaces of the two samples (calibration and full) and assign a weight to each photometric region of the calibration sample equal to the ratio between the densities of galaxies in the full sample and the calibration sample in a given region. In this study we use a nearest neighbour algorithm to compute the weights that we use extensively throughout the paper. A detailed description of the method can be found in \cite{Lima2008}. 

\begin{figure}
\centering
\includegraphics[width=90mm]{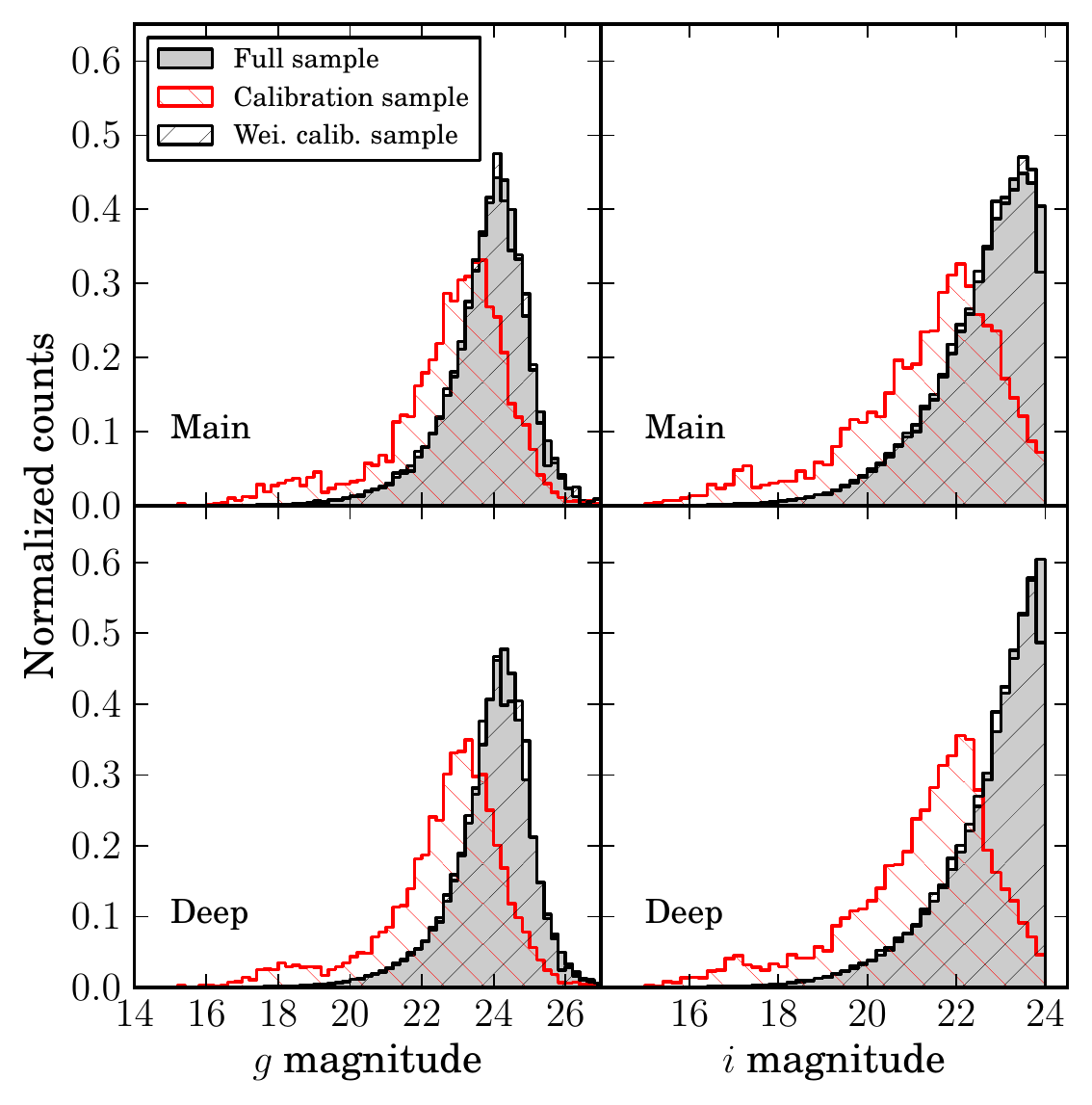}
\caption{$g$ and $i$ magnitude distributions for the full, calibration and weighted calibration sample. The difference between the full and the calibration samples is apparent, the latter being significantly brighter. After applying the weighting procedure described in \citet{Lima2008}, the weighted calibration distributions agree very well with the corresponding DES-SV distributions.} 
\label{fig:magnitudes_weights}
\end{figure}

We apply the weighting technique within a region in the multidimensional space defined by $18 < i_{AB} < 24$; $0 < g-r < 2$; $0 < r-i < 2$. In Fig. \ref{fig:magnitudes_weights} one can check how the weighting procedure is efficiently applied for the sample used in this study. The figure shows, for two DES bands, and the Main and Deep samples, the magnitude distributions for the full sample, the calibration sample and the weighted calibration sample, whose distributions agree very well with those of the full sample.   

\section{Photometric redshifts in the DES-SV calibration samples}
\label{sec:analysis_spec}
In this section we include all the photo-$z$ analyses using the calibration data defined in Sec. \ref{sec:spectroscopic_sample}. The analysis is carried out employing an extensive set of statistics. To construct most of the metrics used in this paper, we first define the bias to be $\Delta z = z_{\rm phot}-z_{\rm spec}$ and the normalized bias by its reported error as $\Delta z' = (z_{\rm phot}-z_{\rm spec})/ \epsilon_{\rm phot}$, where $\epsilon_{\rm phot}$ is the error in the estimation of the photo-$z$'s. We present the standard metrics used to compare the accuracy of the different codes in Table \ref{tab:def_metrics}, together with the DES science requirements for photo-$z$'s, set before the start of the survey. The DES science requirements are driven by the Dark Energy science that DES plans to carry out, in particular by weak lensing and large-scale structure tomographic measurements.

\begin{table*}
\caption{Definition of the metrics used in the text to present the main results. These are computed in the whole redshift range as well as in bins of width $0.1$ in photometric redshift. Detailed definitions can be found in the appendix.}
\label{tab:def_metrics}
\centering
\renewcommand{\footnoterule}{}
\begin{tabular}{lll}
Metric & Description & Requirement\\
\hline
$\overline{\Delta z}$ & mean of the  $\Delta z$ distribution & - \\
$\sigma_{\Delta z}$ & standard deviation of the $\Delta z$ distribution & - \\
$\Delta z_{50}$ & median of the $\Delta z$ distribution & - \\
$\sigma_{68}$ & half width of the interval around $\Delta z_{50}$ containing 68\% of the galaxies & $ < 0.12 $ \\
${\rm out}_{2\sigma}$ & fraction of galaxies with: $|\Delta z - \overline{\Delta z} | > 2\sigma_{\Delta z}$ & $ < 0.1 $\\
${\rm out}_{3\sigma}$ & fraction of galaxies with: $|\Delta z - \overline{\Delta z} | > 3\sigma_{\Delta z}$ & $ < 0.015 $\\
$\overline{\Delta z'}$ & mean of the  $\Delta z' = \Delta z / \epsilon_{\rm phot}$ distribution & - \\
$\sigma_{\Delta z'}$ & standard deviation of the  $\Delta z'$ distribution & - \\
${\rm N}_{\rm poisson}$ & difference between $N(z)^{\rm phot}$ and $N(z)^{\rm spec}$ normalized by Poisson fluctuations & - \\
${\rm KS}$ & Kolmogorov - Smirnov statistic for $N(z)^{\rm phot}$, $N(z)^{\rm spec}$ & - \\
\hline
\end{tabular}
\end{table*}

The photo-$z$ metrics we consider are intended to measure the quality of the
photometric redshifts in terms of their bias, scatter, and outlier fraction
statistics, and also in terms of the fidelity of the photo-$z$ errors and of
the agreement between the photo-$z$ and true redshift distributions.
Detailed definitions of these metrics are given in Appendix A, while
here we briefly summarize and motivate our choice of metrics:

$\bullet$ {\it Bias:} To quantify the overall photo-$z$ bias, we use the 
mean bias $\overline{\Delta z}$ and the median bias $\Delta z_{50}$.

$\bullet$ {\it Scatter:}  To measure the photo-$z$ scatter, we use both the
standard deviation $\sigma_{\Delta z}$ of $\Delta z$ and the 68-percentile 
width $\sigma_{68}$ of $\Delta z$ about the median 
(i.e., $\Delta z_{50} \pm \sigma_{68}$ covers 68\% of the $\Delta z$ 
distribution).  For a Gaussian distribution of $\Delta z$, we would have 
$\sigma_{\Delta z} = \sigma_{68}$. However, in general $\Delta z$ is not 
distributed as a Gaussian, so that $\sigma_{68}$ measures the width of the 
core of the $\Delta z$ distribution, whereas $\sigma_{\Delta z}$ is sensitive 
to the tails.  The DES science requirements specify $\sigma_{68} < 0.12$ for 
a 90\% of the selectable sample of galaxies.

$\bullet$ {\it Outlier Fractions:} To quantify the fraction of objects with 
large $|\Delta z|$, we measure the $2\sigma_{\Delta z}$ and 
$3\sigma_{\Delta z}$ outlier fractions out$_{2\sigma}$ and out$_{3\sigma}$, 
respectively, both defined relative to the mean photo-$z$ bias 
$\overline{\Delta z}$.  The DES science requirements limit these outlier 
fractions to be out$_{2\sigma} < 0.1$ and out$_{3\sigma} < 0.015$.

$\bullet$ {\it Fidelity of Photo-$z$ Errors:} To assess the fidelity of the 
reported photo-$z$ error $\epsilon_{\rm phot}$, we normalize $\Delta z$ by 
$\epsilon_{\rm phot}$ and calculate the resulting error-normalized mean bias
$\overline{\Delta z'}$ and standard deviation $\sigma_{\Delta z'}$.  Ideally,
we would obtain $\overline{\Delta z'} = 0$ and $\sigma_{\Delta z'} = 1$,
as for a Gaussian distribution of zero mean and unit variance.  Deviations 
from these values would indicate inaccuracies in the photo-$z$ errors.

$\bullet$ {\it Redshift Distributions:} Finally, to compare the 
photometric redshift distribution $N(z_{\rm phot})$ to the true redshift 
distribution $N(z_{\rm spec})$, we use two metrics.  The first is 
$N_{\rm poisson}$, which measures the rms difference between $N(z_{\rm phot})$
and $N(z_{\rm spec})$, normalized by Poisson fluctuations 
$\sqrt{N(z_{\rm spec})}$.  The second is the Kolmogorov-Smirnov (KS) metric 
that tests whether $N(z_{\rm phot})$ and $N(z_{\rm spec})$ are consistent with 
being drawn from the same parent distribution.

Some of the key DES science analyses, such as the galaxy angular correlation functions 
for large-scale structure studies, or cosmic shear tomography 
measurements for weak lensing and dark energy constraints, will use DES 
galaxies divided into separate photo-$z$ bins.  
For these photo-$z$ bins, the DES science requirements also 
specify stringent limits on the differences in bias, scatter, and outlier 
fractions between the DES photometric sample and the spectroscopic calibration
sample. For example, there is a requirement that the mean bias 
$|\overline{\Delta z}| < 0.001(1+z)$ in bins of 0.1 in redshift.  Accurate 
characterization of the full $P(z_{\rm spec}|z_{\rm phot})$ distribution, 
i.e., the distribution of true redshift in bins of photo-$z$,
will also be needed for these DES science analyses.  
However, consideration of these more stringent DES photo-$z$ science 
requirements is premature for the present paper, because of the
limited number and sky distribution of the SV spectroscopic calibration fields.
These fields are subject to sample variance effects, i.e., fluctuations in
galaxy densities and redshift distributions due to clustering and 
large-scale structure, and in fact, as detailed in \citet{Cunha2012},
meeting the requirements will necessitate a much larger number of 
widely-distributed spectroscopic calibration fields (e.g., $\sim$100) than 
are currently available from completed redshift surveys.  We will thus postpone
discussion of these issues and tests of these more stringent requirements 
for future DES photo-$z$ papers.  Nonetheless, we do present in a later 
section some example $P(z_{\rm spec}|z_{\rm phot})$ distribution for several 
selected photo-$z$ codes.

All the results shown in this paper have been weighted according to the technique presented in Section \ref{sec:weights}, and they include a cut on the 10$\%$ of the galaxies having larger estimated photo-$z$ error, as given from each particular code (this 10$\%$ cut on \pz error is allowed by the DES science requirements). This cut introduces small differences among the testing samples used by each photo-$z$ code in the comparison. With this we test the ability of each particular code to select the most problematic objects in the sample. In addition, clustering measurements can be affected by severe photo-$z$ quality cuts as presented in \citet{Marti2013}, where they also show a procedure to correct for these effects.

All the errors for the statistics presented in the paper come from bootstrap resampling using 100 samples, unless otherwise stated.

\subsection{Methods}
\label{sec:methods}
Before going in detail into the photo-$z$ analysis we present here a brief description of the different methods we have used to estimate photo-$z$'s, which include most of the relevant photo-$z$ codes available. We have emphasized in the details at the time of running these codes; for an exhaustive description of them see Table \ref{tab:codes}. For template-based methods, a standardized set of filter throughput curves has been used. Most of the codes have been run in standalone mode, while a fair fraction of them has been run within the DES Science Portal, with compatible results. Due to the large number of codes used, the paper, other than showing the DES-SV photo-$z$ capabilities, also serves as a helpful reference to compare different photo-$z$ codes using real data from a deep galaxy survey. 

\begin{table*}
\caption{List of methods used to estimate photo-$z$'s. Code type and main references are given.}
\label{tab:codes}
\centering
\renewcommand{\footnoterule}{}
\begin{tabular}{lll}
Code & Type & Reference\\
\hline
DESDM, Artificial Neural Network & Training-based & \citet{Oyaizu2008a} \\
ANNz, Artificial Neural Network & Training-based & \citet{Collister2004} \\
TPZ, Prediction Trees and Random Forest & Training-based & \citet{CarrascoKind2013a,CarrascoKind2014} \\
RVMz, Relevance Vector Machine & Training-based & \citet{Tipping2001} \\
NIP-kNNz, Normalized Inner Product Nearest Neighbor & Training-based & de Vicente et al., in preparation \\
ANNz2, Machine Learning Methods  & Training-based & Sadeh et al., in preparation \\
ArborZ, Boosted Decision Trees & Training-based & \citet{Gerdes2010}  \\
SkyNet, Classification Artificial Neural Network & Training-based & \citet{Bonnett,SkyNet} \\ 
BPZ, Bayesian Photometric Redshifts & Template-based & \citet{Benitez2000,Coe2006} \\
EAZY, Easy and Accurate Redshifts from Yale & Template-based & \citet{Brammer2008} \\
LePhare & Template-based & \citet{Arnouts2002,Ilbert2006} \\
ZEBRA, Zurich Extragalactic Bayesian Redshift Analyzer & Template-based & \citet{Feldmann2006} \\
Photo-Z & Template-based & \citet{Bender2001} \\
\hline
\end{tabular}
\end{table*}

\subsubsection{DESDM}

The DESDM (default) photo-$z$'s were computed using
the same artificial neural network method that
was applied to the Sloan Digital Sky Survey Data Release 6 (DR6) sample,
as described in detail by \cite{Oyaizu2008a}.  In brief, 
we used a neural network configuration with 10 input nodes, 
consisting of the 5 $grizY$ {\tt MAGAUTO} magnitudes and the 
5 $grizY$ {\tt MAGDETMODEL} magnitudes, followed by 3 hidden layers with
15 nodes per layer.  The formal minimization to determine the
neural network weights was done on the training set data, while choosing
the set of network weights that gave the lowest photo-$z$ scatter
on the testing set, after 300 iterations of the weight minimization.
Moreover, to reduce sensitivity to initial conditions in the
minimization procedure, we repeated the procedure 10 times, starting
each time at a different initial position in the space of weights.
The final photo-$z$ for a galaxy was taken to be the average of the 
photo-$z$'s computed from the optimal weights for each of the 10 network 
minimizations.

We also computed accompanying photo-$z$ errors using the empirical 
\textit{nearest neighbor error} (NNE) technique, described in detail
by \cite{Oyaizu2008b}.  The NNE method estimates the photo-$z$ error for
each galaxy empirically, based on the photo-$z$'s and true redshifts of the 
galaxy's 100 nearest neighbors in the spectroscopic testing set, 
where neighbor distance is defined using a simple flat metric in the space 
consisting of the 10 input magnitudes noted above.  Specifically, the 
NNE photo-$z$ error $\sigma$ is defined so that it corresponds to the width of 68\% of 
the $|z_{phot}-z_{spec}|$ distribution of the nearest neighbors. 

\subsubsection{ANNz}

ANNz (\cite{Collister2004}) is a training-based method that uses a neural network scheme to find a functional relationship between a given set of input parameters (e.g. magnitudes, colors, etc) and outputs a desired quantity (e.g. redshift). The results shown in this paper have been obtained by using a neural network architecture of 2 layers with 10 nodes each, and using as inputs the set of 5 {\tt MAGAUTO} and 5 {\tt MAGDETMODEL} magnitudes. Attempts to use a larger number of nodes as well as colors as inputs resulted in larger photo-$z$ errors. The uncertainties in the photo-$z$ estimation are computed using standard propagation of the errors in the input magnitudes to the error in photo-$z$, by using the functional relationship between these input parameters and the output photo-$z$.  

\subsubsection{TPZ (Trees for Photo-$Z$)}

TPZ\footnote{http://lcdm.astro.illinois.edu/research/TPZ.html} \citep{CarrascoKind2013a,CarrascoKind2014} is a machine learning, parallel algorithm that uses prediction trees and random forest techniques to produce both robust photometric redshift pdfs and ancillary information for a galaxy sample. A prediction tree is built by asking a sequence of questions that recursively split the input data taken from the spectroscopic sample, frequently into two branches, until a terminal leaf is created that meets a stopping criterion (e.g., a minimum leaf size or a variance threshold). The dimension in which the data are divided is chosen to be the one with highest information gain  among the random subsample of dimensions obtained at every point. This process produces less correlated trees and allows the exploration of several configurations within the data. The small region bounding the data in the terminal leaf node represents a specific subsample of the entire data with similar properties. Within this leaf, a model is applied that provides a fairly comprehensible prediction, especially in situations where many variables may exist that interact in a nonlinear manner as is often the case with photo-$z$ estimation. 

By perturbing the data using their magnitude errors and by taking bootstrapping samples, many (600 in this application) uncorrelated trees can be created whose results are aggregated to construct each individual pdf. 
For the application to DES-SV data, we have used both {\tt MAGAUTO} and {\tt MAGDETMODEL} magnitudes in the five DES bands, together with all the corresponding colors as well as their associated errors.

\subsubsection{RVMz}

RVMz is an empirical photo-$z$ code based on the relevance vector machine algorithm from \cite{Tipping2001}, a Bayesian sparse kernel method for regression. The relevance vector machine (RVM) has characteristics similar to the support vector machine, but includes a Bayesian treatment for the determination of the model weights.
This has the advantage that the parameters governing model complexity and noise variance are found in the training run itself, and therefore the RVM does not require cross validation to optimize these parameters.
We use the RVM implementation in the R-package {\tt kernlab} from \cite{Karatzoglou2004}.
To obtain photo-$z$ estimates, we used {\tt MAGDETMODEL} magnitudes ($grizY$) and colors ($g-r$, $r-i$, $i-z$, $z-Y$) as input. 
We reconstruct the pdf by combining the uncertainties in the datasets and the model. In the training set we use the $k$-fold cross validation technique, which consists of partitioning the data in $k$ groups, then $k - 1$ of these groups are used to train the model which is then evaluated on the hold-out group. This process is then repeated for all possible choices of the hold-out group and the resulting mean squared error for the redshift prediction is evaluated. At this stage we obtain the model error as the RMS of the predicted pdf. Details of the method will be described in Rau et. al. (in prep.).

\subsubsection{NIP-kNNz}

NIP-kNNz (Juan De Vicente et al. (2014), in preparation) is a novel technique that computes the photo-$z$ from a Nearest Neighbour approach based on the Normalized Inner Product (NIP). While Euclidean magnitude-distance ensures that close galaxies in magnitude space are assigned the same redshift, it does not considers as neighbors galaxies with the same color but separated in overall magnitude. 
NIP metrics corrects this by considering two galaxies as neighbors, and hence with close redshift, when they have similar colors, rather than magnitudes. The metric is based on the inner product definition:
\begin{equation}
\text{NIP} = \cos \alpha = \frac{\mathbf{M}_t \cdot  \mathbf{M}_p}{M_t M_p} \ ,
\end{equation}
where $\mathbf{M}_t$ and $\mathbf{M}_p$ are the multi-magnitude vectors of training and photometric galaxies respectively. For this particular application, the five {\tt MAGDETMODEL} magnitudes were used and turned into fluxes. The normalized inner product is related to the angle that the two multi-magnitude vectors form. Maximizing NIP is equivalent to minimize the angle between the two vectors.
Regarding the photo-$z$ error, an empirical formula has been derived to account for three different contributions. The first term is the floor error related to the finite spectroscopic redshift precision, which is all that remains In the best case scenario, when the magnitude vectors of the photometric and the spectroscopic galaxies point in the same direcction. The second contribution comes from the characterization of the photo-$z$ errors in the spectroscopic sample. NIP-kNN is run over all galaxies in the spectroscopic sample to obtain their photo-$z$s. One half of the difference between the spectroscopic $z$ and the photo-$z$ of the spectroscopic galaxy is taken as its photo-$z$ error. When NIP-kNN is applied to a galaxy in the photometric sample, it inherits not only the $z$ of the closest spectroscopic galaxy but also its photo-$z$ error. The third term is the metric distance $\sin(\alpha)$ that accounts for neighborhood, multiplied by a constant determined empirically. 
Assuming the spectroscopic sample spans the range of redshifts of the photometric sample, NIP-kNNz achieves, by construction, an accurate reconstruction of the redshift distribution $N(z)$

\subsubsection{ANNz2}

ANNz2 is a new major version of the public photo-$z$ estimation software, ANNz \citep{Collister2004}, which will be made public
in 2015.
The new code incorporates several machine-learning methods, such as artificial neural networks (ANNs), boosted
decision trees (BDTs, \citet{Freund1997}, described below in \ref{arborz}) and k-nearest neighbors (KNNs).
The different algorithms are used in concert in order to optimize the photo-$z$ reconstruction
performance, and to estimate the uncertainties of the photometric solutions.
This is done by generating a wide selection of machine-learning solutions with e.g., different
ANN architectures, initialized by different random seeds. The optimization is performed by ranking the
different solution according to their performance, which is determined by the respective photo-$z$ scatter
of each solution.

The single solution with the best performance is chosen as the nominal photo-z estimator
of ANNz2. In addition, the entire collection of solutions is used in order to derive a photo-z probability density
function (pdf), constructed in two steps. First, each solution is folded with an error distribution,
which is derived using the KNN error estimation method of~\citet{Oyaizu2008a}.
The ensemble of solutions is then combined. This is done by weighing the different estimators, in such a way as
to produce a pdf which describes the underlying photometric errors.
The inputs used in this study were the five \magauto and the five \magdetmodel magnitudes.

\subsubsection{ArborZ}
\label{arborz}
The ArborZ algorithm \citep{Gerdes2010, Sypniewski2014} is a training-set-based, publicly-available \citep{ArborZ-code} photo-$z$ estimator that makes use of boosted decision trees (BDTs). BDTs were developed to classify objects characterized by a vector of observables $\mathbf{x}$ into two categories. 
Decision trees are trained iteratively, with initially misclassified objects given higher weight, or ``boosted'', in the next training cycle. An individual decision tree is a relatively weak classifier. But the ``forest'' of trees generated during the training process, when their outputs are combined in a way that assigns higher weight to trees with lower misclassification rates, collectively constitutes a strong classifier. To adapt a binary classifier to the problem of determining a continuous quantity like redshift, we divide the redshift range of interest into $N$ discrete bins with a width roughly 25-50\% of the expected photo-$z$ resolution, and train a separate BDT classifier for each redshift bin, using a forest size of 50 trees. Each classifier is trained to identify galaxies with a redshift falling in its particular bin as ``signal,'' and to reject galaxies falling more than $3\sigma_{z,phot}$ away from its bin as ``background.'' The $3\sigma_{z,phot}$ exclusion region between signal and background objects is introduced in order to avoid the overtraining that could result from treating a galaxy with, e.g., a redshift of 0.999 as signal and one with 1.001 as background. Each BDT classifier in this ensemble, when presented with a new galaxy, produces a probability that the given galaxy falls within its redshift bin. This collection of probabilities constitutes the ArborZ pdf. ArborZ also provides a single best-estimate photo-$z$ (from the median of the pdf) and its error, $\sigma_{68}$ (from its width); however, the full pdf provides the best characterization of a galaxy's photo-$z$. 

In the DES-SV sample, we train ArborZ using the \magauto and \magdetmodel magnitudes in $grizY$, with 50 fixed-width redshift bins out to $z=1.5$. The results are fairly robust with respect to reasonable variations in the forest size, number of bins, the choice of fixed- or variable-width bins, and the use of either or both sets of magnitudes. 

\subsubsection{SkyNet, Classification Neural Network}

This method, first used in \citet{Bonnett}, consists of using a neural network to classify galaxies in classes, in this case redshift bins. 
A neural network with a $softmax$ transformed output \citep{Pyle1999} is able to estimate the probability that an object belongs to a certain class. 
Given the fact that a galaxy cannot live in more than one redshift bin at the same time, a  neural network with a \textit{softmax} transformation is ideally suited to return a pdf for each galaxy.
Before training the neural network, we bin our data in $n$ redshift bins, the classes.
The neural network is fed the \magauto magnitudes, the \magdetmodel magnitudes and the correct classes. 
The neural network outputs $n$ values between $[0, 1]$  for each galaxy.
These $n$ values sum up to 1 and hence can be interpreted as the probability that a galaxy resides in a  redshift bin.  In this particular run we used $n=40$, resulting in a $\delta z = 0.035$ granularity  in the probability density function. 
The publicly available neural network software SkyNet \citep{SkyNet} was used for this work.
The neural net is trained using 3 hidden layers with respectively 20, 30 and 40 nodes per layer. 

\subsubsection{BPZ}

The BPZ (Bayesian Photometric Redshifts) \pz code from \cite{Benitez2000} and \cite{Coe2006} is a template-based method that returns the whole probability density distribution $p(z|m_i)$ that the galaxy is at redshift $z$ when its magnitudes in each band are $m_i$, and also a single photo-$z$ value computed as the maximum of $p(z|m_i)$. Following Bayes Theorem, $p(z|m_i)$ is the product of a likelihood and a prior probability function that represents our previous knowledge of the redshift and spectral type $t$ distributions of the sample in the analysis. In the likelihood, we use the five \magdetmodel magnitudes.
\begin{itemize}
\item \textit{Templates:} We use the eight spectral templates that BPZ carries by default based on \cite{Coleman1980,Kinney1996}, and add two more interpolated templates between each pair of them by setting the input parameter \texttt{INTERP=2} (option by default). 
\item \textit{Prior:} We explicitly calibrate the prior in each test by fitting the empirical function $\Pi(z,t|m_0)$ proposed in \cite{Benitez2000} to the corresponding DES-SV training sample. With this, we are able to remove most of the catastrophic outliers which for template-based methods tend to constitute a sizable fraction of all galaxies.
\item {\it Training:} No other training or calibration has been attempted.
\end{itemize}

\subsubsection{EAZY}

The EAZY \pz code \citep{Brammer2008} is a template-based maximum likelihood method that has 
been specifically optimized for use when representative spectroscopic redshifts are not available for training set based
estimators. In this implementation we use the five {\tt MAGDETMODEL} magnitudes.
\begin{itemize}
\item {\it Templates:}  
The code uses
a novel non-negative matrix factorization algorithm to construct a minimal set of templates which are linear combinations
of templates derived from semi-analytic models, based on the \citet{Bruzual2003} models but with the star formation histories computed from the semi-analytic models of \citet{DeLucia2007a}. These model templates are likely to be more representative of galaxies, particularly
at high redshifts, compared to the commonly used \citet{Coleman1980} or \citet{BruzualA.1993} templates. The code 
also makes use of a template error function to account for uncertainties in templates over specific wavelength ranges. 
\item {\it Priors:} No priors are used.
\item{\it Training:}
No calibration of the templates is performed using spectroscopic training data. 
Equally, photometric offsets cannot be derived using spectroscopic training sets. In instances where significant offsets need to be applied
to the photometric catalogues, the code is therefore unlikely to perform well relative to template-based codes where these offsets can be
directly estimated from the training data.
\end{itemize}

\subsubsection{LePhare}

LePhare (\cite{Arnouts2002,Ilbert2006}) is a public template fitting code that uses a $\chi^2$ minimization of differences between observed and theoretical magnitudes to find the best template (galaxy type) and redshift. The code also provides for each object upper and lower 1-sigma limits for this estimate, a maximum likelihood estimate for the redshift, K-corrections and absolute magnitudes for each band and a probability distribution function. An adaptive method can be used to improve theoretical magnitudes, as well as an $N(z)$ prior to minimize catastrophic errors. The effect of emission lines on the theoretical magnitudes can be estimated and taken into account. Several sets of SEDs and extinction laws are available in the code to be used. 
For this study we use the five \magdetmodel magnitudes.

\begin{itemize}
\item {\it Templates:} A set of 66 SEDs were used in the analysis of the CFHTLS data by
\cite{Ilbert2006} obtained from interpolation of the largely used \cite{Coleman1980} templates for different Hubble types and \cite{Kinney1996} for starburst galaxies. Since template fitting codes are time consuming, we searched for a reduced group of templates from this large set, yielding essentially the same overall statistics as far as dispersion and outlier fraction are concerned. We performed several tests using the VVDS-02hr sample with available spectroscopic redshifts and found a reduced set of 21 templates encompassing SEDs for 12 Ellipticals, 6 Spirals, 1 Im and 2 starburst with satisfactory results. 
Several tests removing the $u$-band from VVDS and CFHTLS data indicated that the discrepancies of photo-$z$ from the true (spectroscopic) value increase due to galaxy type--extinction degeneracy. From these tests we concluded that in order to minimize this problem we should keep the extinction values E(B-V) less or equal to 0.25 (for types Scd and later) and use only three very late type SEDs (1 Im and 2 starbursts).
\item {\it Prior:} Although the $N(z)$ prior available in LePhare refers to the (B-I) color from the VVDS survey, we verified that applying the procedure to the $g-i$ DES color we achieved a significant minimization of the outlier fraction, and therefore we used this prior.
\item {\it Training:} The adaptive training method available in LePhare was used to obtain a re-calibration (zero-points offsets) in each band. This procedure was first applied to the training sets and the shifts obtained were used when the code was run on the testing samples.
\end{itemize}

\subsubsection{ZEBRA}

The Zurich Extra-galactic Bayesian Redshift Analyser (\cite{Feldmann2006}) is a flexible and powerful photometric redshift code, based around template fitting. The code produces a posterior distribution for each galaxy in redshift and template space, $P(z,T)$, as well as marginalized distributions for $P(z)$ and $p(T)$. For redshift computation, the filters were smoothed over a scale of $100\AA$, and the templates smoothed over $30\AA$. The tests were conducted with a redshift resolution in linear steps of $\Delta z=0.01$. The five {\tt MAGDETMODEL} $grizY$ magnitudes  were used in this analysis.

\begin{itemize}
\item {\it Templates:} $81$ templates were used in ZEBRA's Bayesian mode. These templates were selected from a super-set of SEDs, and consist of the most frequent best-fit templates at $z=z_{{\rm spec}}$ for galaxies in the training sample. The super-set of template SEDs were produced by ZEBRA's template correction module from log interpolations between the \citet{Coleman1980} and \citet{Kinney1996} templates. The module allows the user to define redshift intervals within which the templates are modified to better fit the input photometry. For this stage a photometric sample from the COSMOS field was used. A fraction of the testing set galaxies have counterparts within this photometric sample, but the photometry is independent from the DES-SV data and the specroscopic redshifts were not used.
\item {\it Prior:} In Bayesian mode, ZEBRA constructs a self-consistent iterative prior from the galaxy likelihood functions, $L(z,T)$. The approach is similar to that taken by Brodwin et al. (2006), but operates in 2-dimensional redshift and template space. The prior constructed from the training sample was used for both samples. 
\item {\it Training:} \cite{Bordoloi2010} describe a method of using known (spectroscopic) redshifts to correct the individual marginalized redshift probability distributions, $P(z)$. The method demands that the spectroscopic redshifts sample their respective $P(z)$ fairly, i.e. the distribution of cumulative probabilities between zero and $z_{{\rm spec}}$ should be flat. We apply a simple first pass of their approach in bins of redshift, with width $\Delta z_{{\rm phot}} = 0.1$. Galaxies were assigned to these bins based on their maximum posterior redshift. After correction of the individual $P(z)$, a new $z_{{\rm phot}}$ was computed as the maximum of the corrected $P(z)$.
\end{itemize}

\subsubsection{PhotoZ}

\texttt{PhotoZ} \citep{Bender2001} is a Bayesian template fitting photometric redshift code.
The redshift probability of an object is obtained by multiplying the probability of a $\chi^2$ fit of template SEDs by prior probabilities for redshift and luminosity.
The total probability of a model then reads:
\begin{eqnarray}
P(\vec{\mu}|m)\propto \mathcal{L}(m|\vec{\mu})\cdot P(\vec{\mu}),
\end{eqnarray}
where $m$ denotes the photometric data (in magnitudes or fluxes), and $\vec{\mu}$ are the model parameters, i.e., redshift $z$ and luminosity $M$. In this analysis we used the five  {\tt MAGDETMODEL} $grizY$ magnitudes.
\begin{itemize}
\item {\it Templates:} The template set we use contains templates ranging from starforming (blue) to passively evolving (red) galaxies. It includes model SEDs from \cite{Bender2001}, which were created from spectroscopically observed objects from the Hubble Deep Field North. Another three templates (an S0, Sac, and an Sbc galaxy) are from \cite{Mannucci2001}, and two empirical SEDs (of an Scd and an Sbc galaxy) are from \citet{Coleman1980}. Our model set additionally includes 13 SEDs from \cite{Ilbert2006} which are based on spectra from \citet{Coleman1980} and were optimized to match local star-forming galaxies. This is a combination of template sets already used in the past for photometric redshift estimation (e.g., \citealt{Bender2001} and \citealt{Brimioulle2013}). Furthermore, we incorporate a set of red SEDs in our model set which were created in order to match the colors of luminous red galaxies (LRGs) from SDSS-II \citep{Greisel2013}.
\item {\it Priors:} The redshift and luminosity priors have the form $P(x)\propto\exp\left(-((x-\hat{x})/\sigma)^p\right)$, where $\hat{x}$, $\sigma$, and $p$ are defined individually for each model SED. Setting $\hat{x}$, $\sigma$, and $p$ accordingly, we can, for instance, decrease the probability of observing red models at higher redshifts ($z\gtrsim0.9$), or of galaxies that are too bright in absolute magnitude to exist. In addition to that, we adapt the $z$ and $M$ (absolute magnitude) priors for every model SED in such a way that photometric redshift outliers with $|z_{phot}-z_{spec}|/(1+z)>0.15$ \citep{Ilbert2006} in the Main and Deep DES-SV training sets are less likely. Therefore, we identify their location in the $z$ vs. $M$ space and modify the priors in such a way that they assign smaller probabilities to those regions. This is done solely if the outliers of a template are isolated from good photometric redshift estimates of the same template in the $z$ vs. $M$ space.
\item {\it Training:} 
We iteratively adapt the zero-points for the training catalogs using the median magnitude offsets between the data and the model predictions while optimizing the photometric redshift performance.
\end{itemize}

\subsection{Results of the photo-$z$ analyses}
After the description of the codes, we turn to the study of their performance in a number of tests using different configurations of the data samples. While most of these tests focus on estimating how the photo-$z$ determination will perform for the standard DES data, others look for improvements by using deeper photometry or additional bands. We also check the differences in the results under variations in the calibration data and the weights used. Note that the results presented in this subsection are those considering all the galaxies (with quality cuts), which are represented by one single statistic, later in the paper we analyze some of these results in more detail.
\subsubsection{Test 1: \textit{Main-Main}}
\label{sec:test1}
This test is the most representative of the results shown in this paper, the default case. We use here the Main training sample to train and calibrate the photo-$z$ algorithms and the Main testing sample to validate them, therefore, the test represents the real situation for most of the data collected in the DES survey. 

\begin{figure*}
\centering
\includegraphics[width=150mm]{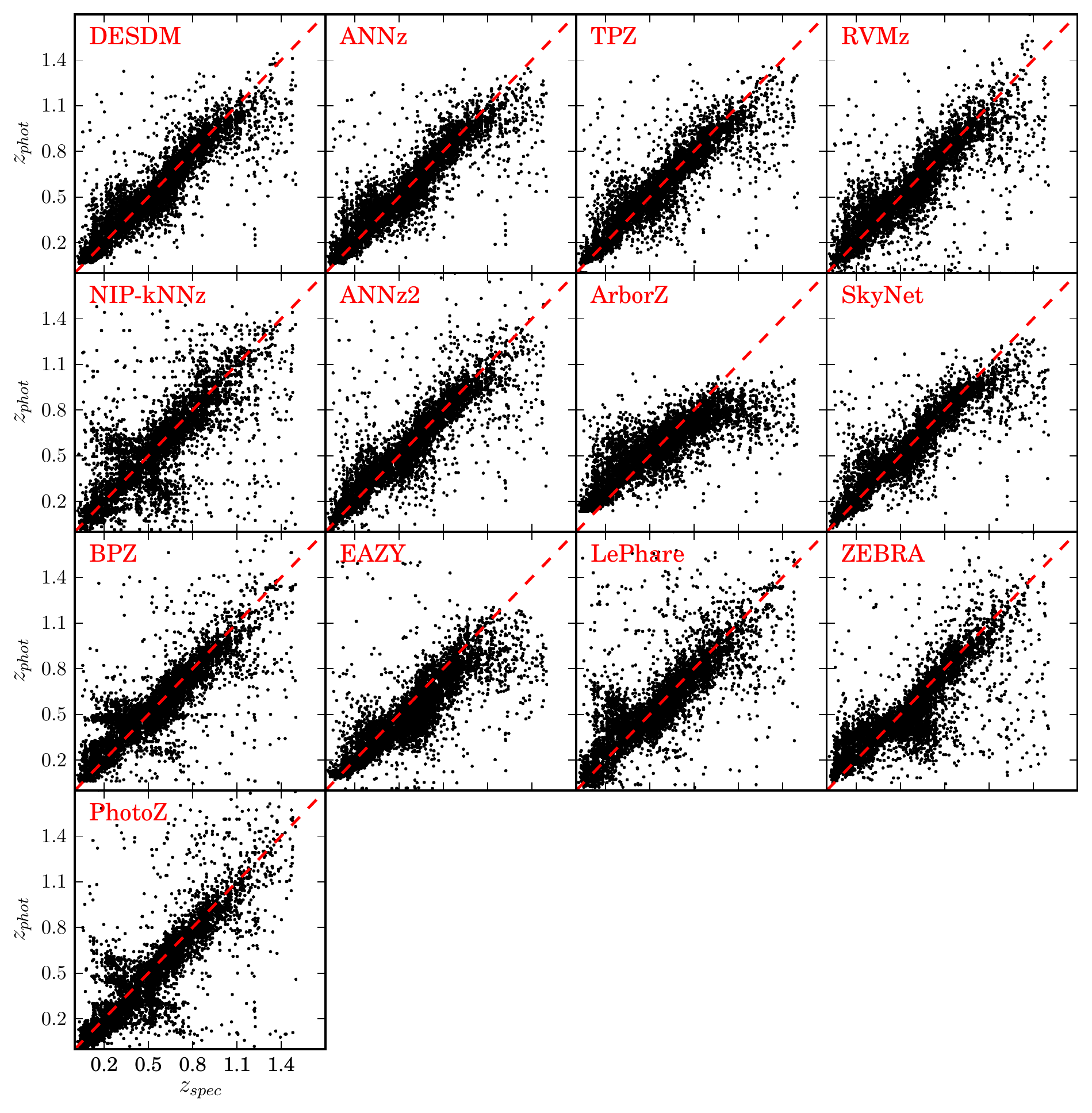}
\caption{$z_{phot}$ vs. $z_{spec}$ scatter plot for all the codes analyzed in Test 1 and listed in Table \ref{tab:codes}.}
\label{fig:scatters}
\end{figure*}

\begin{figure}
\centering
\includegraphics[width=90mm]{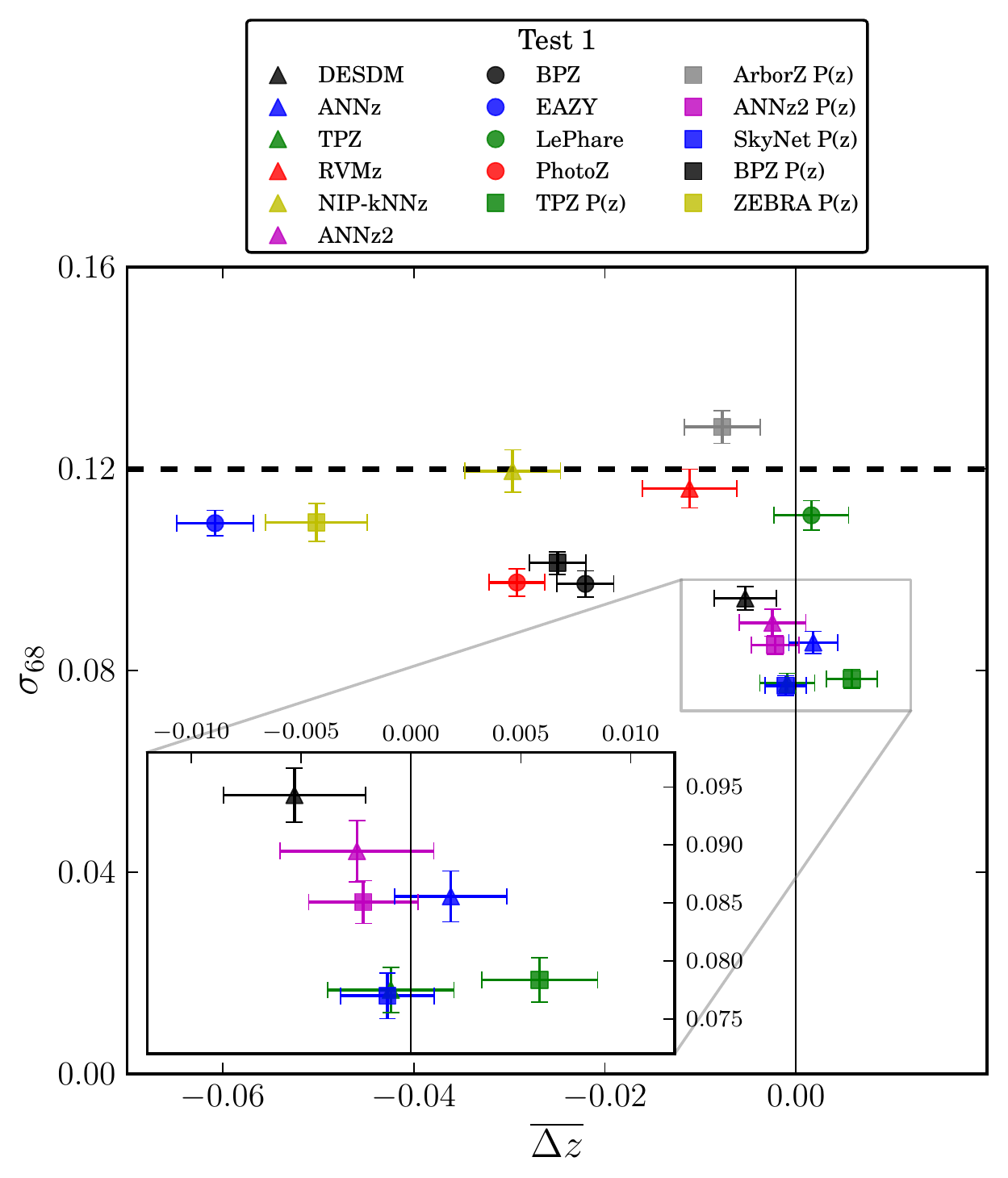}
\caption{$\sigma_{68}$ vs. bias for all the codes analyzed in Test 1. Black dashed lines represent the DES science requirements in this and subsequent figures. Training-based codes have triangles as markers, template-based have circles, and codes producing a probability density function (pdf) for the redshift are marked with a square. This will also be the convention for the next figures. Training-based codes, either producing a single photo-$z$ estimate or a pdf, are the only ones present in the region of best performance (zoom-in).}
\label{fig:sigma68vsbias_test1}
\end{figure}

\begin{figure}
\centering
\includegraphics[width=90mm]{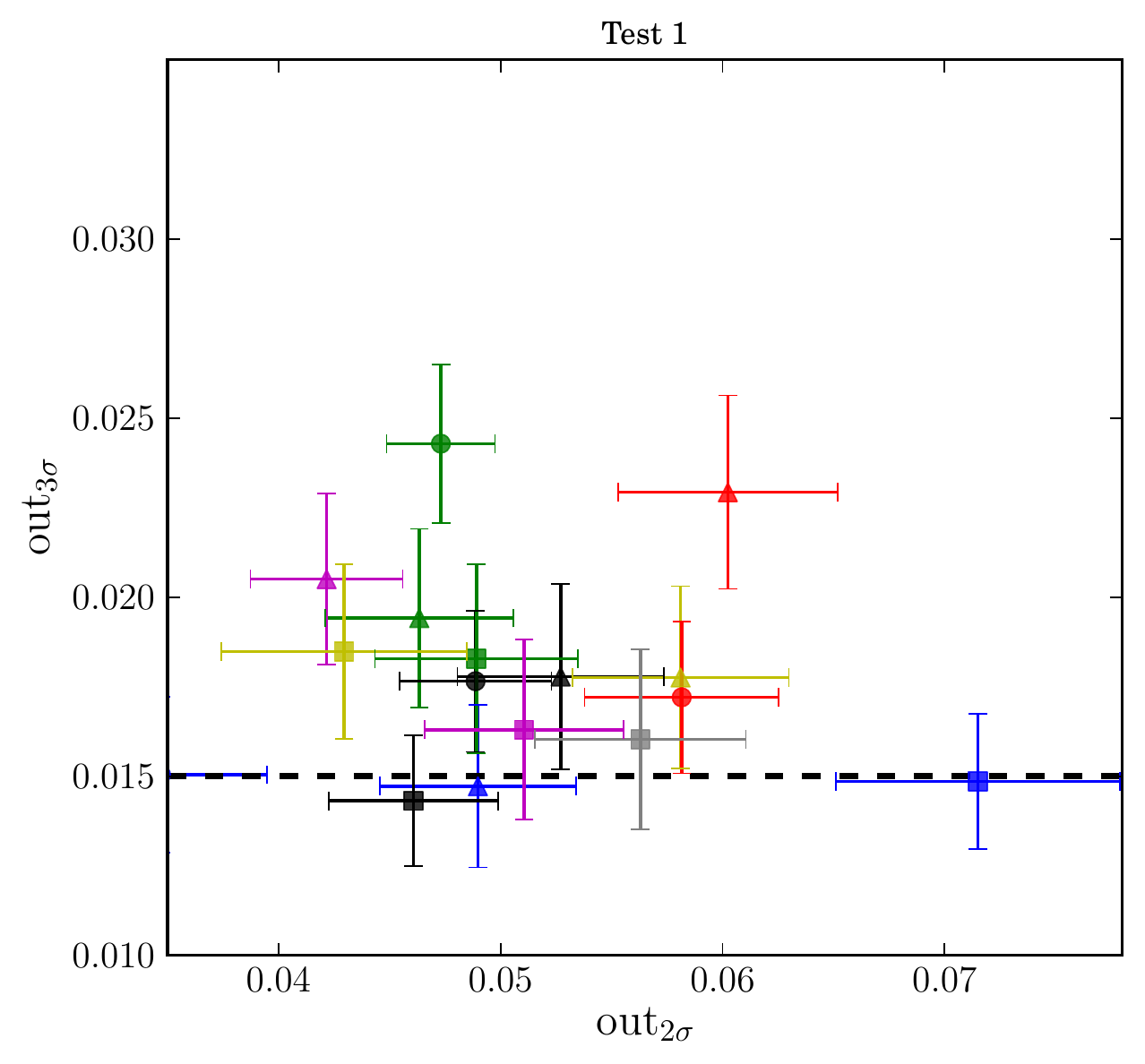}
\caption{$3\sigma$ vs. $2\sigma$ outlier fraction for all the codes analyzed in Test 1. The results are more homogeneous than those for $\sigma_{68}$ vs. bias (Fig. \ref{fig:sigma68vsbias_test1}).}
\label{fig:frac3vsfrac2_test1}
\end{figure}

\begin{figure}
\centering
\includegraphics[width=90mm]{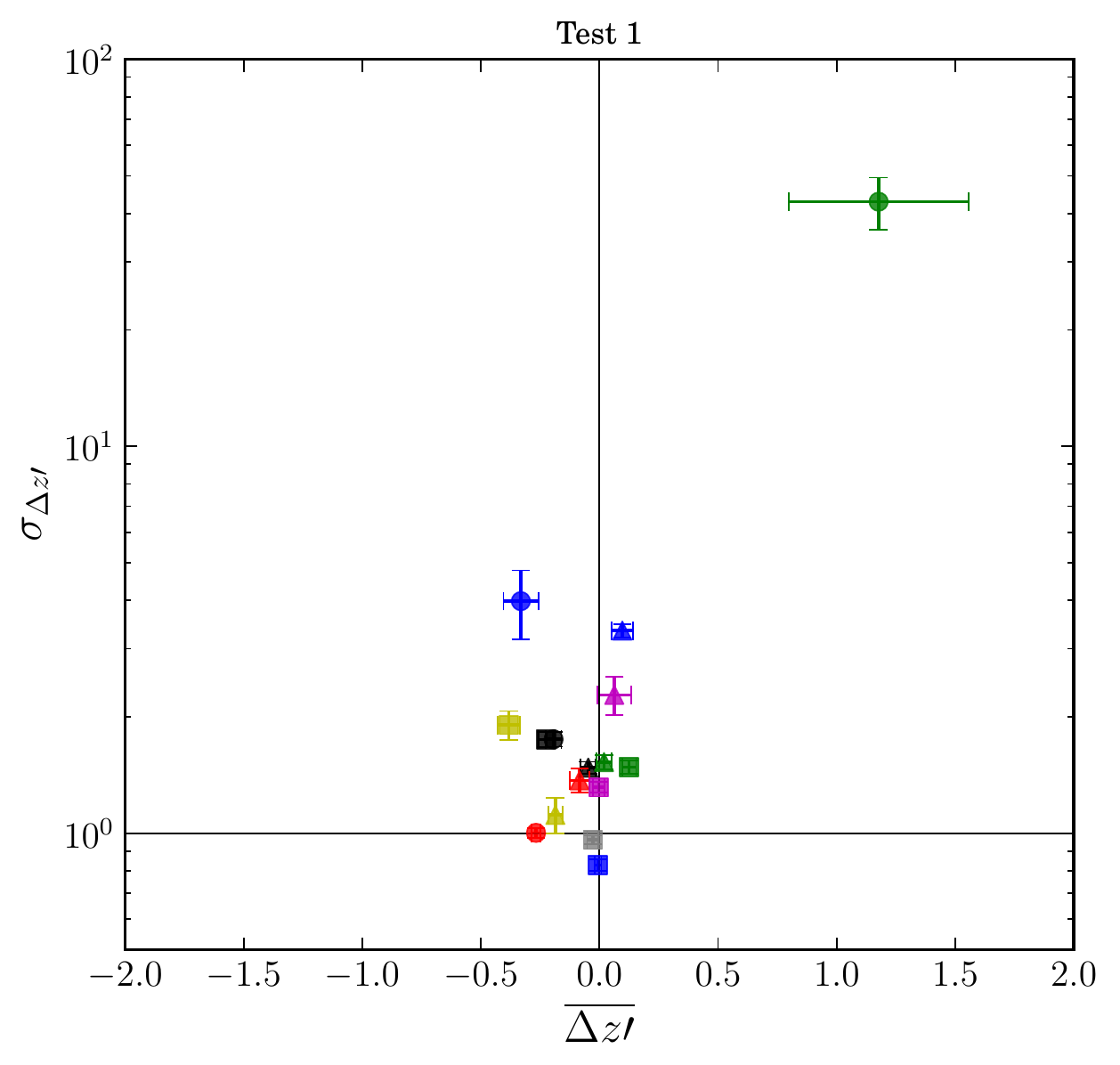}
\caption{RMS vs. bias of the normalized $\Delta z$ distribution $\Delta z' = \Delta z / \epsilon_{\rm phot}$ for all the codes analyzed in Test 1. Most codes underestimate $\epsilon_{\rm phot}$, leading to large values of $\sigma_{\Delta z\prime}$.}
\label{fig:sigmanormvsbiasnorm_test1}
\end{figure}

\begin{figure}
\centering
\includegraphics[width=90mm]{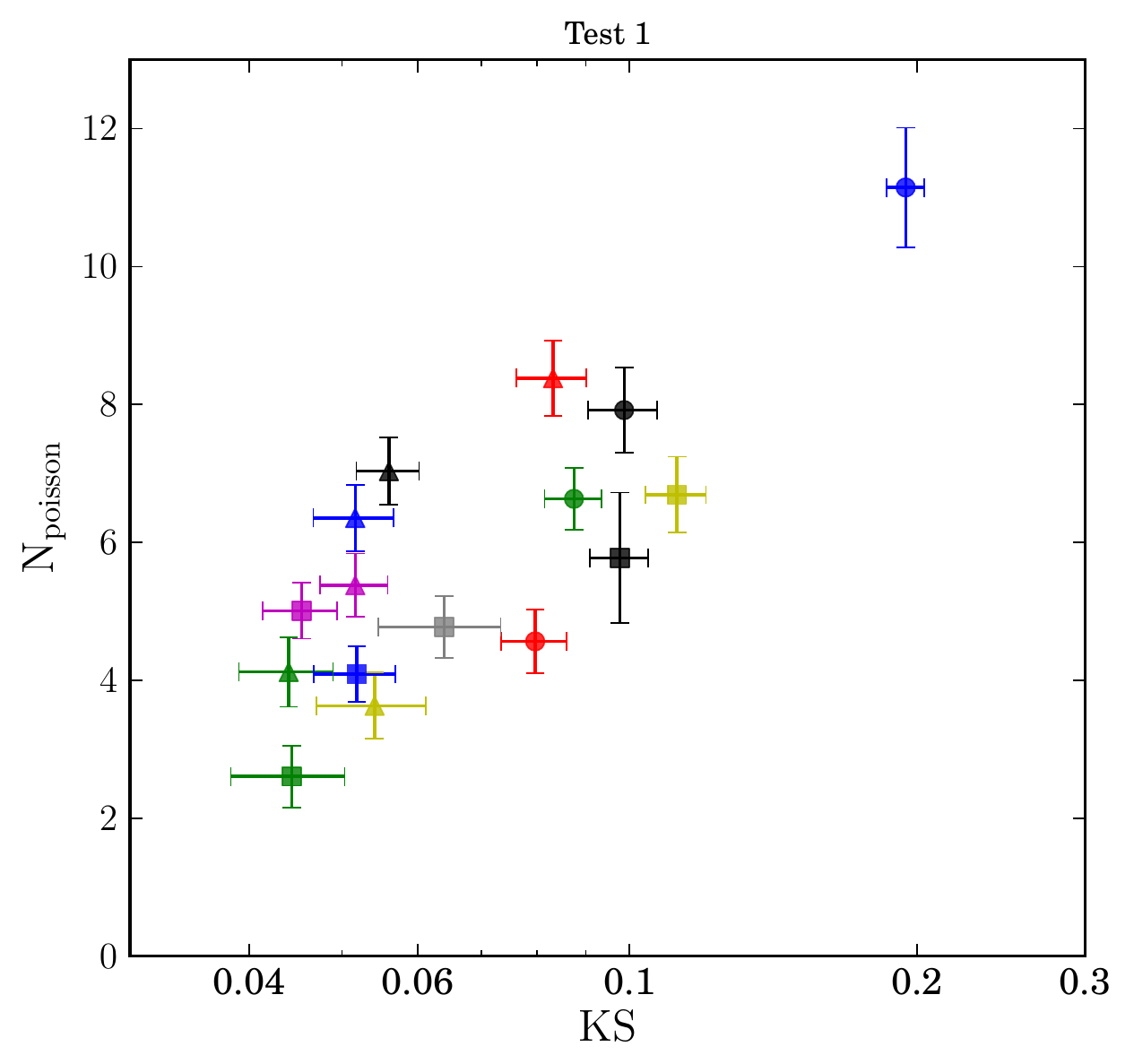}
\caption{N$_{\rm poisson}$ vs. KS statistics for all the codes analyzed in Test 1. Both metrics show how the true galaxy redshift distribution is reconstructed through photo-$z$'s, for each code. The smaller the value of the metric, the better the reconstruction. A strong correlation between the two metrics is observed, as expected.}
\label{fig:npoissonvsks_test1}
\end{figure}

In order to display the performance of all codes, and only for this test in particular, in Fig. \ref{fig:scatters} we show the $z_{phot}$ vs. $z_{spec}$ scatter plot for all the codes listed in Table \ref{tab:codes}. Furthermore, we compute all the metrics presented in Table \ref{tab:def_metrics} and described in Appendix \ref{sec:appendix_a}. The results, using all the objects in the testing sample except for the 10\% quality cut mentioned above, are shown in Table 6 and Figs \ref{fig:sigma68vsbias_test1}-\ref{fig:npoissonvsks_test1}. The legend is only shown in Fig. \ref{fig:sigma68vsbias_test1}, but applies to subsequent figures corresponding to this test. 

Figure \ref{fig:sigma68vsbias_test1} shows $\sigma_{68}$, related to the precision of the photometric redshifts (and defined in Appendix \ref{sec:appendix_a}), versus the mean bias of the photo-$z$'s. The black dashed line sets the DES science requirement on $\sigma_{68}$, and one can check how most of the codes presented in this work are below this line, thus fullfilling this important requirement on precision. Also, among the codes satisfying the $\sigma_{68}$ requirement, there is a subgroup having very low bias as well. In Fig. \ref{fig:sigma68vsbias_test1} we show a zoomed-in of this region of interest, where we can see how training-based codes, either producing a single photo-$z$ estimate or a probability density function, $P(z)$, are the ones showing best performance (all the codes in the zoomed-in region belong to the training-based category). 

Figure \ref{fig:frac3vsfrac2_test1} shows the $3\sigma$ vs. $2\sigma$ outlier fractions for Test 1. The requirement on the $2\sigma$ outlier fraction (0.1) is beyond the range of the plot, meaning that all codes fullfill this. However, the $3\sigma$ outlier fraction requirement, shown as the black dashed line, is only met by a few codes. Among these codes, there are cases from the two types of photo-$z$ codes, training and template-based. Also, there is more homogeneity in this plot: many codes agree with others within error bars. Both of these requirements are set based on the spread of the $\Delta z$ distribution and no with respect to a fixed distance from the mean, therefore these values quantify how sharp the $\Delta z$ distribution is with respect to its center.

Figure \ref{fig:sigmanormvsbiasnorm_test1} shows RMS vs. bias for the $\Delta z$ distribution normalized by its error ($\Delta z'$) of each of the codes analyzed in Test 1 ($\sigma_{\Delta z'}$ vs. $\overline{\Delta z'}$ in our notation). A large fraction of the codes yield very high values of $\sigma_{\Delta z'}$ (expected to be close to 1), meaning that all these codes underestimate their photo-$z$ errors, however, there is a group of codes with normalized $\Delta z$ distributions approaching a Gaussian with mean equals to zero and variance equals to one. Also, we do not see a particular type of photo-$z$ code being problematic here, both training and template-based codes populate good and bad regions of the plot.

One crucial aspect of photo-$z$ studies, which will be discussed in more detail in Section~\ref{subsec:desdm_results}, is the estimation and calibration of the true galaxy redshift distributions. In this paper we use two metrics to compare the reconstruction of the true redshift distribution by the different photo-$z$ algorithms: the N$_{\rm poisson}$ and KS statistics, defined in Appendix A. In both cases, the smaller the value, the closer are the true redshift distribution and its reconstruction through photo-$z$'s. Figure~\ref{fig:npoissonvsks_test1} shows these values for all the codes analyzed in Test 1. As expected, the two metrics are strongly correlated. It can also be seen how having a redshift pdf for each galaxy, instead of a single-estimate photo-$z$, helps a given code to have a better redshift reconstruction. This can be inferred looking at the cases where both the pdf and the single-estimate are displayed (TPZ, ANNz2, BPZ): in all these cases the pdf version of the code obtains better results in terms of these two metrics. As for the results, TPZ and the nearest-neighbor code, NIP-kNNz, show the best performance in this regard.

To summarize the results from Test 1, we note that most of the codes presented in this work fulfill the requirements for $\sigma_{68}$ and $2\sigma$ outlier fraction, while only a few fulfill the $3\sigma$ outlier fraction requirement. Also, training-based codes seem to yield better photo-$z$ precision on average and better $N(z)$ reconstruction, but, when evaluating other quantities like outlier fraction or the estimation of photo-$z$ errors, there is no a clear indication as of which class of photo-$z$ approach show more accurate metrics. As pointed out in \cite{CarrascoKind2014} these results might vary for different regions on the multidimensional photometric space or within the redshift range. Usually, training-based algorithms perform better on areas well populated with training galaxies and poorly on those less dense regions (as in high redshift bins), fact that we can observe from Figure ~\ref{fig:scatters} where training-based methods tend to have tighter distributions at the center while some template-based methods can compute \pzs for galaxies at higher redshift more efficiently.

\subsubsection{Test 2: \textit{Deep-Deep}}

\begin{figure}
\centering
\includegraphics[width=90mm]{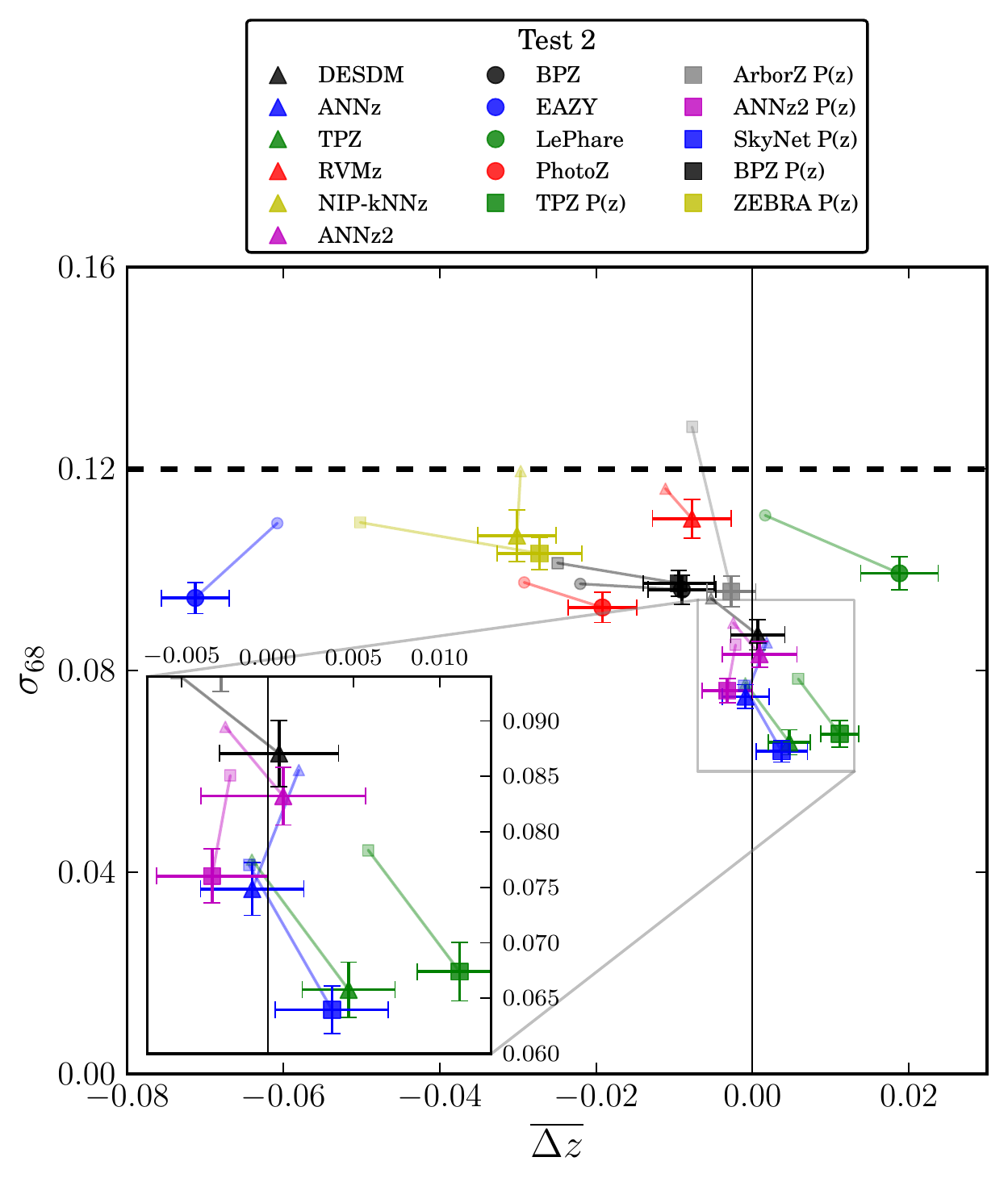}
\caption{$\sigma_{68}$ vs. bias for all the codes analyzed in Test 2. Results of Test 1 for each code are shown as a point without error bars (for simplicity) and connected to the result in this test through a solid line. This convention will also be used in subsequent plots for easier comparison against Test 1, which is the default case. All codes improve in photo-$z$ precision ($\sigma_{68}$) due to the deeper photometry.}
\label{fig:sigma68vsbias_test2}
\end{figure}

In this test we train or calibrate the algorithms using the Deep training sample and we apply them on the Deep testing sample (see results in Table 7). The goal of this test is to check the differences in photo-$z$ performance when we use higher S/N data, with the caveat that this sample has also a slightly different redshift range. In order to enable an easier comparison with Test 1, we include in the plots for this Test 2 the data from the analogous plots in Test 1, and we do it by including the points for Test 1 (without error bars, for simplicity) connected by a straight line to the corresponding point for Test 2 (with error bars). Figure \ref{fig:sigma68vsbias_test2} shows $\sigma_{68}$ versus the mean bias of the computed photo-$z$'s for Test 2. This is analogous to Fig. \ref{fig:sigma68vsbias_test1}. There we can see that there is general improvement in $\sigma_{68}$: most of the codes move to lower values in the plot with respect to Test 1. So we conclude that using higher S/N photometry observations increases the photometric redshift precision of a sample. However, in the case of the bias there is no general trend: some codes improve and some others do not. 

\begin{figure}
\centering
\includegraphics[width=90mm]{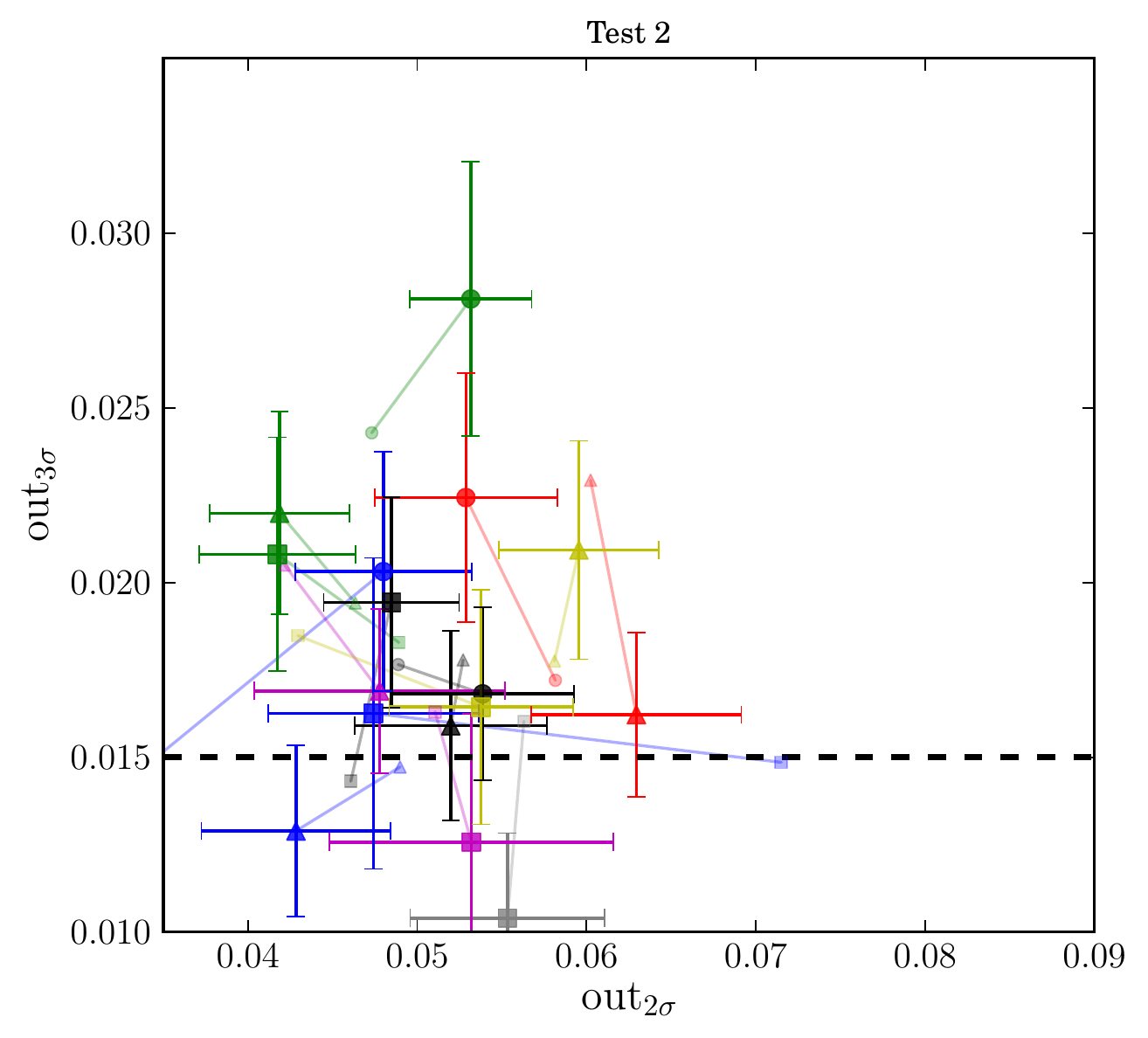}
\caption{$3\sigma$ vs. $2\sigma$ outlier fraction for all the codes analyzed in Test 2. This test shows a larger scatter in the points with respect to the results in Test 1: in general, codes doing well for Test 1 improve in Test 2 while codes doing not so well for Test 1 show a worsening in Test 2.}
\label{fig:frac3vsfrac2_test2}
\end{figure}

\begin{figure}
\centering
\includegraphics[width=90mm]{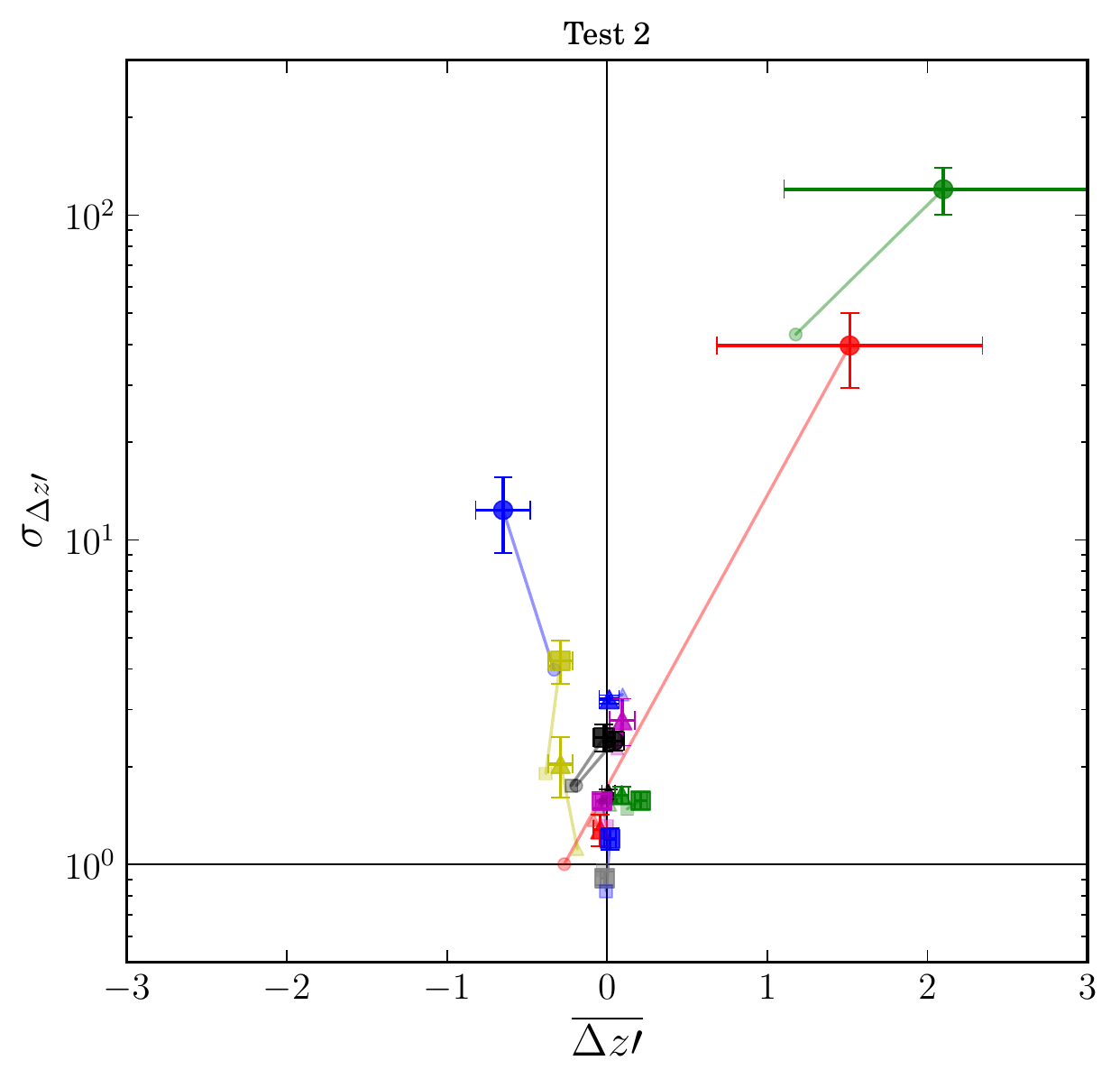}
\caption{RMS vs. bias of the normalized photo-$z$ distribution for all the codes analyzed in Test 2. Here the results are, in general, close to those observed in Test 1.}
\label{fig:sigmanormvsbiasnorm_test2}
\end{figure}

Figures \ref{fig:frac3vsfrac2_test2} and \ref{fig:sigmanormvsbiasnorm_test2} are analogous to Figs. \ref{fig:frac3vsfrac2_test1} and \ref{fig:sigmanormvsbiasnorm_test1}, respectively. In the plot showing the outlier fraction (Fig. \ref{fig:frac3vsfrac2_test2}), Test 2 seems to introduce an additional scattering, i.e. codes that did better in Test 1 seem to improve more in Test 2 than codes that did not do so well in Test 1, which now seem to worsen on average. This fact somehow removes the homogeneity we saw in Fig \ref{fig:frac3vsfrac2_test1}, where many codes produced results compatible within errors. Figure \ref{fig:sigmanormvsbiasnorm_test2} shows the RMS and bias for the normalized $\Delta z$ distribution, which, similarly to Test 1, shows that many codes underestimate the photo-$z$ errors. The behavior here is similar to the one regarding outlier fractions: Test 2 introduces even more differences between codes doing well and codes that are less accurate.

\subsubsection{Test 3: \textit{Deep-Main}}

Test 3 uses the Deep training sample for training or calibration of the algorithms and uses the Main testing sample to compute photo-$z$'s (see results in Table 8 in the Appendix). In contrast to Test 2, this is a realistic case since a Deep training sample already exists and is available to use for photo-$z$ calibration of DES data, with the same depth as the Main testing sample. Therefore, this test explores the possibility of improving the photo-$z$ performance of Test 1 by using higher S/N data for training or calibration. 

\begin{figure}
\centering
\includegraphics[width=90mm]{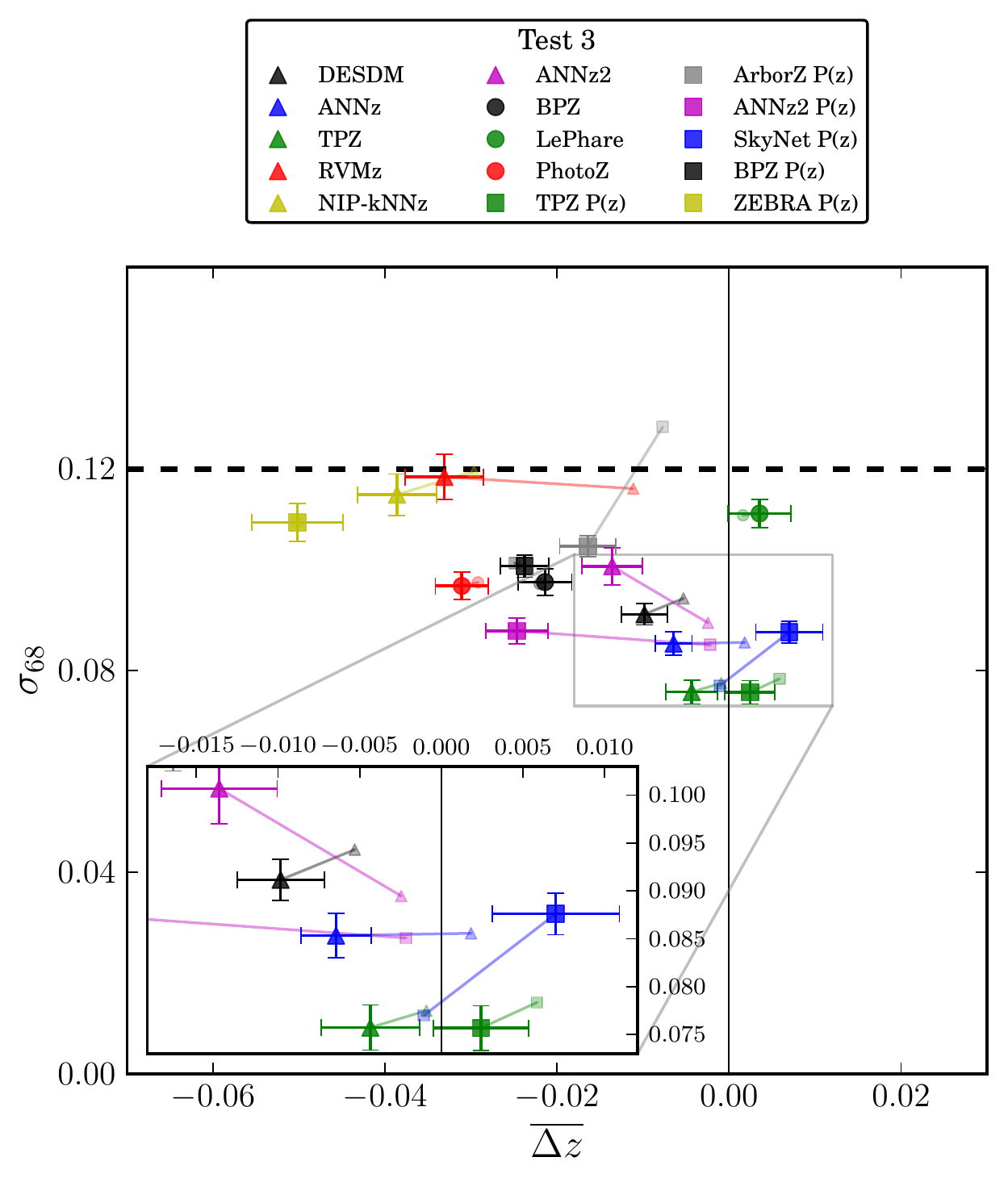}
\caption{$\sigma_{68}$ vs. bias for all the codes analyzed in Test 3. Results are overall compatible with Test 1 although some particular codes, such as ArborZ, show a significant improvement by using deeper data for training.}
\label{fig:sigma68vsbias_test3}
\end{figure}

\begin{figure}
\centering
\includegraphics[width=90mm]{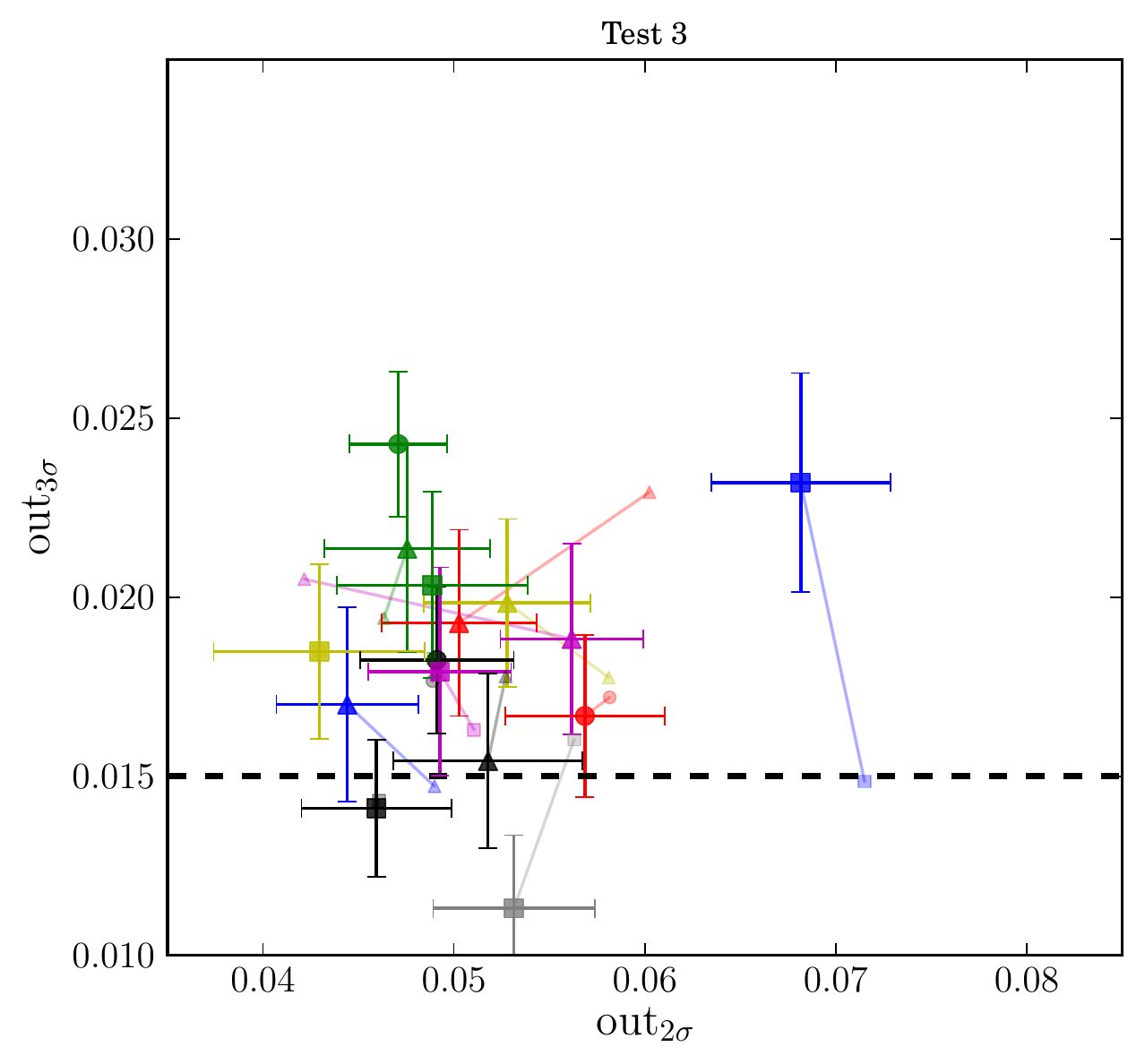}
\caption{$3\sigma$ vs. $2\sigma$ outlier fraction for all the codes analyzed in Test 3. There is a general agreement with the results from Test 1.}
\label{fig:frac3vsfrac2_test3}
\end{figure}

\begin{figure}
\centering
\includegraphics[width=90mm]{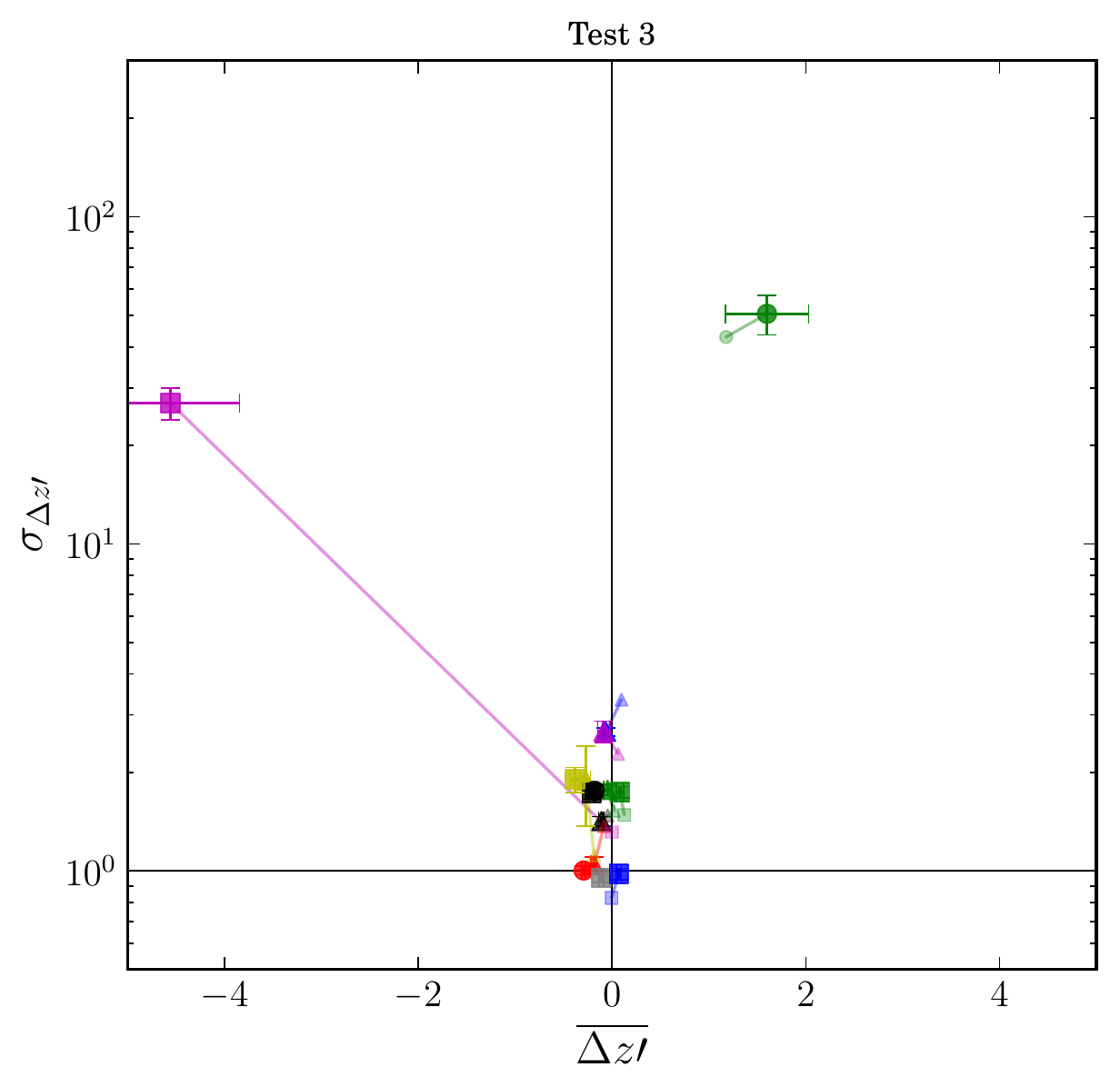}
\caption{RMS vs. bias of the normalized photo-$z$ distribution for all the codes analyzed in Test 3. Again, results generally agree with Test 1.}
\label{fig:sigmanormvsbiasnorm_test3}
\end{figure}

Figure \ref{fig:sigma68vsbias_test3} shows the results on precision and bias for Test 3, and, as before, compares with Test 1. Some template-based codes are not included in this test so less points are shown with respect to Fig. \ref{fig:sigma68vsbias_test1}. In this test, basically all codes fulfill the $\sigma_{68}$ requirement. In addition, some codes show important improvement when using this higher S/N data for training, such as ArborZ. However, there is no general $\sigma_{68}$ improvement as there was for Test 2, and the results are generally comparable with those from Test 1. Figures \ref{fig:frac3vsfrac2_test3} and \ref{fig:sigmanormvsbiasnorm_test3} show also a high degree of compatibility with Test 1, contrary to what we saw in Test 2, where larger differences were appreciated. 

\subsubsection{Importance of the $u$ band}

In this part of the paper we want to adress the effect of adding $u$ band photometry to the photo-$z$ performance. It is important to stress that a $u$ band is available in DECam, although it is not planned to be used in the DES survey.

\begin{figure}
\centering
\includegraphics[width=90mm]{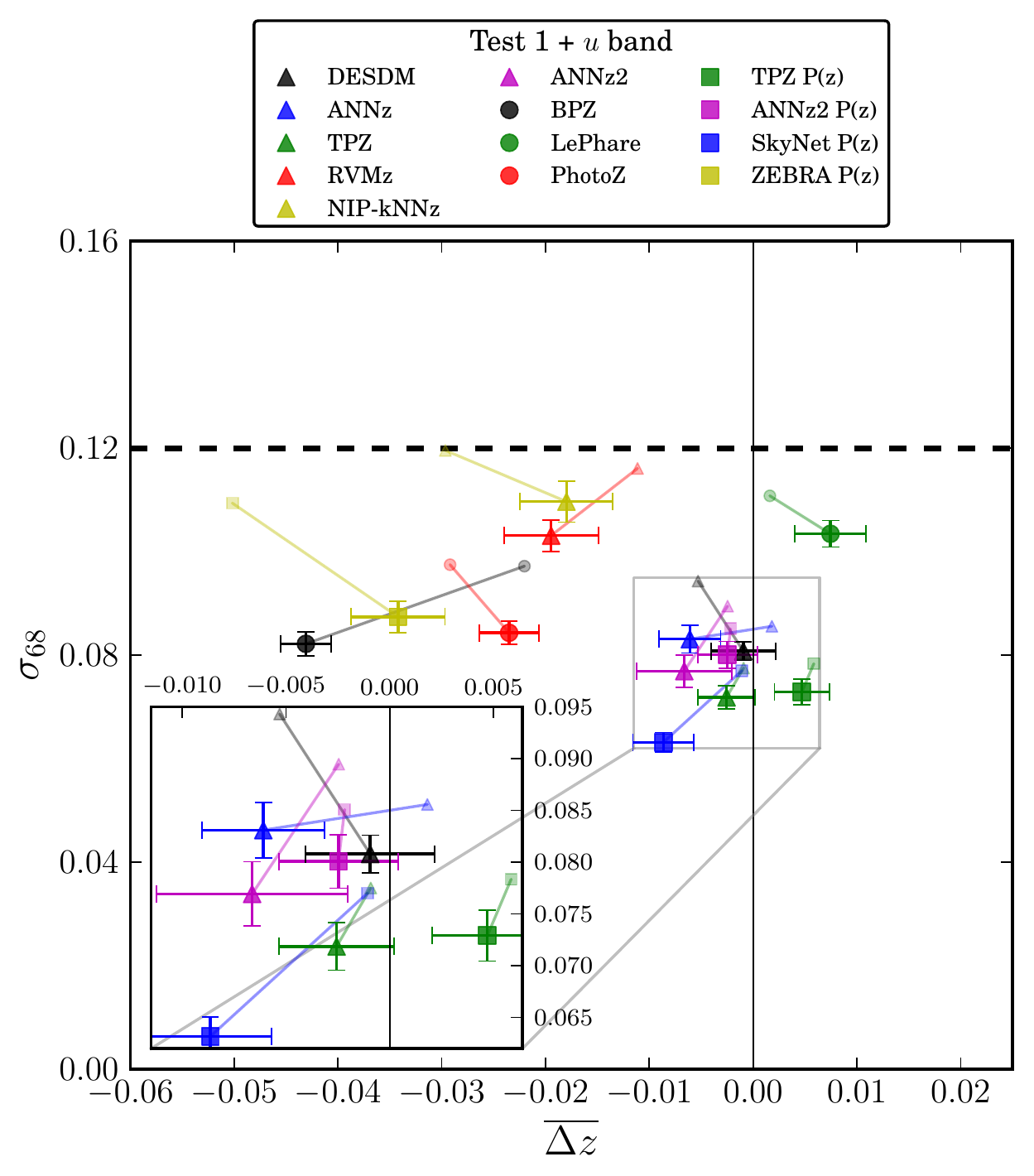}
\caption{$\sigma_{68}$ vs. bias for all the codes analyzed in Test 1 after the addition of the $u$ band. We see how the incorporation of this band significantly improves the photo-$z$ precision.}
\label{fig:sigma68vsbias_test1_u}
\end{figure}

We show the effect of incorporating the $u$ band to Test 1. On one hand, in Fig. \ref{fig:sigma68vsbias_test1_u} we can appreciate this effect in the overall photo-$z$ precision for the whole redshift range, and we clearly see an improvement on $\sigma_{68}$ with respect to Test 1 (shown again as the points without error bars connected through a solid line) for most of the codes present in the plot. 

\begin{figure}
\centering
\includegraphics[width=90mm]{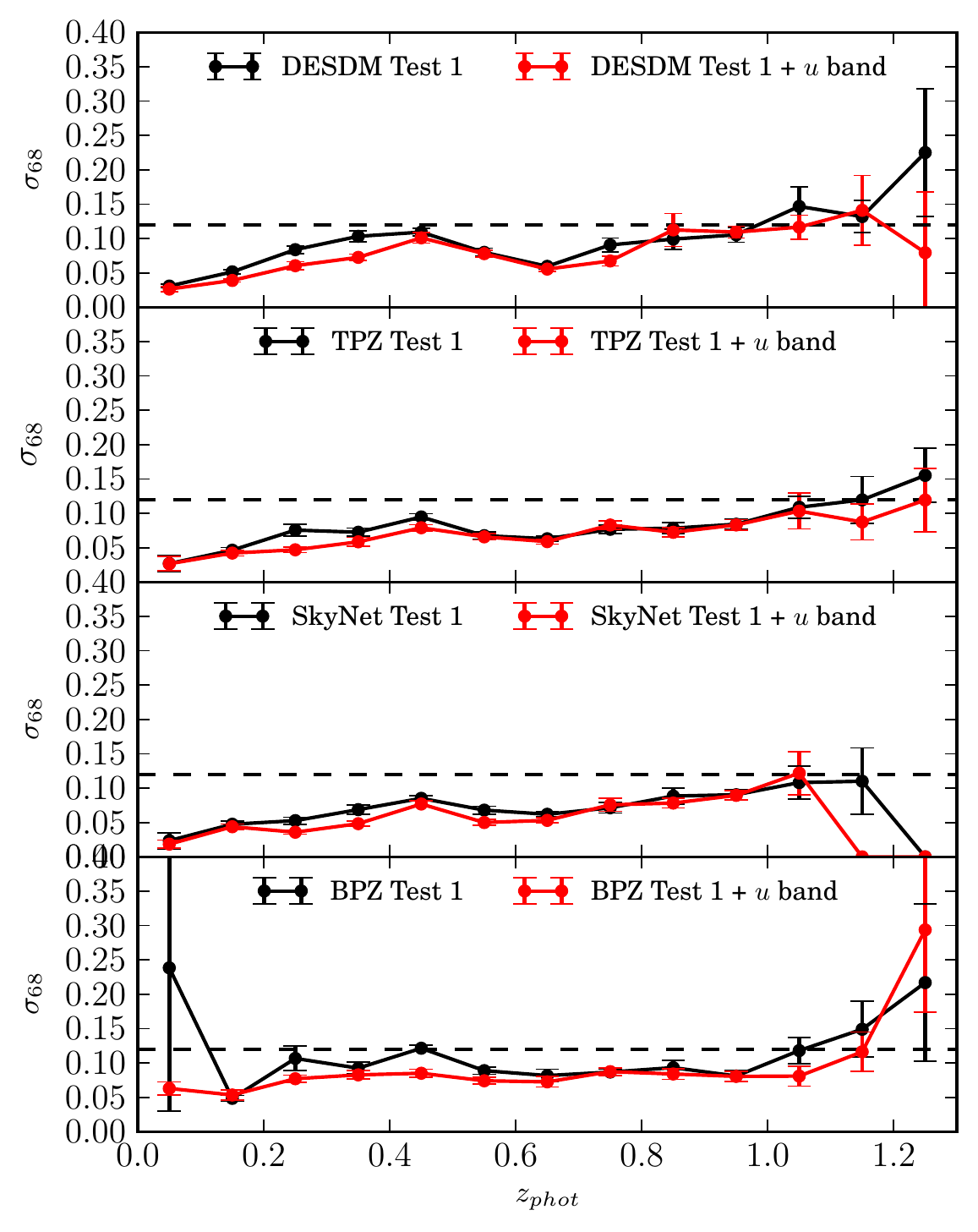}
\caption{$\sigma_{68}$ as a function of photometric redshift for 4 selected codes and for Test 1 + $u$ band. As expected, the improvement in photo-$z$ precision due to the $u$ band addition is more important at low redshift ($<0.5$).}
\label{fig:4est_u}
\end{figure}

\begin{figure}
\centering
\includegraphics[width=90mm]{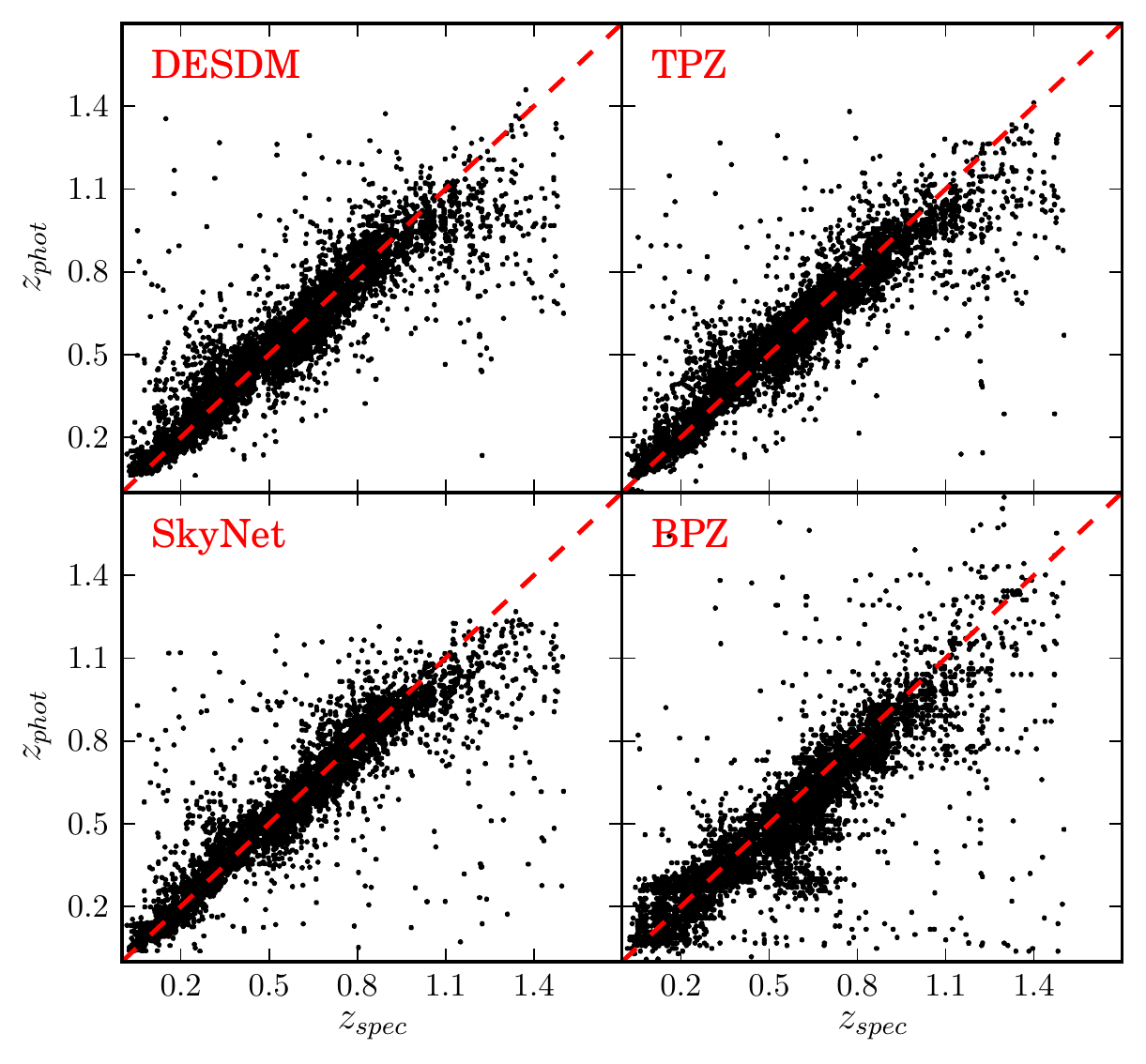}
\caption{$z_{phot}$ vs. $z_{spec}$ scatter plot for 4 selected codes for Test 1 + $u$ band. When comparing with the corresponding plots in Fig. \ref{fig:scatters}, a clear improvement at low redshift, with an important reduction in the number of outliers, can be appreciated.}
\label{fig:scatters_u}
\end{figure}

On the other hand, since $u$ is an ultraviolet band one expects it to be more relevant for low-redshift galaxies. Therefore, in Fig. \ref{fig:4est_u} we show the effect of adding the $u$ band to Test 1 in $\sigma_{68}$ as a function of photometric redshift for 4 selected photo-$z$ codes (the same ones we select for further analysis in \ref{subsec:desdm_results}). There we can clearly observe a signficant improvement in $\sigma_{68}$ at low redshift ($z_{phot} < 0.5$), while the precision at higher redshift is compatible within error bars in the two cases. Furthermore, in order to visually appreciate the improvement at low redshift, in Fig. \ref{fig:scatters_u} we show the $z_{phot}$ vs. $z_{spec}$ scatter plot after adding the $u$ band to Test 1 for the 4 same codes shown before, to be compared directly with Fig. \ref{fig:scatters}.

\subsubsection{Test 4: Importance of different spectrosopic sets}

In all previous tests we have considered a single training sample (in two versions: Main and Deep) constructed with joint data from different spectroscopic surveys with different properties, as described in Section \ref{sec:spectroscopic_sample}. Now we want to study the importance of the different major spectroscopic data sets used in the photo-$z$ performance using different photo-$z$ codes, both training and template-based. In particular, this test, that we call Test 4, consists on selecting from each of the 4 calibration fields depicted in Fig. \ref{fig:fields} the major spectroscopic set used in the field, then training or calibrating the algorithms using only this subset of spectroscopic objects in the training sample, and then apply the algorithms on the full testing sample, including all the spectroscopy available. In this way, we have selected spectroscopy from VVDS Deep in the SN-X3 field, from ACES in the SN-C3 field, from VVDS Wide in the VVDS F14 field and from zCOSMOS in the COSMOS field. For simplicity, we have considered only the Main training and testing samples (i.e. main survey depth photometry), so the results should be compared with Test 1.
 
\begin{table*}
\centering
\caption{$\sigma_{68}$ for the four cases in Test 4, corresponding to training on each of the four major spectroscopic samples and testing on the full Main testing sample. The results for Test 1 are also shown, for comparison.}
\label{tab:test4}
\tabcolsep=0.11cm
\begin{tabular}{l*{6}{c}r}
Codes, $\sigma_{68}$ & Test 1 & Test 4 VVDS Deep & Test 4 VVDS Wide & Test 4 ACES & Test 4 zCOSMOS\\
\hline
DESDM & 0.094 $\pm$ 0.002 & 0.106 $\pm$ 0.002 & 0.139 $\pm$ 0.005 & 0.103 $\pm$ 0.003 & 0.148 $\pm$ 0.008 \\
ANNz & 0.086 $\pm$ 0.002 & 0.101 $\pm$ 0.003 & 0.138 $\pm$ 0.004 & 0.091 $\pm$ 0.003 & 0.104 $\pm$ 0.003 \\
TPZ & 0.078 $\pm$ 0.002 & 0.090 $\pm$ 0.002 & 0.110 $\pm$ 0.005 & 0.093 $\pm$ 0.003 & 0.097 $\pm$ 0.003 \\
NIP-kNNz & 0.120 $\pm$ 0.004 & 0.146 $\pm$ 0.006 & 0.156 $\pm$ 0.007 & 0.127 $\pm$ 0.005 & 0.148 $\pm$ 0.007 \\
ANNz2 & 0.089 $\pm$ 0.003 & 0.099 $\pm$ 0.003 & 0.143 $\pm$ 0.008 & 0.104 $\pm$ 0.004 & 0.137 $\pm$ 0.006 \\
BPZ & 0.097 $\pm$ 0.003 & 0.096 $\pm$ 0.002 & 0.095 $\pm$ 0.002 & 0.095 $\pm$ 0.003 & 0.095 $\pm$ 0.002 \\
LePhare & 0.111 $\pm$ 0.003 & 0.110 $\pm$ 0.003 & 0.110 $\pm$ 0.003 & 0.111 $\pm$ 0.003 & 0.112 $\pm$ 0.003 \\
PhotoZ & 0.097 $\pm$ 0.003 & 0.101 $\pm$ 0.003 & 0.096 $\pm$ 0.002 & 0.096 $\pm$ 0.003 & 0.097 $\pm$ 0.003 \\
TPZ P(z) & 0.078 $\pm$ 0.002 & 0.091 $\pm$ 0.001 & 0.108 $\pm$ 0.004 & 0.093 $\pm$ 0.003 & 0.094 $\pm$ 0.002 \\
BPZ P(z) & 0.101 $\pm$ 0.002 & 0.096 $\pm$ 0.002 & 0.097 $\pm$ 0.002 & 0.097 $\pm$ 0.002 & 0.100 $\pm$ 0.002 \\
ANNz2 P(z) & 0.085 $\pm$ 0.002 & 0.103 $\pm$ 0.003 & 0.217 $\pm$ 0.009 & 0.103 $\pm$ 0.005 & 0.140 $\pm$ 0.009 \\
\end{tabular}
\end{table*}

\begin{figure}
\centering
\includegraphics[width=90mm]{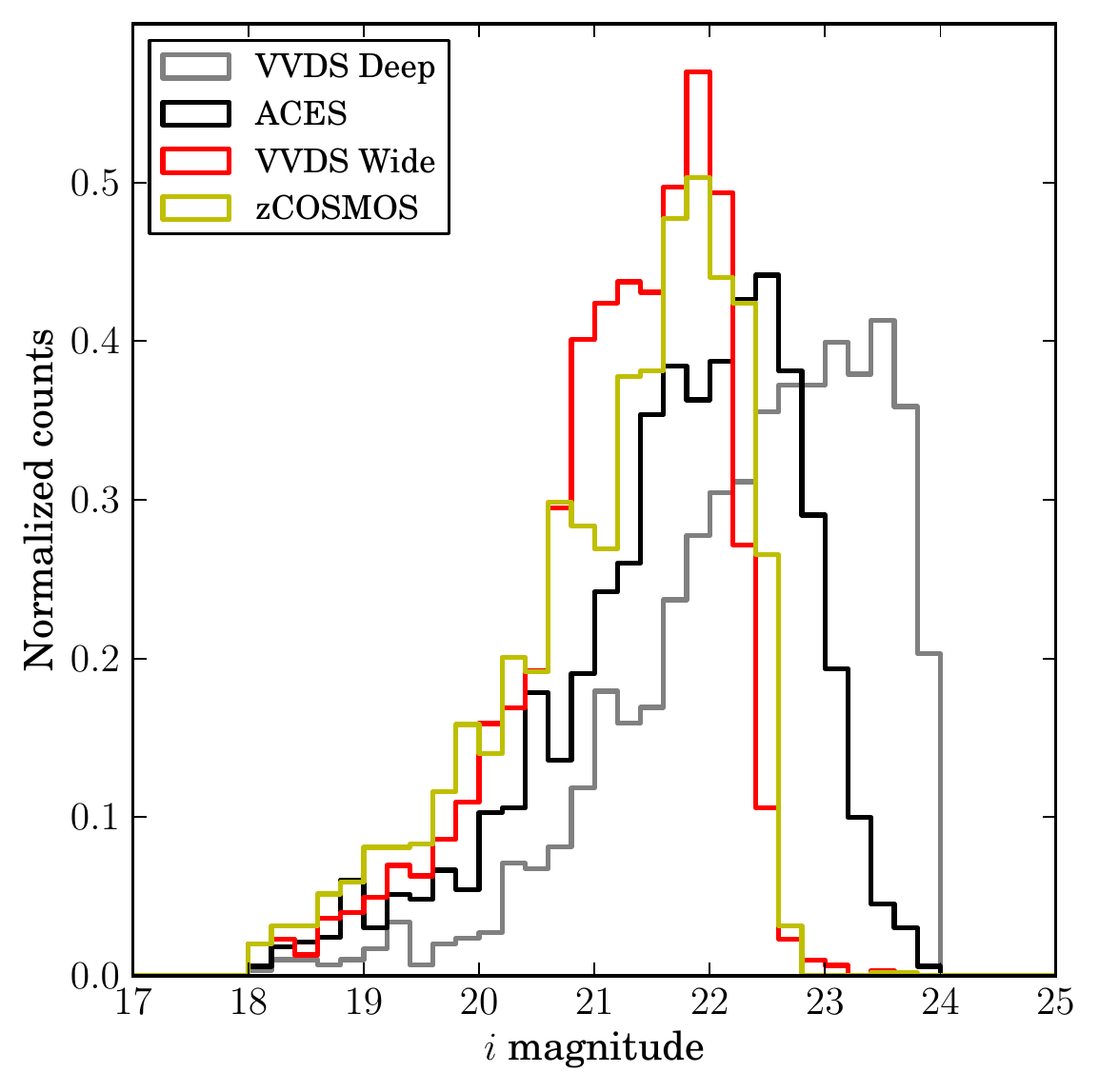}
\caption{$i$ band magnitude distributions for the four training samples used in Test 4, each corresponding only to one of the four major spectroscopic samples used, one from each of the calibration fields.}
\label{fig:magnitudes_test4}
\end{figure}

Table \ref{tab:test4} shows the photo-$z$ precision ($\sigma_{68}$) for the different cases in Test 4 and for Test 1, for comparison. As a first conclusion from this table, we observe how template-based codes, such as LePhare, BPZ or PhotoZ, are not very dependent on the spectroscopic data used for calibration and they show very consistent results for Test 1 and the four cases in Test 4. This is expected since these codes get the photometric information of the galaxies from a given set of predefined templates, either empirical or theoretical, and only use the galaxies in the training sample for calibration of the priors (see \ref{sec:methods} for more information about calibration of the priors for each particular template-based code). On the contrary, and as a second conclusion, we observe in the table how training-based codes are dependent on the data used for their training. For this class of codes (e.g. DESDM, ANNz, TPZ, ...) we can observe how the Test 1 result is generally better than any of the Test 4 results, given that, in this case, all the photometric information of the galaxies comes from the training set and thus having a more complete set helps in the photo-$z$ performance. 

Additionally, there are substantial differences in the photo-$z$ performance depending on the spectroscopic data used for training. In order to understand this, we show in Fig. \ref{fig:magnitudes_test4} the $i$ magnitude distribution for each of the four training sets used in this test. On one hand, the VVDS Deep and the ACES cases of Test 4 give the most similar results to Test 1 since VVDS Deep and ACES are the deepest spectroscopic samples, and the only ones reaching $i_{AB} = 24$ as can be appreciated in Fig. \ref{fig:magnitudes_test4}. On the other hand, Test 4 VVDS Wide gives the poorest results compared to Test 1 due to the fact that the VVDS Wide spectroscopic data is much shallower ($i_{AB} < 22.5$) than DES, hence the training sample is not complete. Finally, using zCOSMOS data as a training sample seems to work better than VVDS Wide but not as well as VVDS Deep and ACES spectroscopy. 

\subsubsection{Stability after removing the highest-weight galaxies}

In every application of a weighting method there exists the danger of the analysis being dependent on a few, highly weighted objects in the sample. Here we demonstrate that the conclusions of this analysis do not change after removing the highest-weight galaxies. We test it only on the photo-$z$ precision of Test 1, for simplicity. 

\begin{table}
\centering
\caption{$\sigma_{68}$ results after removing the 5\% of the galaxies with highest weights in Test 1. Also showing the default Test 1 for comparison.}
\label{fig:weightscut5}
\tabcolsep=0.11cm
\begin{tabular}{l*{3}{c}r}
Codes & Test 1 & Test 1 cut\\
\hline
DESDM & 0.094 $\pm$ 0.002 & 0.088 $\pm$ 0.002 \\
ANNz & 0.086 $\pm$ 0.002 & 0.085 $\pm$ 0.002 \\
TPZ & 0.078 $\pm$ 0.002 & 0.073 $\pm$ 0.002 \\
RVMz & 0.116 $\pm$ 0.004 & 0.103 $\pm$ 0.004 \\
NIP-kNNz & 0.120 $\pm$ 0.004 & 0.110 $\pm$ 0.004 \\
ANNz2 & 0.089 $\pm$ 0.003 & 0.084 $\pm$ 0.003 \\
BPZ & 0.097 $\pm$ 0.003 & 0.095 $\pm$ 0.002 \\
EAZY & 0.109 $\pm$ 0.003 & 0.123 $\pm$ 0.002 \\
LePhare & 0.111 $\pm$ 0.003 & 0.109 $\pm$ 0.003 \\
PhotoZ & 0.097 $\pm$ 0.003 & 0.089 $\pm$ 0.003 \\
TPZ P(z) & 0.078 $\pm$ 0.002 & 0.073 $\pm$ 0.002 \\
ArborZ P(z) & 0.128 $\pm$ 0.003 & 0.117 $\pm$ 0.003 \\
ANNz2 P(z) & 0.085 $\pm$ 0.002 & 0.082 $\pm$ 0.002 \\
SkyNet P(z) & 0.077 $\pm$ 0.002 & 0.067 $\pm$ 0.002 \\
BPZ P(z) & 0.101 $\pm$ 0.002 & 0.097 $\pm$ 0.002 \\
ZEBRA P(z) & 0.109 $\pm$ 0.004 & 0.100 $\pm$ 0.004 \\
\end{tabular}
\end{table}

In Table \ref{fig:weightscut5} we show the $\sigma_{68}$ results for Test 1 together with the same test after removing the 5\% of the galaxies with the highest weights (accounting for almost 30\% of the total weight) in the Main testing sample. Comparing the two results for each of the photo-$z$ codes we clearly see an improvement in photo-$z$ precision when the 5\% cut is applied, but this is expected since removing the highest-weight galaxies also means removing the faintest galaxies in the catalog, which are the most difficult galaxies to get photo-$z$'s of. However, this predicted improvement is both small and uniform among all codes so that the main conclusions reached in the paper remain valid.  

\subsection{Results for DESDM, TPZ, SkyNet and BPZ photo-$z$ codes}
\label{subsec:desdm_results}

So far we have compared a large number of photo-$z$ codes in a variety of situations and configurations. Next we look in greater detail at four photo-$z$ codes: DESDM, TPZ, SkyNet and BPZ.
The DESDM \pz code, a regression artificial neural network, is integrated within the DES Data Management service, so its results will be made available together with all the DES data products, making it a clear choice to be studied in detail here. TPZ and SkyNet are state-of-the-art training-based methods using, respectively, random forests and artificial neural networks to compute \pzs, and yielding the best performance among all the codes utilized in this analysis. Finally, BPZ is the template-based \pz code showing best performance in the tests previously shown, and it has been widely used by other galaxy surveys such as CFHTLenS \citep{Heymans2012,Hildebrandt2012}. All these four codes are public.

\begin{figure}
\centering
\includegraphics[width=90mm]{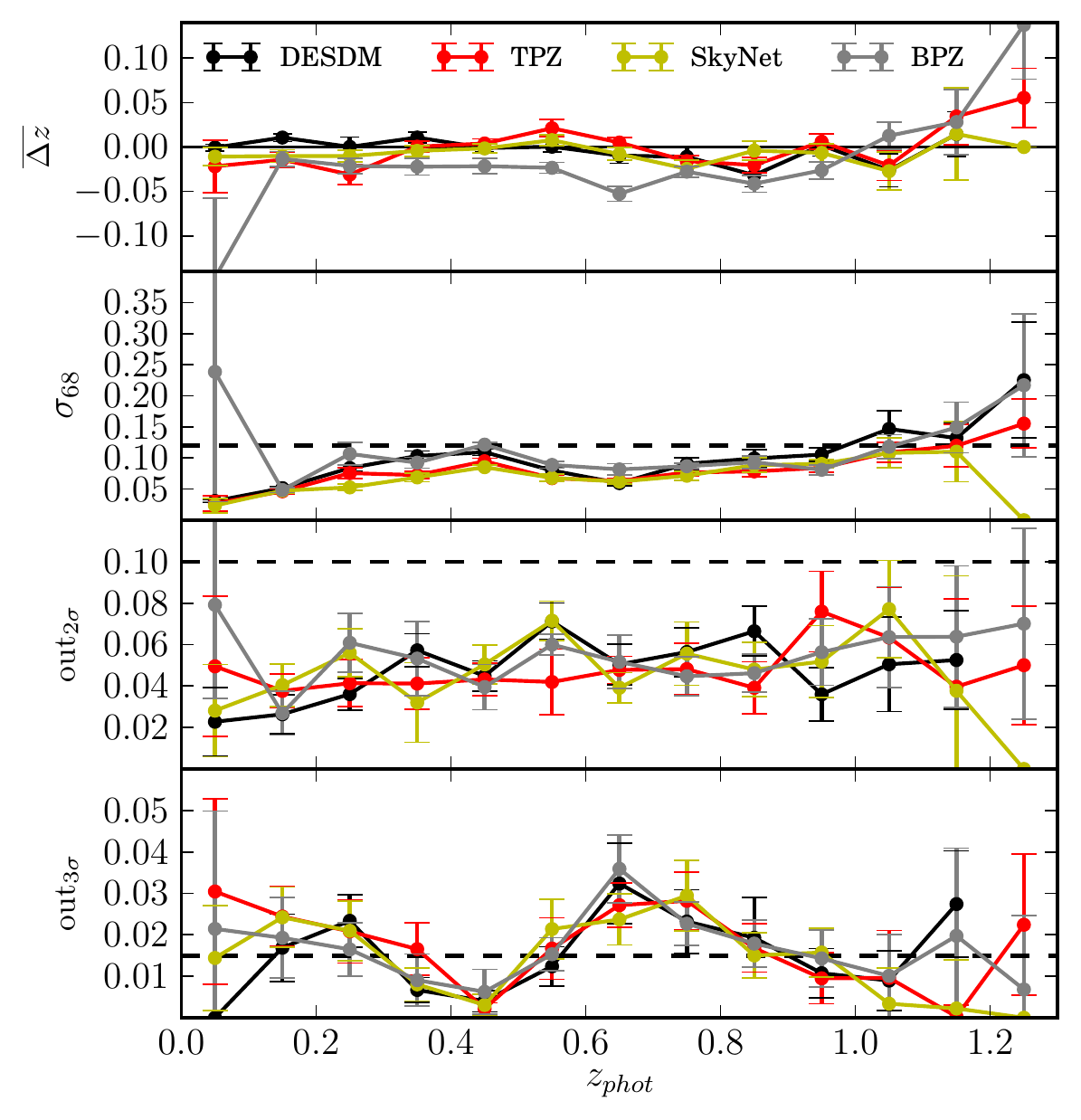}
\caption{Results of Test 1 for the 4 selected photo-$z$ codes. From top to bottom, the photo-$z$ bias, $\sigma_{68}$ and the $2\sigma$ and $3\sigma$ outlier fractions, in bins of $z_{phot}$. DES requirements, displayed as a black dashed line, are shown for the latter 3 metrics.}
\label{fig:4est}
\end{figure}

In \ref{sec:test1} we have studied the default configuration for this analysis, showing the most relevant quantities for all the codes and testing them against the DES requirements. However, we have not looked at the redshift dependence of these quantities, mainly due to difficulties showing that much information for a large number of codes. Figure \ref{fig:4est} shows the photo-$z$ bias, precision and outlier fractions as a function of photometric redshift for the four selected codes. We can see how the four codes behave similarly for all the metrics displayed there. The $\sigma_{68}$ requirement is fulfilled by the codes in most of the redshift range, except at high redshift where error bars are large due to the small number of objects. As we previously observed, the $3\sigma$ outliers fraction is the most difficult requirement to meet, although the results are close to this limit within error bars, while the $2\sigma$ outliers fraction required is met in the whole redshift range for all the codes.  
As mentioned at the begining of section~\ref{sec:analysis_spec}, the requirement on the mean bias in photo-$z$ bins of width 0.1, $|\overline{\Delta z}| < 0.001(1+z)$, is not currently being analyzed, since it necessitates a larger spectroscopic sample in order to be able to calibrate the mean bias away. However, the top plot in Fig.~\ref{fig:4est} shows that for the training-based codes the mean bias in each photo-$z$ bin is compatible with zero within errors, although the current errors are too large to assess whether the requirement is met. In some cases the overall photo-$z$ bias is already at the ~0.001 level, as can be seen in Tables 6--8, although again the errors are large.

\begin{figure}
\centering
\includegraphics[width=90mm]{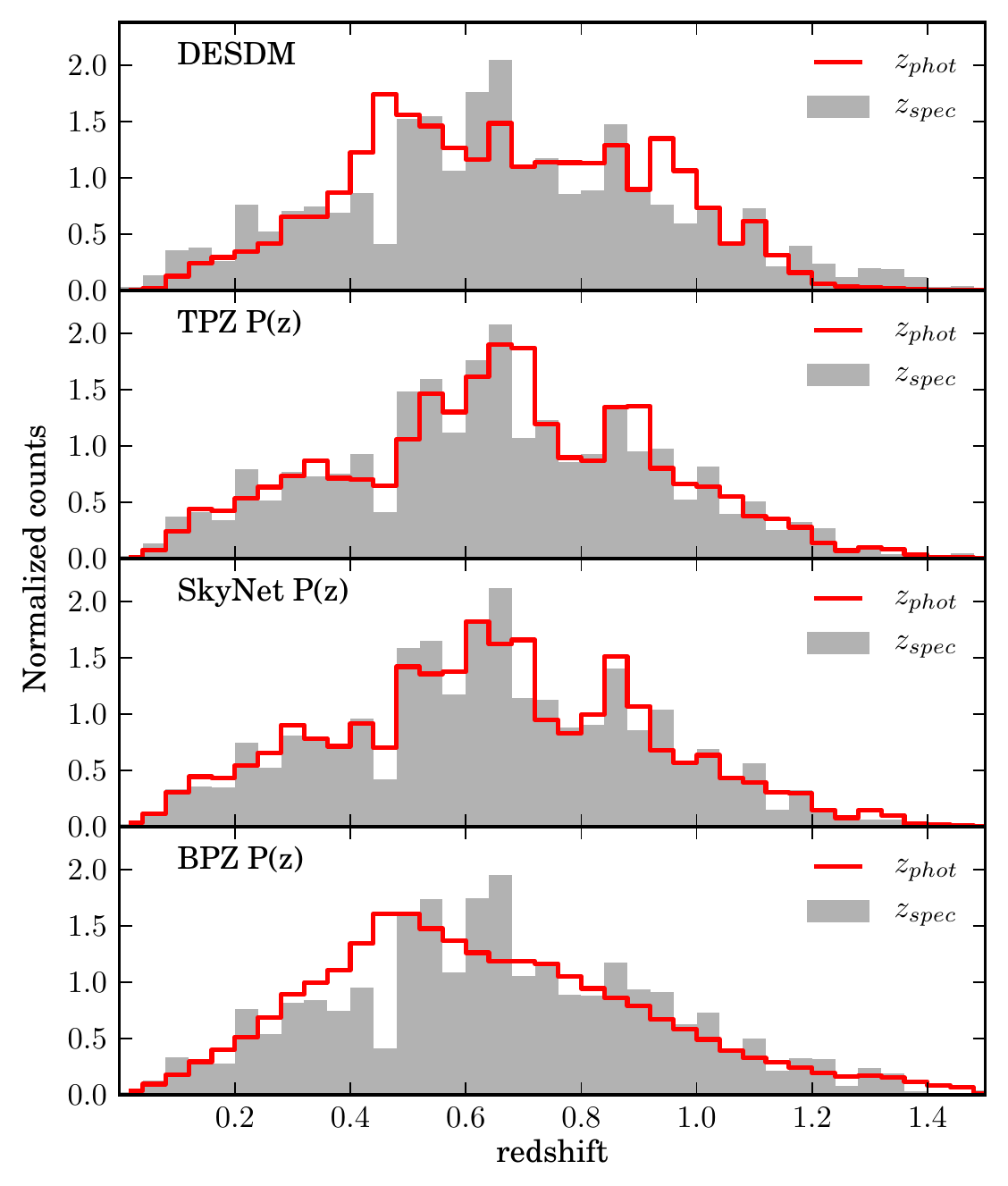}
\caption{Full weighted spectroscopic redshift distribution and its \pz reconstruction using the four selected codes for Test 1. TPZ, SkyNet and BPZ produce redshift pdfs for each galaxy, thus yielding smoother photo-$z$ distributions.}
\label{fig:nofzs_all}
\end{figure}

A very important issue, which is actually the most important result needed from photo-$z$ studies in order to perform many cosmological analyses, is the estimation of true redshift distributions $N(z)$. In Fig. \ref{fig:nofzs_all} we observe how the full redshift distribution reconstructed from the four photo-$z$ codes compares to the spectroscopic distribution. The DESDM code produces one single value for the photo-$z$ of each galaxy in the testing sample while the other three are $P(z)$ codes, so that they return a probability density function (pdf) for each galaxy to be at a given redshift. This is the reason why the $N(z)$ reconstruction looks smoother for TPZ, SkyNet and BPZ, since these are computed from stacking all individual \pz pdfs. Quantitatively, one can measure how good an $N(z)$ reconstruction is by looking at the N$_{\rm poisson}$ and KS metrics in Table 6: the lower these values are, the better is the agreement between the true $N(z)$ and the photo-$z$-reconstructed one. As for the advantage of using $P(z)$ codes, one can observe in Table 6 how the N$_{\rm poisson}$ values for TPZ and BPZ are significantly smaller in their $P(z)$ versions than in their single-estimate photo-$z$ versions. 

\begin{figure*}
\centering
\includegraphics[width=150mm]{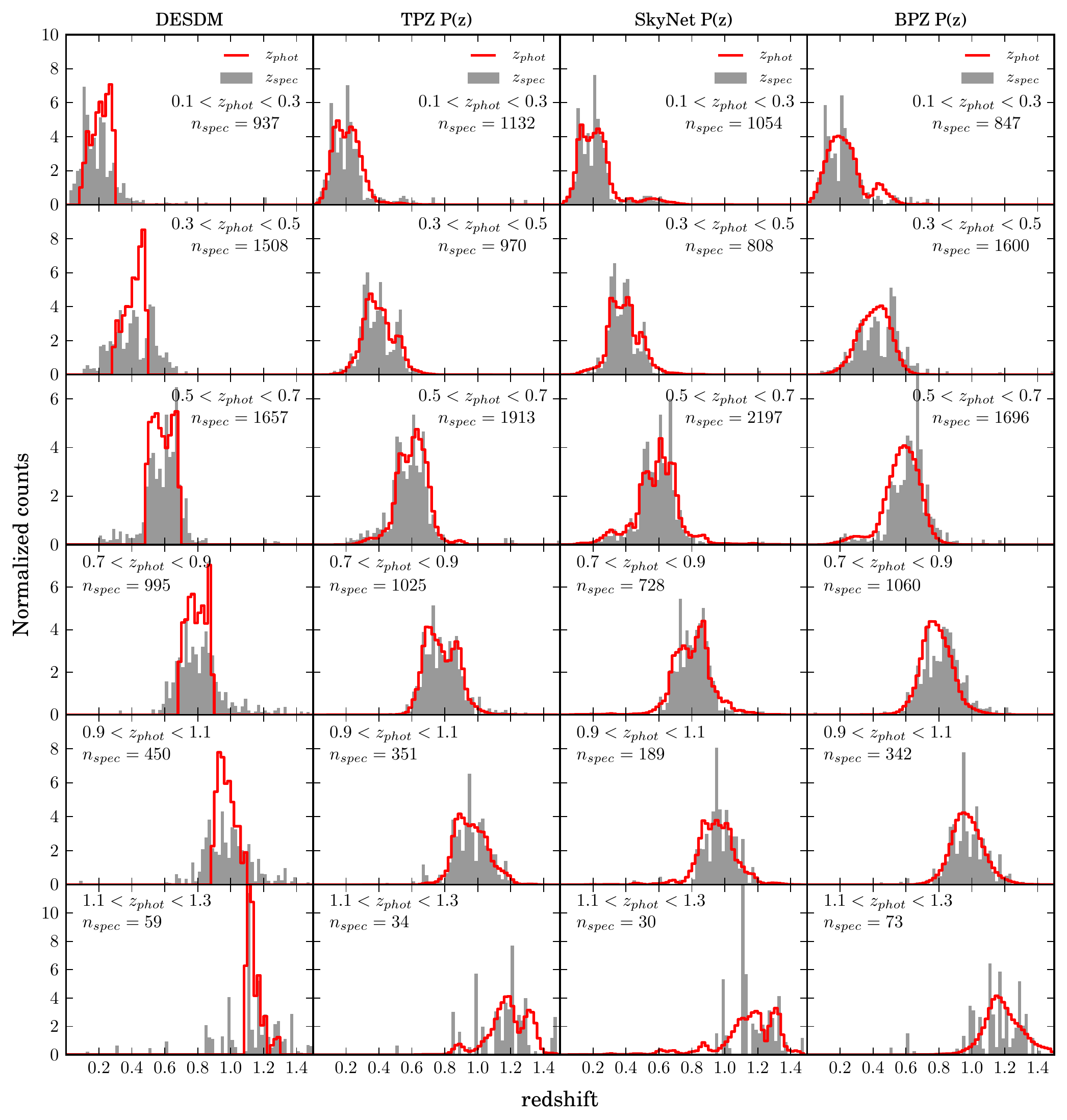}
\caption{Weighted spectroscopic redshift distributions and their \pz reconstruction using the four selected codes, for photo-$z$ bins of width 0.2. The number of spectroscopic galaxies inside each photo-$z$ bin is shown. The DESDM is a single-estimate photo-$z$ code, while TPZ, SkyNet and BPZ are $P(z)$ codes. This is the reason why the photo-$z$ distributions returned by the latter codes can reconstruct the tails of the spectroscopic distributions beyond the photo-$z$ bins. The photo-$z$ bins are defined using the best estimate $z_{phot}$ for each code, while, for TPZ, SkyNet and BPZ the recostructed redshift distributions are obtained by stacking the probability density functions for each galaxy.}
\label{fig:nofzs}
\end{figure*}

On the other hand, although this full redshift distribution is interesting for photo-$z$ analyses, most of the cosmological studies split the galaxy sample into multiple photo-$z$ bins, therefore there is a need to know the true redshift distribution inside each of those photo-$z$ bins. Figure \ref{fig:nofzs} shows the redshift distributions, both spectroscopic and photometric, for six photo-$z$ bins of width 0.2 from $z=0.1$ to $z=1.3$, and for the four photo-$z$ codes selected. The limited number of spectroscopic galaxies available makes the distributions shown in the figure somewhat noisy, especially in the last photo-$z$ bin, where a very small number of galaxies is available. The third and fourth bins in photo-$z$ are the ones presenting the narrowest spectroscopic redshift distributions, which agrees with the fact that the photo-$z$ precision is the highest in this redshift range as can be appreciated in Fig.~\ref{fig:4est}. 

In Fig. \ref{fig:nofzs}, we observe how single-estimate photo-$z$ codes produce a top-hat photo-$z$ distribution for each (photo-$z$ selected) redshift bin. In this case, depicted in the left column of Fig. \ref{fig:nofzs}, the photometric and spectroscopic redshift distributions of each bin are very different and therefore a spectroscopic sample is needed to calibrate the broadening of the redshift bin due to photo-$z$ errors. On the other hand, when using $P(z)$ codes to bin a sample in photometric redshift, one selects a galaxy to be inside a given redshift bin by looking at the position of the median of the pdf (other choices are also possible, e.g. the mode), checking whether it is within the boundaries of the the bin and summing the full pdf of the galaxies inside, including probabilities beyond the bin limits. That makes the photo-$z$ distribution broader than the bin limits and closer to the spectroscopic redshift distribution of the bin, as can be seen in the three rightmost columns in Fig. \ref{fig:nofzs}. We can see on those panels how the tails of the spectroscopic distributions are well represented by the photo-$z$ distributions. This is an important point in favor of $P(z)$ codes since their ability to reproduce the spectroscopic redshift distribution of a photo-$z$ selected bin by stacking their redshift pdfs makes them less dependent on a spectroscopic calibration sample. 

In summary, we have characterized for each code the true redshift distribution inside each photo-$z$ bin. This is the most important quantity to be extracted from any photo-$z$ analysis, since it is the starting point for many cosmological studies such as galaxy clustering and weak lensing. Regarding the performance in such task, the four codes studied in this section show similar spectroscopic redshift distributions for each photo-$z$ bin, but $P(z)$ codes are able to yield a better reconstruction of these distributions by adding up the redshift pdfs for each galaxy which makes them somewhat less reliant in the precise photo-$z$ calibration.

\section{Discussion}
\label{sec:discussion}

We now discuss some of the results and implications of the analyses presented above. First, we consider the results from different types of \pz codes, and afterwards we compare the main outcomes from this study with previous results in the literature.

The \pz codes showing the best performance in the analysis are all training-based methods. Among them, there are various codes using Artificial Neural Networks (ANNs) in different ways and configurations (see Section \ref{sec:methods}), and the similarities and differences between them go beyond the network architecture. DESDM uses hyperbolic tangent activation functions in the hidden layer while ANNz and SkyNet use a sigmoid activation function. ANNz2 results are a mean of several runs where both activations functions are used in the different runs. \citet{Vanzella2004} came to the conclusion that the choice of activation between sigmoid and hyperbolic tangent functions has little effect on the \pz performance. SkyNet was also run with rectified linear units as activation function, which have been shown to outperform more traditional activation functions in object classification \citep{relu}, but no improvement in the \pz problem was observed. SkyNet, ANNz and ANNz2 use weight regularization to avoid over 
fitting while all the four methods monitor the performance on a validation set to prevent overtraining. DESDM and ANNz use first-order gradient information while SkyNet and ANNz2 also use second-order derivative information to train the network. SkyNet and ANNz2 are the only networks adding a constraint to the outputs: ANNz2 does this on a redshift bin per redshift bin basis while SkyNet uses a \textit{softmax} transformation in the final layer, adding a constraint on all redshift bins simultaneously. In conclusion, the neural networks with regularization perform better than the un-regularized DESDM network while the fact that SkyNet uses a \textit{softmax} constrained output in combination with a back-propagation algorithm that uses second order derivative information seems to give it the edge over the rest of ANNs.

Aside from ANNs, TPZ, which is a state-of-the-art \pz code using Prediction Trees and Random Forests, performs remarkably well in all the tests in this work. The prediction trees and random forest techniques used by TPZ have the advantage that they  have fewer hyper parameters to be chosen compared to neural networks. 
Neural networks have, amongst others, to choose the amount of  hidden layers, the amount of nodes  per hidden layer, the learning rate and at least one regularization parameter if  present. 
Random forests used in TPZ have only 2 hyper parameters to choose: the amount of trees used and the size of the subsample set of features used at each split. 
This leaves out the choice of activation function in neural networks and the choice of the measure of information gain at each split in random forests, maximizing its performace.

Furthermore, training-based \pz codes show lower bias compared to that of template-based codes, which indicates possible systematic inaccuracies in the template sets. This can be solved by using adaptive recalibration procedures, which adjust the zero-point offsets in each band using the training sample. Such technique has been succesfully applied by LePhare in this work, as was also the case in~\citet{Hildebrandt2010}.
 
The \pz precision values obtained by template-based methods in this study are all compatible with each other within 10\%. BPZ and PhotoZ yield the highest \pz precision among the methods of this class. The reason for this is probably not the template sets they use, which for BPZ is a combination of \citet{Coleman1980} and \citet{Kinney1996}, and a more complex combination of templates for PhotoZ (see \ref{sec:methods}), since other codes include similar libraries. The fact that they both use Bayesian priors calibrated on DES data, and not only on previous datasets, can be relevant here, although this makes the result more dependent on how representative the training sample is of the full DES data sample. 

In order to set up a context for DES \pz results, it is worth checking the performance obtained by previous similar surveys. Of particular interest is the comparison to the CFHTLS and CFHTLenS (which are two different reductions of the same survey), \pz results \citep{Coupon2008,Hildebrandt2012}, due to the similarities between CFHTLS and DES in terms of survey specifications. However, it is important to stress the differences between DES and CFHTLS. For instance, CFHTLS uses deep $u$ band photometry by default, although they also test the importance of this band by computing their \pzs without it \citep{Ilbert2006}. They find a clear degradation in their \pz precision at $z_{spec}<0.4$, compatible with our findings. Another difference comes from the fact that they do not apply any weighting technique to their calibration sample, thus leaving room for discrepancies between their calibration and full samples. In addition we should point out that their spectroscopic samples for training/calibration are much 
larger than the ones used here by about a factor of 5. Finally, their photometry (S/N$\simeq11$ at $i_{AB}=24$) is deeper than the Main reduction in this paper, being close to our Deep sample (see Fig. \ref{fig:ston}). In \citet{Coupon2008}, CFHTLS \pzs show a precision of about $\sigma_{68} \simeq 0.085$ (roughly translating their result to the metric used here) for $ugriz$ photometry in galaxies up to $i_{AB}<24$. So, despite the differences stated above, their value of the \pz precision is at the level of the DES \pz precision shown in this paper. Moreover, this level of precision is obtained already in Test1, which uses slightly shallower photometry and lacks the $u$ band. Even so, the differences in \pz precision obtained by different algorithms might be relevant here. In particular, the results reported in \citet{Coupon2008} used a template-based \pz code, while the most precise methods reported in this study come from training-based codes. If we do the comparison between template-based codes, the 
results in this study are close but do not exceed the precision reported by CFHTLS.

In terms of outlier fraction, \citet{Coupon2008} report a value for the fraction of galaxies with $\left | \Delta z/(1+z) \right |>0.15$ of $\eta = 10.1\%$. In order to enable the comparison, since the metric used in this paper differs from their approach, we have computed $\eta$ for two cases in this study. On the one hand, we have estimated $\eta$ for a template-based \pz code (BPZ), trying to make a fair comparison since this is the class of algorithm used in \citet{Coupon2008}. In this case, we obtain $\eta = 10.0\%$, in perfect agreement with their result. On the other hand, computing this outlier rate for a template-based code (DESDM) yields $\eta = 6.1\%$, significantly improving the template-based code result. 

We conclude that the overall \pz results shown in this paper are within the expectations for DES if we take into account the performance reported by previous comparable studies for a survey of similar characteristics.

\section{Summary and conclusions}
\label{sec:conclusions}

The Dark Energy Survey succesfully installed DECam during the second semester of 2012, starting its operations in November 2012 with a Science Verification (SV) period lasting until February 2013. Among the 150 sq.~deg.~covered by SV observations, four different fields, of about 3 sq.~deg.~each, overlap with areas with substantial spectroscopic coverage such as VVDS 02hr or COSMOS. Using $grizY$ photometry for galaxies matched to the existing spectroscopic data in these four calibration fields, this paper presents the photometric redshift performance of the DES survey in the SV period. Most of the relevant photo-$z$ codes have been used in the analysis. 

Since spectroscopic galaxy samples are generally shallower, a weighting technique is used to make the calibration sample of galaxies to mimic the DES full sample in magnitude and color space in order to properly estimate the photo-$z$ performance in the DES galaxy sample. 

Calibration and testing samples have been produced with two different depths: Main is the default depth in the DES survey, and Deep corresponds to the depth in SNe fields. Test 1, which uses the Main training and testing samples, represents the default case for photo-$z$ estimation in DES. Results from 13 different codes are analyzed in this case, showing fluctuations in photo-$z$ performance but a general agreement in codes of the same type (machine-learning or template fitting algorithms). In particular, most of the codes analyzed comfortably meet the DES science requirements in terms of photo-$z$ precision and several also meet the requirements on the fractions of outliers. 

In Test 2 we explore the impact of deeper and higher S/N photometry in photo-$z$ calculations, showing that all the codes used improve their results significantly, as expected. In Test 3 we explore the possibilty of using deeper photometry only for training/calibration of the algorithms. In this case we see no general improvement, although there is a significant enhancement using specific codes.   

In an additional test, we consider the incorporation of the $u$ band, which is available in DECam but not used in the DES survey, demonstrating a general improvement in photo-$z$ precision, particularly at low redshift ($<0.5$). This is expected since the $u$ band is crucial for the filter set to bracket the 4000\AA$\:$break in low redshift galaxies. However, due to the high mean redshift of the full DES sample, the impact of not using that band in the overall \pz precision is less important than in previous, shallower surveys such as SDSS. Moreover, we study the importance of the different spectroscopic data sets used, showing how the sets spanning the whole photometric space are crucial for training-based methods, and demonstrated that the results are stable under the removal of the galaxies with highest weights from the analysis.

Generally, training-based \pz codes show the best performance in the tests in terms of \pz precision and bias.  Among them, TPZ, using Prediction Trees and Random Forest, and SkyNet, a state-of-the-art Artificial Neural Network application, seem to yield the most accurate results, achieving a core photometric redshift resolution below $\sigma_{68} = 0.08$. The fact that these two new codes perform better than others extensively used in the literature shows how there is room for improvement in the \pz industry. On the other hand, all template fitting methods employed show consistent results between them, although the use of Bayesian priors specifically calibrated on DES data and adaptive template-recalibration procedures appears to help significantly.

Finally, in the last part of the paper, we choose four photo-$z$ codes, representing different techniques and types, and we present a more detailed analysis of their results. We show some of the most relevant metrics as a function of redshift and, most importantly, we study the estimation of the true redshift distributions $N(z)$ computed using \pzs. For these four codes, we obtain the true redshift distributions in six photo-$z$ bins. Figure \ref{fig:nofzs_all} shows the DES \pz capabilities in such crucial task, and demonstrates the ability to split the DES full sample in tomographic redshift bins of width 0.2 already with these early data. Furthermore, the calibration of the true redshift distribution of a photo-$z$ selection is the most important ingredient for cosmological studies involving galaxy clustering or weak lensing, and it is an important outcome from this paper, enabling further science analyses. 

The \pz analyses carried out in this work using these early stage DES data will serve as a benchmark for future data releases, and as the survey area grows during the observation period, more spectroscopic data will be available allowing a better calibration and a better sampling for training algorithms. Therefore these promising early results will do nothing but improve in the near future, which will allow putting tighter constrains on several cosmological parameters. Furthermore, the 5-band optical and near-infrared photometry of DES can be combined with the infrared $J$ and $K_s$ photometry provided by the VHS survey \citep{reference_vhs} in 90\% of the DES footprint. This should result in improved photometric redshift estimations, particularly at high redshift \citep{Banerji2008}.

\section*{Acknowledgments}

This paper has gone through internal review by the DES collaboration.

We are grateful for the extraordinary contributions of our CTIO colleagues and the DES Camera, Commissioning and Science 
Verification teams in achieving the excellent instrument and telescope conditions that have made this work possible.
The success of this project also relies critically on the expertise and dedication of the DES Data Management organization.

Funding for the DES Projects has been provided by the U.S. Department of Energy, the U.S. National Science Foundation, the Ministry of Science and Education of Spain, 
the Science and Technology Facilities Council of the United Kingdom, the Higher Education Funding Council for England, the National Center for Supercomputing 
Applications at the University of Illinois at Urbana-Champaign, the Kavli Institute of Cosmological Physics at the University of Chicago, Financiadora de Estudos e Projetos, 
Funda{\c c}{\~a}o Carlos Chagas Filho de Amparo {\`a} Pesquisa do Estado do Rio de Janeiro, Conselho Nacional de Desenvolvimento Cient{\'i}fico e Tecnol{\'o}gico and 
the Minist{\'e}rio da Ci{\^e}ncia e Tecnologia, the Deutsche Forschungsgemeinschaft and the Collaborating Institutions in the Dark Energy Survey.

The Collaborating Institutions are Argonne National Laboratories, the University of California at Santa Cruz, the University of Cambridge, Centro de Investigaciones Energeticas, 
Medioambientales y Tecnologicas-Madrid, the University of Chicago, University College London, the DES-Brazil Consortium, the Eidgen{\"o}ssische Technische Hochschule (ETH) Z{\"u}rich, 
Fermi National Accelerator Laboratory, the University of Edinburgh, the University of Illinois at Urbana-Champaign, the Institut de Ciencies de l'Espai (IEEC/CSIC), 
the Institut de Fisica d'Altes Energies, the Lawrence Berkeley National Laboratory, the Ludwig-Maximilians Universit{\"a}t and the associated Excellence Cluster Universe, 
the University of Michigan, the National Optical Astronomy Observatory, the University of Nottingham, The Ohio State University, the University of Pennsylvania, the University of Portsmouth, 
SLAC National Accelerator Laboratory, Stanford University, the University of Sussex, and Texas A\&M University.

The DES participants from Spanish institutions are partially supported by MINECO under grants AYA2009-13936, AYA2012-39559, AYA2012-39620, and FPA2012-39684, which include FEDER funds from the European Union. J.G acknowledges a fellowship from DS-CAPES Program. This work was partially support by STFC grant ST/K00090X/1. Please contact the author(s) to request access to research materials discussed in this paper.

This research uses data from the VIMOS VLT Deep Survey, obtained from the VVDS database operated by Cesam, Laboratoire d'Astrophysique de Marseille, France.

Data for the OzDES spectroscopic survey were obtained with the Anglo-Australian Telescope (program numbers 12B/11 and 13B/12). Parts of this research were conducted by the Australian Research Council Centre of Excellence for All-sky Astrophysics (CAASTRO), through project number CE110001020. 

Funding for SDSS-III has been provided by the Alfred P. Sloan Foundation, the Participating Institutions, the National Science Foundation, and the U.S. Department of Energy Office of Science. The SDSS-III web site is http://www.sdss3.org/.

SDSS-III is managed by the Astrophysical Research Consortium for the Participating Institutions of the SDSS-III Collaboration including the University of Arizona, the Brazilian Participation Group, Brookhaven National Laboratory, Carnegie Mellon University, University of Florida, the French Participation Group, the German Participation Group, Harvard University, the Instituto de Astrofisica de Canarias, the Michigan State/Notre Dame/JINA Participation Group, Johns Hopkins University, Lawrence Berkeley National Laboratory, Max Planck Institute for Astrophysics, Max Planck Institute for Extraterrestrial Physics, New Mexico State University, New York University, Ohio State University, Pennsylvania State University, University of Portsmouth, Princeton University, the Spanish Participation Group, University of Tokyo, University of Utah, Vanderbilt University, University of Virginia, University of Washington, and Yale University.

\bibliographystyle{mn2e}
\bibliography{/Users/csanchez/Dropbox/bibtex/library}

\section*{Appendix}
\renewcommand{\thesubsection}{\Alph{subsection}}

\subsection{Description of the metrics}
\label{sec:appendix_a}

Here we describe the metrics used throughout this work. For each photo-$z$ code and each galaxy we have either the photo-$z$ estimation and its associated error or a probability density function $P(z)$. 
As described in the text, a vector of weights were computed in order to match the spectroscopic and photometric samples in multi-color and magnitude space. On each sample we have a vector $\boldsymbol{\omega}$ of weights corresponding to the $N$ galaxies on each test set, where $\sum_{i=1}^{N} \omega_i = 1$. If no weights are used, the default value $\omega_i = \frac{1}{N}$ is assigned to each galaxy. We define the individual bias as $\Delta z_i = z_{{\rm phot},i} - z_{{\rm spec},i}$ and the statistics used in this work as follows:

\begin{enumerate}
 \item mean bias($\overline{\Delta z}$):
 \begin{equation}
 \overline{\Delta z} = \frac{\sum \omega_i {\Delta z}_i}{\sum \omega_i}
 \end{equation}

 \item $\sigma_{\Delta z}$ :
\begin{equation}
\sigma_{\Delta z} = \left(\frac{\sum \omega_i \left({\Delta z}_i -  \overline{\Delta z} \right)^2 }{\sum \omega_i}\right)^{\frac{1}{2}}
\end{equation}

 \item median (${\Delta z}_{50}$), the median of the $\Delta z$ distribution, fulfilling:
\begin{equation}
 P_{50}=P (\Delta z \leq {\Delta z}_{50} ) = \int_0^{{\Delta z}_{50}} \omega(\Delta z) d(\Delta z) = \frac{1}{2} 
\end{equation}

 \item $\sigma_{68}$, half of the width of the distribution, measured with respect to the median, where 68\% of the data are enclosed. This is computed as:
\begin{equation}
 \sigma_{68} = \frac{1}{2}\left(P_{84} - P_{16}\right)
 \end{equation}
 \item ${\rm out}_{2\sigma}$, the fraction of outliers above the $2\sigma_{\Delta z}$ level:
\begin{equation}
 {\rm out}_{2\sigma} = \frac{\sum W_i}{\sum \omega_i}
\end{equation}
 where,
 $$W_i = \begin{cases} \omega_i, & \mbox{if }|{\Delta z}_i - \overline{\Delta z}| > 2 \sigma_{\Delta z}  \\ 0 , & \mbox{if }  |{\Delta z}_i - \overline{\Delta z}| \leq 2 \sigma_{\Delta z} \end{cases}$$
 
  \item ${\rm out}_{3\sigma}$, the fraction of outliers above the $3\sigma_{\Delta z}$ level:
\begin{equation}
 {\rm out}_{3\sigma} = \frac{\sum W_i}{\sum \omega_i}
\end{equation}
 where,
 $$W_i = \begin{cases} \omega_i, & \mbox{if }|{\Delta z}_i - \overline{\Delta z}| > 3 \sigma_{\Delta z}  \\ 0 , & \mbox{if }  |{\Delta z}_i - \overline{\Delta z}| \leq 3 \sigma_{\Delta z} \end{cases}$$

 \item $\overline{\Delta z'}$, the mean of the distribution of  $\Delta z$ is normalized by their estimated errors. Ideally this distribution should resemble a normal distribution with zero mean and unit variance. We define $\Delta z'_i = \Delta z_i / \epsilon_{{\rm phot},i}$ where $\epsilon_{{\rm phot},i}$ is the computed error of the photometric redshift for galaxy $i$. Then:
\begin{equation}
 \overline{\Delta z'} = \frac{\sum \omega_i {\Delta z'}_i}{\sum \omega_i}
 \end{equation}

 \item $\sigma_{\Delta z'}$: 
\begin{equation}
 \sigma_{\Delta z'} = \left(\frac{\sum \omega_i \left({\Delta z'}_i -  \overline{\Delta z'} \right)^2 }{\sum \omega_i}\right)^{\frac{1}{2}}
 \end{equation}
 
\item $N_{\rm poisson}$, a metric that quantifies how close the distribution of photometric redshifts $N(z_{\rm phot})$ is to the distribution of spectroscopic redshifts $N(z_{\rm spec})$. For each photometric redshift bin $j$ of width 0.1, we compute the difference of $N(z_{\rm phot}) - N(z_{\rm spec})$ normalized by the Poisson fluctuations on $N(z_{\rm spec})$:
$$
n_{{\rm poisson},j} = \frac{\left(\sum\limits_{z_{{\rm phot},i} \, \epsilon \, {\rm bin}_j} \omega_i N - \sum\limits_{z_{{\rm spec},i} \, \epsilon \, {\rm bin}_j} \omega_i N \right)}{\sqrt{\sum\limits_{z_{{\rm spec},i} \, \epsilon \, {\rm bin}_j} \omega_i N}}
$$
Then $N_{\rm poisson}$ is computed as the RMS of the previous quantity:

\begin{equation}
N_{\rm poisson} = \left( \frac{1}{n_{bins}} \sum\limits_{j=1}^{n_{bins}} n_{{\rm poisson},j}^2 \right)^{\frac{1}{2}}
\end{equation}

\item KS is the Kolmogorov-Smirnov test that quantifies whether the two redshift distributions ($N(z_{\rm phot})$ and $N(z_{\rm spec})$) are compatible with being drawn from the same parent distribution, independently of binning. It is defined as the maximum distance between both empirical cumulative distributions. The lower this value, the closer are both distributions. The empirical cumulative distribution function is calculated as:

$$
F_{\rm spec} (z) = \frac{ \sum\limits_{i=1}^N \Omega_{z_{{\rm spec},i} < z}}{\sum \omega_i}
$$
 where,
 $$
  \Omega_{z_{{\rm spec},i} < z} = \begin{cases} \omega_i, & \mbox{if } z_{{\rm spec},i} < z  \\ 0 , & \mbox{otherwise } \end{cases}
 $$
Similarly, the empirical cumulative distribution function $F_{\rm phot} (z)$ is computed for $N(z_{\rm phot})$. Then the KS statistic is computed as:

\begin{equation}
{\rm KS} = \max_z \left(\lvert F_{\rm phot} (z) - F_{\rm spec} (z)\rvert \right)
\end{equation}

 \end{enumerate}

\noindent For the submissions with a $P(z)$ for each galaxy, these cumulative distributions are computed taking into account the $p(z)$ distribution for each galaxy.

\section*{Affiliations}

{\small
$^{1}$ Institut de F\'{i}sica d'Altes Energies, Universitat Aut\`{o}noma de Barcelona, E-08193 Bellaterra (Barcelona), Spain\\
$^{2}$ Department of Astronomy, University of Illinois, Urbana, IL 61820, USA\\
$^{3}$ Fermi National Accelerator Laboratory, P.O. Box 500, Batavia, IL 60510, USA\\
$^{4}$ Instituci\'o Catalana de Recerca i Estudis Avan\c{c}ats, E-08010 Barcelona, Spain\\
$^{5}$ Department of Physics \& Astronomy, University College London, Gower Street, London WC1E 6BT, UK\\
$^{6}$ ETH Zurich, Institut fur Astronomie, HIT J 11.3, Wolfgang-Pauli-Str. 27, 8093 Zurich, Switzerland\\
$^{7}$ Institute of Cosmology and Gravitation, University of Portsmouth, Dennis Sciama Building, Burnaby Road, Portsmouth, PO1 3FX, UK\\
$^{8}$ Observat\'{o}rio Nacional (ON/MCT), Rua General Jos\'{e} Cristino, 77, Rio de Janeiro 20921-400 - RJ, Brazil\\
$^{9}$ Laborat\'{o}rio Nacional de e-Astronomia, Rua General Jos\'{e} Cristino, 77, 20921-400 S\~ao Crist\'{o}vao, Rio de Janeiro, RJ, Brazil\\
$^{10}$ Institut de Ci\`encies de l'Espai (ICE, IEEC/CSIC), E-08193 Bellaterra (Barcelona), Spain\\
$^{11}$ Kavli Institute for Particle Astrophysics and Cosmology, 452 Lomita Mall, Stanford University, Stanford, CA 94305, USA\\
$^{12}$ Department of Physics, University of Michigan, Ann Arbor, MI 48109, USA\\
$^{13}$ University Observatory Munich, Scheinerstrasse 1, 81679 Munich, Germany\\
$^{14}$ Max Planck Institute for Extraterrestrial Physics, Giessenbachstrasse, 85748 Garching, Germany\\
$^{15}$ University of Nottingham, School of Physics and Astronomy, Nottingham NG7 2RD, UK\\
$^{16}$ Departamento de F\'{i}sica Matem\'{a}tica, Instituto de F\'{i}sica, Universidade de Sao Paulo, Sao Paulo, SP CP 66318, CEP 05314-970, Brazil\\
$^{17}$ Centro de Investigaciones Energ\'{e}ticas Medioambientales y Tecnol\'{o}gicas, Av.Complutense 40, 28040 Madrid, Spain\\
$^{18}$ Cerro Tololo Inter-American Observatory, National Optical Astronomy Observatory, Casilla 603, La Serena, Chile\\
$^{19}$ Space Telescope Science Institute (STScI), 3700 San Martin Drive, Baltimore, MD  21218\\
$^{20}$ National Optical Astronomy Observatory, 950 N. Cherry Ave., Tucson, AZ 85719, USA\\
$^{21}$ Dept. of Physics \& Astronomy, University of Pennsylvania, Philadelphia PA 19104, USA\\
$^{22}$ Argonne National Laboratory, 9700 South Cass Avenue, Lemont, IL 60439, USA\\
$^{23}$ SLAC National Accelerator Laboratory, Menlo Park, CA 94025, USA\\ 
$^{24}$ Research School of Astronomy and Astrophysics, The Australian National University, Canberra, ACT 2611, Australia\\
$^{25}$ ARC Centre of Excellence for All-Sky Astrophysics (CAASTRO)\\
$^{26}$ School of Mathematics and Physics, University of Queensland, QLD, 4072, Australia\\
$^{27}$ George P. and Cynthia Woods Mitchell Institute for Fundamental Physics and Astronomy, and Department of Physics and Astronomy, Texas A\&M University, College Station, TX 77843, USA\\
$^{28}$ Fellow, Radcliffe Institute for Advanced Study, Harvard University, Byerly Hall, 8 Garden Street, Cambridge, MA 02138, USA\\
$^{29}$ Department of Physics, Ludwig-Maximilians-Universit\"{a}t, Scheinerstr.\ 1, 81679 M\"{u}nchen, Germany\\
$^{30}$ Excellence Cluster Universe, Boltzmannstr.\ 2, 85748 Garching, Germany\\
$^{31}$ Department of Astronomy, University of Michigan, Ann Arbor, MI 48109, USA\\
$^{32}$ Institut d'Astrophysique de Paris, Univ. Pierre et Marie Curie \& CNRS UMR7095, F-75014 Paris, France\\
$^{33}$ Centre for Astrophysics and Supercomputing, Swinburne University of Technology, Hawthorn, VIC 3122, Australia\\
$^{34}$ Department of Physics, The Ohio State University, Columbus, OH 43210, USA\\
$^{35}$ Lawrence Berkeley National Laboratory, 1 Cyclotron Road, Berkeley, CA 94720, USA\\
$^{36}$ Australian Astronomical Observatory, North Ryde, NSW 2113, Australia\\
$^{37}$ ICRA, Centro Brasileiro de Pesquisas F\'isicas, Rua Dr. Xavier Sigaud 150, CEP 22290-180, Rio de Janeiro, RJ, Brazil\\
$^{38}$ Instituto de F\'isica, UFRGS, Caixa Postal 15051, Porto Alegre, RS - 91501-970, Brazil\\
$^{39}$ National Center for Supercomputing Applications, 1205 West Clark St., Urbana, IL 61801, USA\\
$^{40}$ SEPnet, South East Physics Network, (www.sepnet.ac.uk)\\
$^{41}$ Jodrell Bank Center for Astrophysics, School of Physics and Astronomy,
University of Manchester, Oxford Road, Manchester, M13 9PL, UK
}

\bsp
\label{lastpage}
\begin{landscape}
\begin{table}
\label{table:test1}
\caption{Results of all the photo-$z$ metrics listed in Appendix A for all the codes analyzed in Test 1. The errors are computed from bootstrap resampling with 100 samples. The weighting procedure has been applied, together with a cut on the 10\% of the galaxies with the highest estimated photo-$z$ errors for each code.}
\tabcolsep=0.11cm
\centering
\begin{tabular}{l*{11}{c}r}
Test 1 & $\overline{\Delta z}$ & $\Delta z_{50}$ & $\sigma_{68}$ & $\sigma_{\Delta z}$ & ${\rm out}_{2\sigma}$ & ${\rm out}_{3\sigma}$ & $\overline{\Delta z\prime}$ & $\sigma_{\Delta z\prime}$ & ${\rm N}_{\rm poisson}$ & ${\rm KS}$ \\
\hline
DESDM & -0.005 $\pm$ 0.003 & -0.003 $\pm$ 0.002 & 0.094 $\pm$ 0.002 & 0.135 $\pm$ 0.005 & 0.053 $\pm$ 0.005 & 0.018 $\pm$ 0.003 & -0.047 $\pm$ 0.032 & 1.479 $\pm$ 0.052 & 7.035 $\pm$ 0.486 & 0.056 $\pm$ 0.004 \\
ANNz & 0.002 $\pm$ 0.003 & -0.001 $\pm$ 0.003 & 0.086 $\pm$ 0.002 & 0.118 $\pm$ 0.004 & 0.049 $\pm$ 0.004 & 0.015 $\pm$ 0.002 & 0.096 $\pm$ 0.046 & 3.341 $\pm$ 0.134 & 6.355 $\pm$ 0.480 & 0.052 $\pm$ 0.005 \\
TPZ & -0.001 $\pm$ 0.003 & 0.004 $\pm$ 0.002 & 0.078 $\pm$ 0.002 & 0.122 $\pm$ 0.006 & 0.046 $\pm$ 0.004 & 0.019 $\pm$ 0.002 & 0.019 $\pm$ 0.032 & 1.529 $\pm$ 0.063 & 4.122 $\pm$ 0.505 & 0.044 $\pm$ 0.005 \\
RVMz & -0.011 $\pm$ 0.005 & -0.004 $\pm$ 0.002 & 0.116 $\pm$ 0.004 & 0.180 $\pm$ 0.008 & 0.060 $\pm$ 0.005 & 0.023 $\pm$ 0.003 & -0.084 $\pm$ 0.041 & 1.371 $\pm$ 0.098 & 8.382 $\pm$ 0.548 & 0.083 $\pm$ 0.007 \\
NIP-kNNz & -0.030 $\pm$ 0.005 & -0.011 $\pm$ 0.002 & 0.120 $\pm$ 0.004 & 0.197 $\pm$ 0.009 & 0.058 $\pm$ 0.005 & 0.018 $\pm$ 0.003 & -0.186 $\pm$ 0.030 & 1.116 $\pm$ 0.117 & 3.633 $\pm$ 0.482 & 0.054 $\pm$ 0.007 \\
ANNz2 & -0.002 $\pm$ 0.003 & -0.003 $\pm$ 0.002 & 0.089 $\pm$ 0.003 & 0.151 $\pm$ 0.009 & 0.042 $\pm$ 0.003 & 0.021 $\pm$ 0.002 & 0.063 $\pm$ 0.071 & 2.280 $\pm$ 0.255 & 5.381 $\pm$ 0.458 & 0.052 $\pm$ 0.004 \\
BPZ & -0.022 $\pm$ 0.003 & -0.021 $\pm$ 0.002 & 0.097 $\pm$ 0.003 & 0.137 $\pm$ 0.006 & 0.049 $\pm$ 0.003 & 0.018 $\pm$ 0.002 & -0.194 $\pm$ 0.032 & 1.750 $\pm$ 0.075 & 7.919 $\pm$ 0.622 & 0.099 $\pm$ 0.008 \\
EAZY & -0.061 $\pm$ 0.004 & -0.063 $\pm$ 0.003 & 0.109 $\pm$ 0.003 & 0.153 $\pm$ 0.010 & 0.035 $\pm$ 0.005 & 0.015 $\pm$ 0.002 & -0.331 $\pm$ 0.074 & 3.982 $\pm$ 0.804 & 11.148 $\pm$ 0.868 & 0.195 $\pm$ 0.009 \\
LePhare & 0.002 $\pm$ 0.004 & -0.007 $\pm$ 0.003 & 0.111 $\pm$ 0.003 & 0.171 $\pm$ 0.008 & 0.047 $\pm$ 0.002 & 0.024 $\pm$ 0.002 & 1.177 $\pm$ 0.379 & 42.883 $\pm$ 6.603 & 6.632 $\pm$ 0.444 & 0.087 $\pm$ 0.006 \\
PhotoZ & -0.029 $\pm$ 0.003 & -0.029 $\pm$ 0.002 & 0.097 $\pm$ 0.003 & 0.142 $\pm$ 0.006 & 0.058 $\pm$ 0.004 & 0.017 $\pm$ 0.002 & -0.268 $\pm$ 0.020 & 1.003 $\pm$ 0.030 & 4.565 $\pm$ 0.464 & 0.080 $\pm$ 0.006 \\
TPZ P(z) & 0.006 $\pm$ 0.003 & 0.011 $\pm$ 0.002 & 0.078 $\pm$ 0.002 & 0.119 $\pm$ 0.006 & 0.049 $\pm$ 0.005 & 0.018 $\pm$ 0.003 & 0.125 $\pm$ 0.030 & 1.484 $\pm$ 0.059 & 2.607 $\pm$ 0.449 & 0.044 $\pm$ 0.006 \\
ArborZ P(z) & -0.008 $\pm$ 0.004 & 0.001 $\pm$ 0.004 & 0.128 $\pm$ 0.003 & 0.153 $\pm$ 0.005 & 0.056 $\pm$ 0.005 & 0.016 $\pm$ 0.003 & -0.028 $\pm$ 0.024 & 0.962 $\pm$ 0.025 & 4.774 $\pm$ 0.449 & 0.064 $\pm$ 0.009 \\
ANNz2 P(z) & -0.002 $\pm$ 0.002 & -0.002 $\pm$ 0.002 & 0.085 $\pm$ 0.002 & 0.118 $\pm$ 0.004 & 0.051 $\pm$ 0.004 & 0.016 $\pm$ 0.003 & -0.004 $\pm$ 0.024 & 1.315 $\pm$ 0.042 & 5.010 $\pm$ 0.404 & 0.045 $\pm$ 0.004 \\
SkyNet P(z) & -0.001 $\pm$ 0.002 & 0.001 $\pm$ 0.002 & 0.077 $\pm$ 0.002 & 0.104 $\pm$ 0.003 & 0.072 $\pm$ 0.006 & 0.015 $\pm$ 0.002 & -0.006 $\pm$ 0.014 & 0.829 $\pm$ 0.027 & 4.091 $\pm$ 0.404 & 0.052 $\pm$ 0.005 \\
BPZ P(z) & -0.025 $\pm$ 0.003 & -0.025 $\pm$ 0.003 & 0.101 $\pm$ 0.002 & 0.132 $\pm$ 0.006 & 0.046 $\pm$ 0.004 & 0.014 $\pm$ 0.002 & -0.224 $\pm$ 0.033 & 1.750 $\pm$ 0.080 & 5.776 $\pm$ 0.948 & 0.098 $\pm$ 0.007 \\
ZEBRA P(z) & -0.050 $\pm$ 0.005 & -0.030 $\pm$ 0.002 & 0.109 $\pm$ 0.004 & 0.177 $\pm$ 0.012 & 0.043 $\pm$ 0.006 & 0.018 $\pm$ 0.002 & -0.383 $\pm$ 0.047 & 1.906 $\pm$ 0.166 & 6.692 $\pm$ 0.550 & 0.112 $\pm$ 0.008 \\
\end{tabular}
\end{table}
\end{landscape}

\begin{landscape}
\begin{table}
\label{table:test2}
\caption{Results of all the photo-$z$ metrics for all the codes analyzed in Test 2. The errors are computed from bootstrap resampling with 100 samples. The weighting procedure has been applied, together with a cut on the 10\% of the galaxies with the highest estimated photo-$z$ errors for each code.}
\tabcolsep=0.11cm
\centering
\begin{tabular}{l*{11}{c}r}
Test 2 & $\overline{\Delta z}$ & $\Delta z_{50}$ & $\sigma_{68}$ & $\sigma_{\Delta z}$ & ${\rm out}_{2\sigma}$ & ${\rm out}_{3\sigma}$ & $\overline{\Delta z\prime}$ & $\sigma_{\Delta z\prime}$ & ${\rm N}_{\rm poisson}$ & ${\rm KS}$ \\
\hline
DESDM & 0.001 $\pm$ 0.003 & -0.002 $\pm$ 0.002 & 0.087 $\pm$ 0.003 & 0.123 $\pm$ 0.005 & 0.052 $\pm$ 0.006 & 0.016 $\pm$ 0.003 & 0.008 $\pm$ 0.041 & 1.648 $\pm$ 0.059 & 8.822 $\pm$ 0.716 & 0.058 $\pm$ 0.008 \\
ANNz & -0.001 $\pm$ 0.003 & -0.003 $\pm$ 0.002 & 0.075 $\pm$ 0.002 & 0.105 $\pm$ 0.007 & 0.043 $\pm$ 0.006 & 0.013 $\pm$ 0.002 & 0.013 $\pm$ 0.062 & 3.228 $\pm$ 0.099 & 7.692 $\pm$ 0.679 & 0.057 $\pm$ 0.006 \\
TPZ & 0.005 $\pm$ 0.003 & 0.004 $\pm$ 0.002 & 0.066 $\pm$ 0.002 & 0.101 $\pm$ 0.005 & 0.042 $\pm$ 0.004 & 0.022 $\pm$ 0.003 & 0.091 $\pm$ 0.042 & 1.637 $\pm$ 0.099 & 4.123 $\pm$ 0.483 & 0.045 $\pm$ 0.007 \\
RVMz & -0.008 $\pm$ 0.005 & -0.005 $\pm$ 0.003 & 0.110 $\pm$ 0.004 & 0.168 $\pm$ 0.008 & 0.063 $\pm$ 0.006 & 0.016 $\pm$ 0.002 & -0.043 $\pm$ 0.041 & 1.283 $\pm$ 0.144 & 10.166 $\pm$ 0.750 & 0.079 $\pm$ 0.008 \\
NIP-kNNz & -0.030 $\pm$ 0.005 & -0.013 $\pm$ 0.003 & 0.107 $\pm$ 0.005 & 0.169 $\pm$ 0.009 & 0.060 $\pm$ 0.005 & 0.021 $\pm$ 0.003 & -0.291 $\pm$ 0.076 & 2.040 $\pm$ 0.429 & 4.463 $\pm$ 0.652 & 0.054 $\pm$ 0.007 \\
ANNz2 & 0.001 $\pm$ 0.005 & -0.005 $\pm$ 0.002 & 0.083 $\pm$ 0.003 & 0.140 $\pm$ 0.010 & 0.048 $\pm$ 0.007 & 0.017 $\pm$ 0.002 & 0.094 $\pm$ 0.080 & 2.784 $\pm$ 0.459 & 7.737 $\pm$ 0.701 & 0.051 $\pm$ 0.006 \\
BPZ & -0.009 $\pm$ 0.004 & -0.015 $\pm$ 0.002 & 0.096 $\pm$ 0.003 & 0.143 $\pm$ 0.007 & 0.054 $\pm$ 0.005 & 0.017 $\pm$ 0.002 & 0.044 $\pm$ 0.062 & 2.400 $\pm$ 0.160 & 8.205 $\pm$ 0.750 & 0.053 $\pm$ 0.005 \\
EAZY & -0.071 $\pm$ 0.004 & -0.069 $\pm$ 0.003 & 0.094 $\pm$ 0.003 & 0.155 $\pm$ 0.010 & 0.048 $\pm$ 0.005 & 0.020 $\pm$ 0.003 & -0.650 $\pm$ 0.171 & 12.351 $\pm$ 3.233 & 11.076 $\pm$ 0.839 & 0.210 $\pm$ 0.010 \\
LePhare & 0.019 $\pm$ 0.005 & -0.008 $\pm$ 0.002 & 0.099 $\pm$ 0.003 & 0.183 $\pm$ 0.010 & 0.053 $\pm$ 0.004 & 0.028 $\pm$ 0.004 & 2.099 $\pm$ 0.994 & 120.100 $\pm$ 19.658 & 7.509 $\pm$ 1.169 & 0.071 $\pm$ 0.007 \\
PhotoZ & -0.019 $\pm$ 0.004 & -0.029 $\pm$ 0.003 & 0.093 $\pm$ 0.003 & 0.147 $\pm$ 0.007 & 0.053 $\pm$ 0.005 & 0.022 $\pm$ 0.004 & 1.514 $\pm$ 0.829 & 39.656 $\pm$ 10.272 & 4.932 $\pm$ 0.601 & 0.054 $\pm$ 0.007 \\
TPZ P(z) & 0.011 $\pm$ 0.002 & 0.011 $\pm$ 0.002 & 0.067 $\pm$ 0.003 & 0.099 $\pm$ 0.004 & 0.042 $\pm$ 0.005 & 0.021 $\pm$ 0.003 & 0.210 $\pm$ 0.038 & 1.576 $\pm$ 0.090 & 3.702 $\pm$ 0.451 & 0.051 $\pm$ 0.006 \\
ArborZ P(z) & -0.003 $\pm$ 0.003 & 0.001 $\pm$ 0.002 & 0.096 $\pm$ 0.003 & 0.119 $\pm$ 0.006 & 0.055 $\pm$ 0.006 & 0.010 $\pm$ 0.002 & -0.017 $\pm$ 0.023 & 0.907 $\pm$ 0.043 & 5.470 $\pm$ 0.499 & 0.055 $\pm$ 0.004 \\
ANNz2 P(z) & -0.003 $\pm$ 0.003 & -0.003 $\pm$ 0.002 & 0.076 $\pm$ 0.002 & 0.110 $\pm$ 0.007 & 0.053 $\pm$ 0.008 & 0.013 $\pm$ 0.004 & -0.032 $\pm$ 0.041 & 1.570 $\pm$ 0.081 & 6.682 $\pm$ 0.464 & 0.050 $\pm$ 0.005 \\
SkyNet P(z) & 0.004 $\pm$ 0.003 & -0.002 $\pm$ 0.002 & 0.064 $\pm$ 0.002 & 0.110 $\pm$ 0.008 & 0.047 $\pm$ 0.006 & 0.016 $\pm$ 0.004 & 0.017 $\pm$ 0.035 & 1.200 $\pm$ 0.092 & 9.807 $\pm$ 0.709 & 0.079 $\pm$ 0.006 \\
BPZ P(z) & -0.009 $\pm$ 0.005 & -0.019 $\pm$ 0.002 & 0.097 $\pm$ 0.003 & 0.146 $\pm$ 0.008 & 0.048 $\pm$ 0.004 & 0.019 $\pm$ 0.003 & -0.020 $\pm$ 0.067 & 2.463 $\pm$ 0.234 & 6.169 $\pm$ 0.888 & 0.058 $\pm$ 0.007 \\
ZEBRA P(z) & -0.027 $\pm$ 0.005 & -0.019 $\pm$ 0.003 & 0.103 $\pm$ 0.003 & 0.163 $\pm$ 0.008 & 0.054 $\pm$ 0.005 & 0.016 $\pm$ 0.003 & -0.292 $\pm$ 0.075 & 4.247 $\pm$ 0.649 & 8.400 $\pm$ 1.427 & 0.071 $\pm$ 0.007 \\
\end{tabular}
\end{table}
\end{landscape}

\begin{landscape}
\begin{table}
\label{table:test3}
\caption{Results of all the photo-$z$ metrics for all the codes analyzed in Test 3. The errors are computed from bootstrap resampling with 100 samples. The weighting procedure has been applied, together with a cut on the 10\% of the galaxies with the largest estimated photo-$z$ errors for each code.}
\tabcolsep=0.11cm
\centering
\begin{tabular}{l*{11}{c}r}
Test 3 & $\overline{\Delta z}$ & $\Delta z_{50}$ & $\sigma_{68}$ & $\sigma_{\Delta z}$ & ${\rm out}_{2\sigma}$ & ${\rm out}_{3\sigma}$ & $\overline{\Delta z\prime}$ & $\sigma_{\Delta z\prime}$ & ${\rm N}_{\rm poisson}$ & ${\rm KS}$ \\
\hline
DESDM & -0.010 $\pm$ 0.003 & -0.010 $\pm$ 0.002 & 0.091 $\pm$ 0.002 & 0.127 $\pm$ 0.004 & 0.052 $\pm$ 0.005 & 0.015 $\pm$ 0.002 & -0.111 $\pm$ 0.027 & 1.421 $\pm$ 0.047 & 6.819 $\pm$ 0.523 & 0.076 $\pm$ 0.006 \\
ANNz & -0.006 $\pm$ 0.002 & -0.009 $\pm$ 0.002 & 0.085 $\pm$ 0.002 & 0.114 $\pm$ 0.003 & 0.044 $\pm$ 0.004 & 0.017 $\pm$ 0.003 & -0.063 $\pm$ 0.036 & 2.664 $\pm$ 0.078 & 7.867 $\pm$ 0.556 & 0.096 $\pm$ 0.007 \\
TPZ & -0.004 $\pm$ 0.003 & -0.001 $\pm$ 0.002 & 0.076 $\pm$ 0.002 & 0.124 $\pm$ 0.006 & 0.048 $\pm$ 0.004 & 0.021 $\pm$ 0.003 & -0.047 $\pm$ 0.040 & 1.770 $\pm$ 0.067 & 3.812 $\pm$ 0.837 & 0.038 $\pm$ 0.005 \\
RVMz & -0.033 $\pm$ 0.005 & -0.016 $\pm$ 0.003 & 0.118 $\pm$ 0.005 & 0.177 $\pm$ 0.009 & 0.050 $\pm$ 0.004 & 0.019 $\pm$ 0.003 & -0.186 $\pm$ 0.029 & 1.044 $\pm$ 0.059 & 8.100 $\pm$ 0.594 & 0.088 $\pm$ 0.008 \\
NIP-kNNz & -0.039 $\pm$ 0.005 & -0.011 $\pm$ 0.003 & 0.115 $\pm$ 0.004 & 0.191 $\pm$ 0.009 & 0.053 $\pm$ 0.004 & 0.020 $\pm$ 0.002 & -0.270 $\pm$ 0.047 & 1.889 $\pm$ 0.516 & 3.921 $\pm$ 0.372 & 0.068 $\pm$ 0.007 \\
ANNz2 & -0.014 $\pm$ 0.004 & -0.012 $\pm$ 0.002 & 0.101 $\pm$ 0.004 & 0.144 $\pm$ 0.005 & 0.056 $\pm$ 0.004 & 0.019 $\pm$ 0.003 & -0.087 $\pm$ 0.063 & 2.668 $\pm$ 0.203 & 6.517 $\pm$ 0.514 & 0.071 $\pm$ 0.007 \\
BPZ & -0.021 $\pm$ 0.003 & -0.021 $\pm$ 0.002 & 0.098 $\pm$ 0.003 & 0.138 $\pm$ 0.006 & 0.049 $\pm$ 0.004 & 0.018 $\pm$ 0.002 & -0.182 $\pm$ 0.032 & 1.759 $\pm$ 0.073 & 7.895 $\pm$ 0.626 & 0.096 $\pm$ 0.008 \\
LePhare & 0.004 $\pm$ 0.004 & -0.008 $\pm$ 0.003 & 0.111 $\pm$ 0.003 & 0.170 $\pm$ 0.007 & 0.047 $\pm$ 0.003 & 0.024 $\pm$ 0.002 & 1.599 $\pm$ 0.429 & 50.489 $\pm$ 6.968 & 6.371 $\pm$ 0.433 & 0.089 $\pm$ 0.006 \\
PhotoZ & -0.031 $\pm$ 0.003 & -0.032 $\pm$ 0.002 & 0.097 $\pm$ 0.003 & 0.140 $\pm$ 0.006 & 0.057 $\pm$ 0.004 & 0.017 $\pm$ 0.002 & -0.294 $\pm$ 0.020 & 1.003 $\pm$ 0.029 & 4.992 $\pm$ 0.514 & 0.088 $\pm$ 0.007 \\
TPZ P(z) & 0.002 $\pm$ 0.003 & 0.006 $\pm$ 0.002 & 0.076 $\pm$ 0.002 & 0.119 $\pm$ 0.006 & 0.049 $\pm$ 0.005 & 0.020 $\pm$ 0.003 & 0.082 $\pm$ 0.038 & 1.740 $\pm$ 0.071 & 3.321 $\pm$ 0.366 & 0.030 $\pm$ 0.004 \\
ArborZ P(z) & -0.016 $\pm$ 0.003 & -0.011 $\pm$ 0.003 & 0.105 $\pm$ 0.002 & 0.128 $\pm$ 0.005 & 0.053 $\pm$ 0.004 & 0.011 $\pm$ 0.002 & -0.112 $\pm$ 0.023 & 0.953 $\pm$ 0.030 & 4.958 $\pm$ 0.380 & 0.081 $\pm$ 0.006 \\
ANNz2 P(z) & -0.025 $\pm$ 0.004 & -0.015 $\pm$ 0.002 & 0.088 $\pm$ 0.003 & 0.128 $\pm$ 0.007 & 0.049 $\pm$ 0.004 & 0.018 $\pm$ 0.003 & -4.557 $\pm$ 0.712 & 26.898 $\pm$ 2.972 & 7.402 $\pm$ 0.541 & 0.098 $\pm$ 0.008 \\
SkyNet P(z) & 0.007 $\pm$ 0.004 & 0.007 $\pm$ 0.003 & 0.088 $\pm$ 0.002 & 0.129 $\pm$ 0.006 & 0.068 $\pm$ 0.005 & 0.023 $\pm$ 0.003 & 0.071 $\pm$ 0.025 & 0.980 $\pm$ 0.039 & 8.149 $\pm$ 0.865 & 0.093 $\pm$ 0.006 \\
BPZ P(z) & -0.024 $\pm$ 0.003 & -0.024 $\pm$ 0.003 & 0.101 $\pm$ 0.002 & 0.130 $\pm$ 0.006 & 0.046 $\pm$ 0.004 & 0.014 $\pm$ 0.002 & -0.211 $\pm$ 0.033 & 1.734 $\pm$ 0.079 & 5.325 $\pm$ 0.503 & 0.096 $\pm$ 0.007 \\
ZEBRA P(z) & -0.050 $\pm$ 0.005 & -0.030 $\pm$ 0.002 & 0.109 $\pm$ 0.004 & 0.177 $\pm$ 0.012 & 0.043 $\pm$ 0.006 & 0.018 $\pm$ 0.002 & -0.383 $\pm$ 0.047 & 1.906 $\pm$ 0.166 & 6.692 $\pm$ 0.550 & 0.112 $\pm$ 0.008 \\
\end{tabular}
\end{table}
\end{landscape}

\end{document}